\def\jsp{1}
\if0\jsp 
\documentclass[onecolumn,smallextended,final]{svjour2}
\pdfoutput=1
\else
\documentclass[a4paper]{article}
\pdfoutput=1
\usepackage[a4paper,margin=2.4cm]{geometry}
\fi

\usepackage{upgreek}
\usepackage{times}
\usepackage[T1]{fontenc}
\usepackage{calc}
\usepackage{mathptmx}
\usepackage{amssymb}
\usepackage{amsmath}
\usepackage{graphicx}
\usepackage{longtable}
\usepackage[figuresright]{rotating}
\usepackage{placeins}
\usepackage{algorithmic}
\usepackage{ncalgorithmic}
\algsetup{indent=1.5em}
\usepackage{stmaryrd}
\usepackage{tikz}
\usetikzlibrary{shapes.geometric}
\usetikzlibrary{trees}
\usetikzlibrary{arrows}

\if0\jsp

{}

\else

\usepackage{hyperref}
\hypersetup{
    pdfauthor={Nathan Clisby},
    pdftitle={Efficient implementation of the pivot algorithm for self-avoiding walks},
    breaklinks = true,
    colorlinks = true,
    linkcolor = blue,
    anchorcolor = blue,
    citecolor = blue,
    filecolor = blue,
    urlcolor = blue
    }

\fi

\providecommand{\url}[1]{{#1}}
\providecommand{\urlprefix}{URL }
\expandafter\ifx\csname urlstyle\endcsname\relax
  \providecommand{\doi}[1]{DOI~\discretionary{}{}{}#1}\else
  \providecommand{\doi}{DOI~\discretionary{}{}{}\begingroup
  \urlstyle{rm}\Url}\fi

\newcommand{\ptr}{\ensuremath{\hspace{-2.5pt} \shortrightarrow \hspace{-2.5pt}}}

\providecommand{\e}[1]{\ensuremath{\times 10^{#1}}}

\newcommand{\Z}{{\mathbb Z}}

\newcommand\tstrut{\rule{0pt}{2.4ex}}
\newcommand\bstrut{\rule[-1.0ex]{0pt}{0pt}}

\def\walklll{
  \draw[fill] (1,2) circle (3pt); 
  \draw[fill] (1,1) circle (3pt);
  \draw (1,2) -- (1,1);
}

\def\walkllr{
  \draw[fill] (2,1) circle (3pt); 
  \draw[fill] (3,1) circle (3pt);
  \draw (2,1) -- (3,1);
}

\def\walklrl{
  \draw[fill] (3,0) circle (3pt); 
  \draw[fill] (3,-1) circle (3pt);
  \draw (3,0) -- (3,-1);
}

\def\walklrr{
  \draw[fill] (4,-1) circle (3pt); 
  \draw[fill] (5,-1) circle (3pt);
  \draw (4,-1) -- (5,-1);
}

\def\walkrll{
  \draw[fill] (5,0) circle (3pt); 
  \draw[fill] (5,1) circle (3pt);
  \draw (5,0) -- (5,1);
}

\def\walkrlr{
  \draw[fill] (4,1) circle (3pt); 
  \draw[fill] (4,2) circle (3pt);
  \draw (4,1) -- (4,2);
}

\def\walkrrl{
  \draw[fill] (3,2) circle (3pt); 
  \draw[fill] (3,3) circle (3pt);
  \draw (3,2) -- (3,3);
}

\def\walkrrr{
  \draw[fill] (4,3) circle (3pt); 
  \draw[fill] (5,3) circle (3pt);
  \draw (4,3) -- (5,3);
}

\def\walkll{
\walklll
\walkllr
\draw (1,1)--(2,1);
}

\def\walklr{
\walklrl
\walklrr
\draw (3,-1)--(4,-1);
}

\def\walkrl{
\walkrll
\walkrlr
\draw (5,1)--(4,1);
}

\def\walkrr{
\walkrrl
\walkrrr
\draw (3,3)--(4,3);
}

\def\walkl{
\walkll
\walklr
\draw (3,1)--(3,0);
}

\def\walkr{
\walkrl
\walkrr
\draw (4,2)--(3,2);
}

\def\walk{
\walkl
\walkr
\draw (5,-1)--(5,0);
}

\begin{document}

\title{Efficient implementation of the pivot algorithm for
  self-avoiding walks}
\if0\jsp
\journalname{Journal of Statistical Physics}
\author{Nathan Clisby}
\institute{ARC Centre of Excellence for Mathematics
  and Statistics of Complex Systems,  
  Department of Mathematics and Statistics, 
  The University of Melbourne, Victoria 3010, Australia. \\
  \email{n.clisby@ms.unimelb.edu.au}}

\else

\author{Nathan Clisby \\
ARC Centre of Excellence for Mathematics and Statistics of Complex Systems,\\
Department of Mathematics and Statistics, \\
The University of Melbourne, Victoria 3010, Australia. \\
n.clisby@ms.unimelb.edu.au}

\fi

\date{May 7, 2010}

\maketitle

\begin{abstract}
The pivot algorithm for self-avoiding walks has been implemented
in a manner which is dramatically faster than previous
implementations, enabling
extremely long walks to be efficiently simulated.
We explicitly describe
the data structures and algorithms used, and
provide a heuristic argument that the mean time per attempted pivot
for $N$-step self-avoiding walks is $O(1)$ for the square and simple
cubic lattices. Numerical experiments conducted for self-avoiding
walks with up to 
268 million steps are consistent with $o(\log N)$ behavior for the square
lattice and $O(\log N)$ behavior for the simple cubic lattice.
Our method
can be adapted to other models of polymers with 
short-range interactions, on the lattice or in the continuum, and hence
promises to be widely useful.
\if0\jsp
\else
\fi
\end{abstract}

\section{Introduction and results}
\label{sec:introduction}

The self-avoiding walk (SAW) model is an important model in
statistical physics~\cite{Madras1993}. It models the excluded-volume
effect observed in real polymers, and exactly captures universal features
such as critical exponents and amplitude ratios. It is also an
important model in the study of 
critical phenomena, as it is the $n \rightarrow 0$ limit of
the $n$-vector model, which includes the Ising model ($n = 1$) as
another instance. 
Indeed, one can straightforwardly simulate SAWs in
the infinite volume limit, which makes this model particularly
favorable for the calculation of critical parameters.
Exact results are known for self-avoiding walks in two
dimensions~\cite{Nienhuis1982,Lawler2004} and
for $d \geq 4$ (mean-field behavior has been proved for $d \geq
5$~\cite{Hara1992}), but not for the most physically interesting
case of $d=3$. 

The pivot algorithm is a powerful and oft-used approach to the
study of self-avoiding walks, invented by
Lal~\cite{Lal1969} and later elucidated and popularized by Madras and
Sokal~\cite{Madras1988}. 
The pivot algorithm uses pivot moves as the transitions in a Markov
chain which proceeds as follows.
From an initial SAW of length $N$, such as a straight rod, new
$N$-step walks
are successively
generated by choosing
a site of the walk at random, and attempting to apply a lattice
symmetry operation,
or pivot, to one of the parts of the walk; if the
resulting walk is self-avoiding the move is
accepted, otherwise the move is rejected and the original walk is
retained.
Thus a Markov chain is formed in the
ensemble of SAWs of fixed length;
this chain satisfies detailed balance and is ergodic,
ensuring that SAWs are sampled uniformly at random. 

One typical use of the pivot algorithm is to calculate observables
which characterize the size of the 
SAWs: the squared end-to-end distance $R_{\mathrm{e}}^2$,
the squared radius of gyration $R_{\mathrm{g}}^2$, and the mean-square
distance of a
monomer from its endpoints $R_{\mathrm{m}}^2$.
To leading order we expect the mean values of
these observables over all 
SAWs of $N$ steps, with each SAW is given equal weight, to be $\langle
R_x^2 \rangle_N \sim D_x N^{2\nu}$ ($x \in \{\mathrm{e,g,m}\}$), with
$\nu$ a universal critical exponent. 

For $N$-step SAWs, the implementation of
the pivot algorithm due to Madras and Sokal has estimated mean time
per attempted pivot of $O(N^{0.81})$ 
on $\Z^2$ and $O(N^{0.89})$ on $\Z^3$; performance was significantly
improved by Kennedy~\cite{Kennedy2002a} to $O(N^{0.38})$ and
$O(N^{0.74})$ respectively.

In this article, we give a detailed description of a new
data structure we call the SAW-tree. This data structure allows us to
 implement the pivot 
algorithm in a highly efficient manner: we present a heuristic
argument that the mean time
per attempted pivot is $O(1)$ on $\Z^2$ and $\Z^3$, and numerical
experiments which show that for walks of up to $N = 2^{28} - 1 \approx
2.7\e{8}$ steps the algorithmic complexity is well approximated by $O(\log N)$.
This improvement enables the rapid simulation of walks
with many millions of steps. 

In a companion article~\cite{Clisby2010}, we describe the
algorithm in 
general terms, and demonstrate the power of the method by applying it
to the problem of calculating the critical exponent $\nu$ for
three-dimensional 
self-avoiding walks.

Thus far the SAW-tree has been implemented for $\Z^2$, $\Z^3$, and $\Z^4$, but
it can be straightforwardly adapted to other lattices and the continuum,
as well as polymer models with short-range interactions.
Other possible extensions would be to allow for branched polymers,
confined polymers, or simulation of polymers in solution.

We intend to implement the SAW-tree and associated methods as an open
source software library for use by researchers in the field of polymer
simulation. 

\subsection{Pivot algorithm}
\label{sec:introductionpivot}

Madras and Sokal~\cite{Madras1988} demonstrated, through strong
heuristic arguments
and numerical experiments, that the pivot algorithm results in a Markov chain
with short integrated autocorrelation time for global observables.
The pivot algorithm is far more efficient than
Markov chains which utilize local moves; see
\cite{Madras1988,Li1995,Sokal1994,Sokal1996} for 
detailed discussion.

The implementation of the pivot algorithm by Madras and Sokal utilized
a hash table to record 
the location of each site of the walk. They showed that for $N$-step
SAWs the probability of a pivot move being accepted is $O(N^{-p})$,
with $p$
dimension-dependent but close to zero ($p \lesssim 0.2$). As accepted
pivots typically result in a large change in global observables such as
$R_{\mathrm{e}}^2$, this leads to the conclusion that the pivot
algorithm has integrated autocorrelation time $O(N^p)$, with possible
logarithmic corrections. In
addition, they argued convincingly that the CPU time per successful
pivot is $O(N)$ for their implementation. Throughout this article we
work with the mean time per attempted pivot, $T(N)$, which for the
Madras and Sokal implementation is $O(N^{1-p})$.

Madras and Sokal argued that $O(N)$ per successful pivot is best
possible because it takes 
time $O(N)$ to merely write down an $N$-step SAW.
Kennedy~\cite{Kennedy2002a}, however, recognized that it is \emph{not}
necessary to 
write down the SAW for each successful pivot, and developed a data
structure and algorithm which cleverly utilized 
geometric constraints to break the $O(N)$
barrier. In this paper, we develop methods which further improve the
use of geometric constraints to obtain a highly efficient
implementation of the pivot algorithm.

\subsection{Results}
\label{sec:introductionresults}

We have efficiently implemented the pivot algorithm
via a data structure we call the SAW-tree, which
allows rapid Monte Carlo simulation of SAWs with millions of
steps. 
This new implementation can also be adapted to other models of
polymers with short-range interactions, on the lattice and in the
continuum, and hence promises to be widely useful.

The heart of our implementation of the algorithm involves performing
intersection tests between 
``bounding boxes'' of different sub-walks when a pivot is attempted.
In \cite{Clisby2010} we generated large samples of walks with up to
$2^{25} - 1 \approx 3.3\e{7}$ steps, but for the purpose of
determining the complexity of our algorithm we have also
generated smaller samples of walks of up to $2^{28} - 1 \approx
2.7\e{8}$ steps. 
For $N = 2^{28}-1$, the mean number of intersection tests needed per attempted
pivot is remarkably low: 39 for $\Z^2$, 158 for $\Z^3$, and 449 for $\Z^4$. 

In Sec.~\ref{sec:complexity} we present heuristic arguments 
for the asymptotic behavior of the mean time per attempted pivot for
$N$-step SAWs, $T(N)$, and test these predictions with computer
experiments for $N \leq
2.7\e{8}$. We summarize our results in
Table~\ref{tab:performance}; note that $O(f(N))$ indicates $T$ is
bounded above by $f$ asymptotically, $o(f(N))$ indicates $f$
dominates $T$, $\upomega(f(N))$ indicates $T$ dominates $f$, and
$\Theta(f(N))$ indicates $f$ bounds $T$ both above and below. 
For comparison, we
also give the algorithmic complexity of the implementations of Madras
and Sokal~\cite{Madras1988}, and Kennedy~\cite{Kennedy2002a}. 
In Sec.~\ref{sec:complexityhighdim}, we develop an argument for the
complexity of our algorithm on $\Z^4$; this same argument leads to an
estimate 
for the performance of the implementation of Madras 
  and Sokal on $\Z^4$. We do not know the complexity of
  Kennedy's implementation for $\Z^4$ and $\Z^d$ with $d > 4$, but we suspect
  it is $O(N^q)$ with $0.74 < q < 1$, with possible logarithmic
  corrections.
\begin{table}[!ht]
\caption{$T(N)$, the mean time per attempted pivot for $N$-step
  SAWs. A tighter bound for $\Z^4$ is reported in
  Sec.~\ref{sec:complexityhighdim}, but this relies on an untested assumption.}
\label{tab:performance}
\begin{center}
\begin{tabular}{ccccc}
\hline\hline 
Lattice  & Madras and Sokal & Kennedy & \multicolumn{2}{c}{This work \tstrut} \\
  & & & Predicted & Observed \bstrut \\
\hline
$\Z^2$ & $O(N^{0.81})$ & $O(N^{0.38})$ & $O(1)$ & $o(\log N)$ \tstrut\\
$\Z^3$ & $O(N^{0.89})$ & $O(N^{0.74})$ & $O(1)$ & $O(\log N)$ \\
$\Z^4$ & $o(N)$ & ? & $o(\log N)$ & $\upomega(\log N)$\\
$\Z^d$, $d > 4$ & $O(N)$ & ? & $\Theta(\log N)$ & ? \bstrut\\
\hline\hline
\end{tabular}
\end{center}
\end{table}

Our implementation is also fast in
practice: 
for simulations of walks of length $2^{20} \approx 10^6$ on $\Z^3$, our
implementation is
almost 400 times faster when compared with Kennedy's, and close to
four thousand times faster
when compared with that of
Madras and Sokal. We have measured $T(N)$ for each implementation over
a wide range of $N$ on $\Z^2$, $\Z^3$, and $\Z^4$, and report these
results in Sec.~\ref{sec:comparison}.

\subsection{Outline of paper}
\label{sec:introductionoutline}

In Sec.~\ref{sec:implementation}, we give a detailed description of
the SAW-tree data structure and associated methods which are
required for implementing the pivot algorithm.

In Sec.~\ref{sec:complexity} we present heuristic arguments 
that $T(N)$ for
self-avoiding walks on $\Z^2$ and $\Z^3$ is $O(1)$, and numerical
evidence which shows that for walks of up to $N = 2^{28} - 1 \approx
2.7\e{8}$ steps $T(N)$ is $o(\log N)$ for $\Z^2$ and $O(\log N)$ for $\Z^3$.
We also discuss
the behavior of our implementation for higher dimensions.

In
Sec.~\ref{sec:initialization} we discuss initialization of the Markov
chain, including details of how many data points are discarded. We
also explain why it is highly desirable to 
have a procedure such as \textbf{Pseudo\-\_dimerize} for
initialization (pseudo-code in 
Sec.~\ref{sec:implementationhigh}) when studying very long walks, and
show that the 
expected running time of \textbf{Pseudo\-\_dimerize} is $\Theta(N)$.

In Sec.~\ref{sec:autocorrelation} we discuss the autocorrelation
function for
the pivot algorithm, and show that the batch method for estimating 
confidence intervals is accurate, provided the batch size is large
enough. 
This confirms the accuracy of the confidence intervals for our data
published in \cite{Clisby2010}. 

Finally, in Sec.~\ref{sec:comparison} we compare the performance of 
our implementation
with previous implementations of the pivot
algorithm~\cite{Madras1988,Kennedy2002a}. We show that the SAW-tree
implementation is not only dramatically faster for long walks, it is
also faster than the other implementations for walks with
as few as 63 steps.

\section{Implementation details}
\label{sec:implementation}

Self-avoiding walks (SAWs) are represented as binary trees (see
e.g. \cite{Sedgewick1998Books}) via a recursive definition;
we describe here the SAW-tree data structure and associated methods using
pseudo-code.

These methods can be extended to include translations,
splitting of walks, joining of walks, and testing for intersection
with surfaces. Indeed, for SAW-like models (those with short range
interactions), it should be possible to implement a wide variety of
global moves and tests for SAWs of $N$ steps in time $O(\log N)$ or
better. It is also
possible to parallelize code by, for example, performing intersection
testing for a variety of proposed pivot moves
simultaneously. Parallelization of the basic operations is also
possible, but would be considerably more difficult to implement.

In this section we give precise
pseudo-code definitions of the data structure and algorithms.
For reference, R-trees~\cite{Guttman1984} and
bounding volume hierarchies (see
e.g. \cite{Klosowski1998}) are data structures which arise in the
field of computational geometry which are related to the SAW-tree.

\subsection{The self-avoiding walk}
\label{sec:implementationsaw}

For self-avoiding walks, the self-avoidance condition is enforced on
sites rather than bonds, and this means that the SAW-tree is naturally
defined in terms of sites.
This representation also has the advantage that the basic objects, sites,
have physical significance as they correspond to the monomers in a polymer.
The only consequences of this choice are notational: a SAW-tree of $n$
sites has $n-1$ steps. We adopt this notation for the remainder of
this section. When discussing the complexity of various algorithms we
will still use $N$ rather than $n$ in order to be consistent with the companion
article and other sections of the present work.

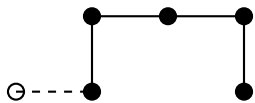
\begin{figure}[!ht]
\begin{center}
  \begin{tikzpicture}[
      thick
    ]
    \draw (0,0) circle (3pt);
    \draw[fill] (1,0) circle (3pt);
    \draw[fill] (1,1) circle (3pt);
    \draw[fill] (2,1) circle (3pt);
    \draw[fill] (3,1) circle (3pt);
    \draw[fill] (3,0) circle (3pt);
    \draw[dashed] (0,0) -- (1,0);
    \draw (1,0) -- (1,1) -- (2,1) -- (3,1) -- (3,0);
  \end{tikzpicture}
\end{center}
\caption{A self-avoiding walk of 5 sites, which we will refer to as $\omega_a$.}
\label{fig:example}
\end{figure}

An $n$-site SAW on $\Z^d$ is a mapping
$\omega : \{0,1,\ldots, n-1\} \to \Z^d$ with $|\omega(i+1)-\omega(i)|=1$
for each $i$
($|x|$ denotes the Euclidean norm of $x$),
and with $\omega(i) \neq \omega(j)$ for all $i \neq j$.
SAWs may be either rooted or unrooted; our convention is that the SAWs
are rooted at the site which is at $\hat{\mathbf{x}}_1$ (unit vector
in the first coordinate
direction), i.e $\omega(0) = \hat{\mathbf{x}}_1$. This convention
simplifies some of the algebra involved in
merging sub-walks, and is represented visually, e.g. in
Fig.~\ref{fig:example}, 
by indicating a dashed
bond from the origin to the first site of the walk.

We denote the group of symmetries of $\Z^d$ as $\mathrm{G}_d$, which
corresponds to the 
dihedral group for $d=2$, and the octahedral group for $d=3$. This
group acts on coordinates by permuting any of the $d$ coordinate
directions ($d!$ choices), and independently choosing the orientation
of each of these coordinates ($2^d$ choices); thus $\mathrm{G}_d$ has
$2^d d!$ elements.
The group of lattice symmetries for $\Z^3$ therefore has 48 elements, and we
use all of them except the identity as potential pivot operations;
other choices are possible.
We can
represent the symmetry group elements as $d \times d$ orthogonal
matrices, and the symmetry group elements act on the coordinates
written as column vectors.

We also define the (non-unique) pivot sequence representation of a self-avoiding
random walk on $\Z^d$ as  
a mapping from the integers to $\mathrm{G}_d$, 
$q_{\omega} : \{0,1,\ldots, n-1\} \to \mathrm{G}_d$. 
The sequence elements $q(i)$ represent changes in the symmetry
operator from 
site $i-1$ to site $i$, while
$q_{\mathrm{abs}}(i)$ represent absolute symmetry operations, i.e. relative to
the first site of the walk.
We can
relate this to the previous definition of a self-avoiding walk in
terms of sites via the recurrence relations 
\begin{align}
q_{\mathrm{abs}}(i) &= q_{\mathrm{abs}}(i-1) q(i), \label{eq:qrecurrence} \\
\omega(i) &= \omega(i-1) + q_{\mathrm{abs}}(i)
\hat{\mathbf{x}}_1, \label{eq:omegarecurrence} 
\end{align}
with $1 \leq i \leq n-1$, and initial conditions $q(0) = I$,
$q_{\mathrm{abs}}(0) = I$, and $\omega(0) = \hat{\mathbf{x}}_1$. 

As noted by Madras and Sokal (footnote 10, p132 in
\cite{Madras1988}), for the pivot sequence representation it 
is possible to perform a pivot of 
the walk in time $O(1)$ by choosing a site $i$ uniformly at random,
and multiplying 
$q(i)$ by a (random) symmetry 
group element.  However, the pivot sequence representation does no
better than the hash table implementation of Madras and Sokal if we
wish to determine if this change results in a
self-intersection, or if we wish to calculate global observables such
as $R_{\mathrm{e}}^2$ for the updated walk.

Forgetting for the moment the self-avoidance condition, and using the
fact that $\mathrm{G}_d$ has $2^dd!$ elements, we see that for random
walks of $n$ sites there are 
$(2^dd!)^{n-1}$ possible pivot sequences, while there are only
$(2d)^{n-1}$ random walks. This suggests that each random walk is
represented by $(2^{d-1}(d-1)!)^{n-1}$ pivot sequences. This can be
derived directly by noting that 
given a pivot sequence $q(1), q(2), q(3), \cdots, q(k), q(k+1), \cdots,
q(n-1)$, we can insert a pivot $q$ which preserves the vector
$\hat{\mathbf{x}}_1$, between two elements $q(k)$ and $q(k+1)$ as
follows 
\begin{displaymath}
q(0), q(1), q(2), q(3), \cdots, q(k) q, q^{-1}q(k+1), \cdots, q(n-1),
\end{displaymath} 
without altering the
walk. The number of 
symmetry group elements which preserve $\hat{\mathbf{x}}_1$ is $2^d
d!/(2d) = 2^{d-1} (d-1)!$, and there are $n-1$ locations where these
symmetry group elements can be inserted, leading to
$(2^{d-1}(d-1)!)^{n-1}$ equivalent pivot representations for a random
walk of $n$ sites. For $d=2$, given $\omega(i)$ the
recurrence relations in Eqs.~\ref{eq:qrecurrence} and
\ref{eq:omegarecurrence} only fix one of the two non-zero elements in
$q(i)$, leaving the choice of sign for the other non-zero element free.
For our example walk $\omega_a$ we have 
\begin{align}
\omega_a &= \left((1,0), (1,1), (2,1), (3,1), (3,0) \right).
\end{align}
We give three of the 16 equivalent choices for the pivot representation of
$\omega_a$, the first
involving only
proper rotations, the second with improper rotations for $q(i)$
with $1 \leq i \leq 4$, and the third with proper and
improper rotations alternating:
\begin{align}
q^{(1)}_{\omega_a} &= \left(\left(\begin{array}{rr} 1 & 0 \\ 0 &
  1 \end{array}\right),
\left(\begin{array}{rr} 0 & -1 \\ 1 & 0 \end{array}\right),
\left(\begin{array}{rr} 0 & 1 \\ -1 & 0 \end{array}\right),
\left(\begin{array}{rr} 1 & 0 \\ 0 & 1 \end{array}\right),
\left(\begin{array}{rr} 0 & 1 \\ -1 & 0 \end{array}\right)\right), \\
q^{(2)}_{\omega_a} &= \left(\left(\begin{array}{rr} 1 & 0 \\ 0 & 1 \end{array}\right),
\left(\begin{array}{rr} 0 & 1 \\ 1 & 0 \end{array}\right),
\left(\begin{array}{rr} 0 & 1 \\ 1 & 0 \end{array}\right),
\left(\begin{array}{rr} 1 & 0 \\ 0 &-1 \end{array}\right),
\left(\begin{array}{rr} 0 & 1 \\ 1 & 0 \end{array}\right)\right),\\
q^{(3)}_{\omega_a} &= \left(\left(\begin{array}{rr} 1 & 0 \\ 0 & 1 \end{array}\right),
\left(\begin{array}{rr} 0 & 1 \\ 1 & 0 \end{array}\right),
\left(\begin{array}{rr} 0 & -1 \\ 1 & 0 \end{array}\right),
\left(\begin{array}{rr} 1 & 0 \\ 0 &-1 \end{array}\right),
\left(\begin{array}{rr} 0 & 1 \\ -1 & 0 \end{array}\right)\right).
\end{align}

The non-uniqueness of the pivot representation for SAWs is due to the
fact that the monomers (occupied sites) are invariant under the
symmetry group $\mathrm{G}_d$, i.e. it is not possible to distinguish
the different orientations of a single site. 
The non-uniqueness is of no practical concern, but perhaps hints that
it may be possible to derive a more succinct and elegant
representation of walks than the mapping to $\mathrm{G}_d$ defined
here. 

The merge operation is the fundamental operation on SAWs which allows
for the binary tree 
data structure we call the SAW-tree. This is related to the
concatenation operation defined, for example, in Sec.~1.2 of
\cite{Madras1993}; 
for concatenation the number
of bonds is conserved, whereas for the merge operation the number of
sites is preserved.
Merging two SAWs with $n$ and $m$ sites respectively results in a SAW
with $n+m$ sites.
It is convenient to also include a pivot 
operation, $q$, when merging the walks, and the result of merging two
walks  $\omega_1 =
(\omega_1(0), \omega_1(1), \cdots,
\omega_1(n-1))$ and $\omega_2 =
(\omega_2(0), \omega_2(1), \cdots, \omega_2(m-1))$ is 
\begin{align}
\text{merge}(\omega_1, q, \omega_2) &=
(\omega_1(0), \omega_1(1), \cdots, \omega_1(n-1), \omega_1(n-1) +
q \omega_2(0), \nonumber \\
& \omega_1(n-1) + q \omega_2(1), \cdots, \omega_1(n-1)
+ q \omega_2(m-1)).
\label{eq:merge}
\end{align}

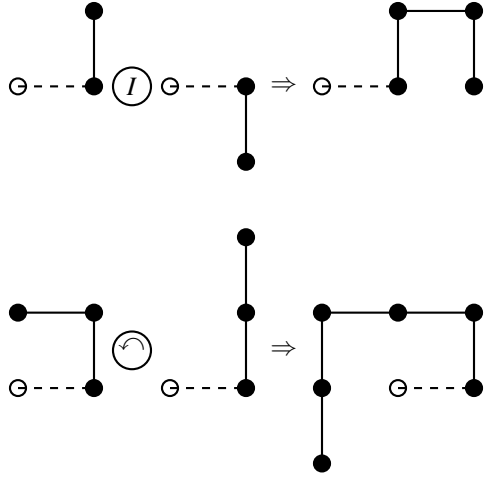
\begin{figure}[!ht]
\begin{center}
  \begin{tikzpicture}[
      thick,
      q/.style={circle,minimum size=5mm,inner sep=0.5pt,draw}
    ]
    \draw (0,0) circle (3pt);
    \draw[fill] (1,0) circle (3pt);
    \draw[fill] (1,1) circle (3pt);
    \draw[dashed] (0,0) -- (1,0);
    \draw (1,0) -- (1,1);

    \draw (1.5,0) node[q] {$I$};

    \draw (2,0) circle (3pt);
    \draw[fill] (3,0) circle (3pt);
    \draw[fill] (3,-1) circle (3pt);
    \draw[dashed] (2,0) -- (3,0);
    \draw (3,0) -- (3,-1);

    \draw (3.5,0.0) node {$\Rightarrow$};

    \draw (4,0) circle (3pt);
    \draw[fill] (5,0) circle (3pt);
    \draw[fill] (5,1) circle (3pt);
    \draw[fill] (6,1) circle (3pt);
    \draw[fill] (6,0) circle (3pt);
    \draw[dashed] (4,0) -- (5,0);
    \draw (5,0) -- (5,1) -- (6,1) -- (6,0);

    \begin{scope}[yshift=-4.0cm]
      \draw (0,0) circle (3pt);
      \draw[fill] (1,0) circle (3pt);
      \draw[fill] (1,1) circle (3pt);
      \draw[fill] (0,1) circle (3pt);
      \draw[dashed] (0,0) -- (1,0);
      \draw (1,0) -- (1,1) -- (0,1);

      \draw (1.5,0.5) node[q] {$\phantom{I}$};
      \draw (1.5,0.58) node {$\curvearrowleft$};

      \draw (2,0) circle (3pt);
      \draw[fill] (3,0) circle (3pt);
      \draw[fill] (3,1) circle (3pt);
      \draw[fill] (3,2) circle (3pt);
      \draw[dashed] (2,0) -- (3,0);
      \draw (3,0) -- (3,1) -- (3,2);

      \draw (3.5,0.5) node {$\Rightarrow$};

      \draw (5,0) circle (3pt);
      \draw[fill] (6,0) circle (3pt);
      \draw[fill] (6,1) circle (3pt);
      \draw[fill] (5,1) circle (3pt);
      \draw[fill] (4,1) circle (3pt);
      \draw[fill] (4,0) circle (3pt);
      \draw[fill] (4,-1) circle (3pt);
      \draw[dashed] (5,0) -- (6,0);
      \draw (6,0) -- (6,1) -- (5,1) -- (4,1) -- (4,0) -- (4,-1);
      \end{scope}
  \end{tikzpicture}
\end{center}
\caption{Examples of the merge operation on SAWs: the open
  circle (empty site) of the right-hand sub-walks are fixed to the ends
  of the corresponding left-hand sub-walks, and the symmetry is then
  applied to the right-hand sub-walk. At the top,
  an identity symmetry operation is applied, while at the
  bottom the symmetry operation is a $180^\circ$ rotation.}
\label{fig:merge}
\end{figure}

The merge operation is represented visually in
Fig.~\ref{fig:merge}. To merge two sub-walks, pin the open circle of 
the left-hand sub-walk to
the origin, and then pin the open circle of the right-hand sub-walk to the
tail end of the left-hand sub-walk. Finally,
apply the symmetry $q$ to the right-hand sub-walk, using the second pin as the
pivot.

\subsection{Definitions}
\label{sec:implementationdefinitions}

Here we define various quantities which are necessary for implementing
our data structure and for calculating observables such as the
mean-square end-to-end distance, $R_{\mathrm{e}}^2$.

We first define various quantities which will be used to calculate
observables which measure the size of a walk:
\begin{align}
\mathbf{X}_{\mathrm{e}}(\omega) &\equiv \omega(n-1) & \text{(vector);} \\
\mathbf{X}(\omega) & \equiv \sum_{i=0}^{n-1} \omega(i) & \text{(vector);}\\
X_2(\omega) & \equiv \sum_{i=0}^{n-1} \omega(i)
\cdot \omega(i) & \text{(scalar).}
\end{align}

A bounding box of a walk is a convex shape 
which completely contains the walk.
The obvious choice of shape for $\Z^d$ is the rectangular prism
with faces formed from the coordinate planes $\mathbf{x}_i =
\mathrm{const}$, $1\leq i \leq d$, with the constants chosen so that
the faces of the prism touch the walk, i.e. the bounding box has
minimum extent. Other choices are possible, e.g. other planes can be
used such as $\mathbf{x}_i \pm \mathbf{x}_j = \mathrm{const}$, $1\leq
i < j \leq d$, and have the advantage of matching the shape of the
walk more closely, but at the expense of more computational overhead
and memory consumption. With closer fitting bounding boxes,
fewer intersection tests need to be performed to ascertain whether two
walks intersect. However, in practice, the
coordinate plane rectangular prism implementation was fastest on our
computer hardware (by a narrow margin), and has the benefit that
it is straightforward to implement.
The choice of bounding box for continuum models is not as obvious;
possibilities include spheres and
oriented rectangular prisms.

We note that the choice of bounding box shape determines the 
maximum number of sites, $b$, a SAW can have so that it is guaranteed that
its bounding box contains the sites of the SAW and no others. 
Suppose we are given two SAWs for which the bounding boxes overlap: if
each of the 
walks has $b$ or fewer sites,
we can be certain that the two walks intersect, while if at least
one of the walks has more than $b$ sites, it may be that the
walks do not intersect.
The value of $b$ determines the cut-off for intersection testing for
the function \textbf{Intersect} in Sec.~\ref{sec:implementationuser}.
For
$\Z^d$ with $d \geq 2$, the bounding box with faces formed from
the coordinate planes leads to the maximum number of sites being two,
as there are counter-examples with three sites (e.g. the bounding box
of $\omega =
\left((0,0), (1,0), 
 (1,1)\right)$ also contains $(0,1)$). For the bounding box with the faces
being the coordinate 
planes and $\mathbf{x}_i \pm \mathbf{x}_j = \mathrm{const}$, the
maximum number of sites is three (as the bounding box of
$\omega = \left((0,0), (1,0),  (2,0), (2,1)\right)$ 
also contains $(1,1)$). It is possible to push this one step
further so 
that the maximum number of sites is four, but five is not possible as 
we can see that $\omega_a$ in
Fig.~\ref{fig:example} has five sites, and an
unvisited site on its convex hull, which must also therefore
be interior to any bounding box.

We write bounding boxes as a product of closed intervals, in the form
$B(\omega) = \times [ \inf{\mathbf{x}}_i : {\mathbf{x}} \in\omega,
    \sup{\mathbf{x}}_i : {\mathbf{x}} \in\omega]$, where the product
is taken over $1 \leq i 
\leq d$. Consider a walk $\omega$, with bounding box $B$, which is
split into left- and 
right-hand sub-walks, $\omega^l = \left(\omega(0),\cdots,\omega(k-1)\right)$ and $\omega^r =
\left(\omega(k),\cdots,\omega(n-1)\right)$,
with bounding boxes $B^l=\times[a_i,b_i]$, and
$B^r=\times[c_i,d_i]$ respectively.
We can then
define the union operation on bounding boxes,
\begin{align}
B &= B^l \cup B^r \nonumber \\
&\equiv \times \left([a_i,b_i]\cup[c_i,d_i]\right)
\nonumber \\
&= \times [\inf\{a_i,c_i\},\sup\{b_i,d_i\}].
\end{align}
The intersection operation is defined as
\begin{align}
B^l \cap B^r &= \times \left([a_i,b_i]\cap[c_i,d_i]\right) \nonumber \\
&= \times [\sup\{a_i,c_i\},\inf\{b_i,d_i\}].
\end{align}
There is no guarantee that $\sup\{a_i,c_i\} \leq \inf\{b_i,d_i\}$, and
we adopt the convention that an interval $[e,f]$ is considered empty
if $e > f$. If any interval is empty, then the corresponding bounding
box is also empty as it contains no interior sites.
A quantity associated with the bounding box which we will find useful 
is the sum of the dimensions of the bounding box, \textbf{Perim}.
If $B = \times[a_i,b_i]$, then we define
\begin{align}
  \textbf{Perim}(B) &= \sum_{i=1}^d (b_i - a_i + 1).
  \label{eq:perim}
\end{align}

For $\omega_a$ (in Fig.~\ref{fig:example}) we have
the following 
values for the various parameters:
\begin{align}
n &= 5; \\
\mathbf{B}(\omega_a) &= [1,3]\times[0,1]; \\
\mathbf{X}_{\mathrm{e}}(\omega_a) &= (3,0); \\
\mathbf{X}(\omega_a) &= (1,0) + (1,1) + (2,1) + (3,1) + (3,0) \nonumber \\
&= (10,3);\\
X_2(\omega_a) &= (1,0)\cdot(1,0) + (1,1)\cdot(1,1) + (2,1)\cdot(2,1) \nonumber \\
&\mathrel{\phantom{=}} + (3,1)\cdot(3,1) + (3,0)\cdot(3,0) \nonumber \\
&= 1 + 2 + 5 + 10 + 9 \nonumber \\
&= 27.
\end{align}

The observables $R_x^{2k}$, with $x \in \{\mathrm{e,g,m}\}, 1 \leq k
\leq 5$, may be straightforwardly calculated from
$\mathbf{X}_{\mathrm{e}}$, $\mathbf{X}$, and $X_2$.
We give expressions for  $R_x^{2}$ with $x
\in \{\mathrm{e,g,m}\}$, and note that higher Euclidean-invariant
moments can be obtained 
via $R_x^{2k} = \left(R_x^2\right)^k$ ($2 \leq k \leq 5$) (these
moments are calculated for $\Z^2$ in
\cite{Caracciolo2005} and for $\Z^3$ in \cite{Clisby2010}). In addition
we introduce another observable, $\mathcal{R}_{\mathrm{m}}^2$, which
measures the mean-square deviation of the walk from the endpoint
$\omega(n-1)$. 
{ \allowdisplaybreaks
\begin{align}
R_{\mathrm{e}}^2 &= |\omega(n-1)-\omega(0)|^2 \nonumber \\
 &= \left(\mathbf{X}_{\mathrm{e}} - \hat{\mathbf{x}}_1\right) \cdot
  \left(\mathbf{X}_{\mathrm{e}} - \hat{\mathbf{x}}_1\right) \\
R_{\mathrm{g}}^2 &=
\frac{1}{2n^2}\sum_{i,j=0}^{n-1}|\omega(i)-\omega(j)|^2 \nonumber \\
&= \frac{1}{n} X_2 - \frac{1}{n^2} \mathbf{X} \cdot \mathbf{X} \\
R_{\mathrm{m}}^2 &= \frac{1}{2n}\sum_{i=0}^{n-1}\left[|\omega(i)-\omega(0)|^2 +
  |\omega(i)-\omega(n-1)|^2 \right] \nonumber \\
&= \frac{1}{2} + \frac{1}{2}\mathbf{X}_{\mathrm{e}} \cdot \mathbf{X}_{\mathrm{e}}
 - \frac{1}{n} \hat{\mathbf{x}}_1 \cdot \mathbf{X}
 - \frac{1}{n} \mathbf{X}_{\mathrm{e}} \cdot \mathbf{X}
 + \frac{1}{n} X_2 \\
\mathcal{R}_{\mathrm{m}}^2 &= \frac{1}{n}\sum_{i=0}^{n-1}
  |\omega(i)-\omega(n-1)|^2 \nonumber \\
 &= \mathbf{X}_{\mathrm{e}} \cdot \mathbf{X}_{\mathrm{e}}
 - \frac{2}{n} \mathbf{X}_{\mathrm{e}} \cdot \mathbf{X}
 + \frac{1}{n} X_2
\label{eq:calrm}
\end{align}
}
In~\cite{Clisby2010}, we chose to calculate
$\mathcal{R}_{\mathrm{m}}^2$ rather than 
$R_{\mathrm{m}}^2$, as it has a slightly simpler 
expression, and relied on the identity
$\langle \mathcal{R}_{\mathrm{m}}^2 \rangle = \langle
R_{\mathrm{m}}^2 \rangle$. Compared with $R_{\mathrm{m}}^2$,
$\mathcal{R}_{\mathrm{m}}^2$ has 
larger variance but smaller integrated autocorrelation time (for the
pivot algorithm). Before performing the computational experiment
in~\cite{Clisby2010}, we 
believed that given the same number of pivot attempts the
confidence intervals for $\langle \mathcal{R}_{\mathrm{m}}^2 \rangle$ and
$\langle R_{\mathrm{m}}^2 \rangle$ would be comparable.
We have since confirmed that working directly with
$R_{\mathrm{m}}^2$ results in a standard error which is of the order
of 17\% smaller for $d=3$, an amount which is not negligible; in
future experiments we will calculate $R_{\mathrm{m}}^2$ directly.

\subsection{Guide to the interpretation of pseudo-code}
\label{sec:implementationpseudo}

Here follow some comments to aid in the interpretation of the
pseudo-code description of the SAW-tree data structure and associated
algorithms.

\begin{itemize}
\setlength{\itemsep}{1pt}
\setlength{\parskip}{0pt}
\setlength{\parsep}{0pt}
\item{All calls are by value, following the C programming language convention. Data structures
  are passed to methods via pointers.}
\item{Pointers: the walk $w$ is a data structure whose 
  member variables
  can be accessed via pointers, e.g. the vector for the end-to-end
  distance for the walk $w$ is $w \ptr \mathbf{X}_{\mathrm{e}}$. The
  left-hand sub-walk of $w$ is indicated by $w^l$, and the
  right-hand sub-walk by $w^r$. This notation is further extended by
  indicating $w^{ll}$ for the left-hand sub-walk of $w^l$, $w^{lr}$
  for the right-hand sub-walk of $w^l$, etc..}
\item{Suggestive notation for member variables used to improve readability;
  all quantities, such as ``$\mathbf{X}_{\mathrm{e}}$'' (the
  end-to-end vector) must
  correspond to a particular walk $w$.  e.g.
 $\mathbf{X}_{\mathrm{e}} \equiv w \ptr
  \mathbf{X}_{\mathrm{e}}$, $\mathbf{X}_{\mathrm{e}}^l \equiv w^l
  \ptr \mathbf{X}_{\mathrm{e}}$ (i.e., superscript $l$ indicates that
  $\mathbf{X}_{\mathrm{e}}^l$ is the end-to-end vector for the left
  sub-walk), $q^l \equiv w^l \ptr q$, $n^l \equiv w^l \ptr n$. }
\item{Variables with subscript $t$ are used
  for temporary storage only.}
\item{Comments are enclosed between the symbols /* and */ following
  the C convention.}
\item{Boolean negation is indicated via the symbol ``!'', e.g. 
  ! TRUE = FALSE.}
\end{itemize}

\subsection{SAW-tree data structure}
\label{sec:implementationdatastructure}

The key insight which has enabled a dramatic improvement in the
implementation of the pivot algorithm is the recognition that
\emph{sequences} of sites and pivots can be replaced by \emph{binary
  trees}. 
The $n$ leaves of the tree are individual sites of the walk, and thus 
encode no information, while each of the $n-1$ (internal) nodes of the
tree contain 
aggregate information about all sites which are below them in the
tree. We call this data structure the SAW-tree, 
which may be defined recursively: 
a SAW-tree of $n$ sites either has $n=1$ and is a leaf, or
has a left child SAW-tree with $0<k<n$ sites,
and a right child SAW-tree with the remaining $n-k$ sites.

Our implementation of the SAW-tree node is
introduced in Table~\ref{tab:definition}. 
A SAW-tree consists of one or more SAW-tree nodes in a binary tree
structure; the pointers $w^l$
and $w^r$ allow traversal 
from the root of the tree to the leaves, while $w^p$ allows for
traversal from the leaves of the tree to the root.  
SAW-trees are created by merging other SAW-trees, with a symmetry
operation acting on the right-hand walk. In particular, any
\emph{internal node} $w$ may be 
expressed in terms of its left child $w^l$, a symmetry operation $q
\equiv w \rightarrow q$, and its right child $w^r$ via a merge operation:
\begin{align}
w &= \text{merge}(w^l, q, w^r).
\end{align}

\begin{table}[htb]
\caption{Definition of the SAW-tree node data structure; see
  Sec.~\ref{sec:implementationdefinitions} for definitions of each of
  the member variables. The node contains
  variables which are required for traversal of the SAW-tree ($w^p$,
  $w^l$, $w^r$), intersection testing ($n$, $q$, $\mathbf{X}_{\mathrm{e}}$, 
  $\mathbf{B}$), and calculation of observables $R_x^2$ with $x \in
  \{\mathrm{e,g,m}\}$ ($n$,
  $\mathbf{X}_{\mathrm{e}}$, $\mathbf{X}$, $X_2$). If the only
  observable to be calculated is $R_{\mathrm{e}}^2$, then  $\mathbf{X}$
  and $X_2$ can be omitted.}
\label{tab:definition}
\begin{center}
\begin{tabular}{lll}
\hline\hline 
\multicolumn{3}{c}{SAW-tree node data structure \tstrut} \\
Type & Name & Description \bstrut\\
\hline
\emph{integer} & $n$ & Number of sites \tstrut\\
\emph{SAW-tree ptr} & $w^p$ & Parent\\
\emph{SAW-tree ptr} & $w^l$ & Left-hand sub-walk\\
\emph{SAW-tree ptr} & $w^r$ & Right-hand sub-walk\\
\emph{matrix} & q & Symmetry group element\\
\emph{vector} & $\mathbf{X}_{\mathrm{e}}$ & $\omega(n-1)$\\
\emph{vector} & $\mathbf{X}$ & $\mathbf{X} = \sum_{i=0}^{n-1} \omega(i)$\\
\emph{integer} & $X_2$ & $X_2 = \sum_{i=0}^{n-1} \omega(i) \cdot \omega(i)$\\
\emph{bounding box} & $\mathbf{B}$ & Convex region \bstrut\\
\hline\hline
\end{tabular}
\end{center}
\end{table}

The leaves of the SAW-tree correspond to sites in a SAW, and are thus
labeled from 0 to
$n-1$. A binary tree with $n$ leaves has $n-1$ internal nodes, and we
label these nodes from 1 to $n-1$, so that the symmetry $q(i)$ is to
the left of $\omega(i),\omega(i+1),\cdots,\omega(n-1)$. The symmetry
$q(0)$ is not part of the SAW-tree as it is applied to the whole walk,
and thus cannot be used in a merge
operation. For some applications it may be necessary to keep track of
$q(0)$, e.g. when studying polymers in a confined region, but in 
\cite{Clisby2010} this was not necessary.

Assume that the end-to-end vectors, $\mathbf{X}_{\mathrm{e}}$, and
symmetry group elements, $q(i)$, for a 
SAW-tree and its left and right children are given. If we know the
location of the anchor site of the parent node, $x_{\mathrm{abs}}$, along with the
overall \emph{absolute} symmetry group element $q_{\mathrm{abs}}$ being
applied to the walk, we can then find the same information for the
left and right children as follows:
\begin{displaymath}
\begin{array}{lllll}
\text{Left:}& & & x_{\mathrm{abs}} &\Leftarrow x_{\mathrm{abs}} \\
& & & q_{\mathrm{abs}} &\Leftarrow q_{\mathrm{abs}} \\
\text{Right:}& & & x_{\mathrm{abs}} &\Leftarrow x_{\mathrm{abs}} + q_{\mathrm{abs}} \mathbf{X}^l_{\mathrm{e}} \\
& & & q_{\mathrm{abs}} &\Leftarrow q_{\mathrm{abs}} q^r \\
\end{array}
\end{displaymath}
Thus $\omega(i)$ can be determined for any site $i$
by iteratively performing this calculation while following
the (unique) path from the root of the SAW-tree to the appropriate
leaf. N.B.: $x_{\mathrm{abs}}$ must be updated before $q_{\mathrm{abs}}$.

We give explicit examples of SAW-trees in
Appendix~\ref{sec:examplesawtrees}. In
Fig.~\ref{fig:sawtree_sequence}, we give a SAW-tree representation
of a SAW with $n$ sites which is precisely equivalent to the pivot
sequence representation. We also give two equivalent
representations of 
$\omega_a$ (shown in Fig.~\ref{fig:example}) in
Figs.~\ref{fig:sawtree_exampleaa} and \ref{fig:sawtree_exampleab}. 

Conceptually we distinguish single-site walks (individual sites), which
reside in the leaves of the tree, from multi-site 
walks.\footnote{Technical note: as the leaves are all
  identical, in practice we use a sentinel node for the leaves, saving
on memory usage.} In particular, the symmetry group element of a single site has
no effect, and in the case where all monomers are identical then all
single sites are identical.

If the SAW-tree structure remains fixed
it is not possible to
rotate part of the walk 
by updating a single symmetry group element, in contrast to the pivot
sequence representation.
This is because when we change a symmetry
group element in a given node, it only alters the position of sites
which are in the right child of the node. To rotate the part of the
walk with sites labeled $i+1$ and greater, we choose the
$i^{\mathrm{th}}$ internal node of the SAW-tree from the left.
We then need
to alter the symmetry group element of this node, and also all nodes
which are above and to the right of it in the SAW-tree.
If we select a random node
then it will likely be near the leaves of the tree, 
and assuming that the SAW-tree is balanced
this means that on
average $O(\log N)$ symmetry group elements will need to be altered.

However, we note that the
root node at the top of the tree has no parents, and therefore only
one symmetry 
group element needs to be altered to rotate the right-hand part of the
walk in this case. By utilizing
tree-rotation operations, which alter the structure of the tree while
preserving node ordering, it is possible to
move the $i^{\mathrm{th}}$ node to the root of the SAW-tree. Once this
has been done,
it \emph{is} then possible to implement a rotation of part of the
walk by updating a
single symmetry group element. 
On average, $O(\log N)$ of these tree-rotation operations are required.

Binary trees are a standard data structure in computer science. By
requiring trees to be balanced, i.e. so that the height of a tree with
$N$ nodes is bounded by a constant times $\log N$, optimal bounds can
be derived for operations such as insertion and deletion of nodes from
the tree. We refer the interested reader to
Sedgewick~\cite{Sedgewick1998Books} for various implementations of
balanced trees, such as red-black balanced trees. 
We have the advantage that our SAW-tree is, essentially, static, which
means that we can make it perfectly balanced without the additional
overhead of maintaining a balanced tree.

\subsection{Primitive operations}
\label{sec:implementationprimitive}

Included in this subsection are the primitive operations, which would
generally not be called from the main program.

Left and right tree-rotations are modified versions of standard tree
operations; for binary trees, only ordering needs to be preserved,
while for SAW-trees the sequence of sites needs to be preserved, which
means that 
symmetry group elements and other variables need to be modified.

\begin{ncalgorithmic}
\STATE \textbf{Procedure:}
\PROC{\textbf{Merge}(\emph{SAW-tree ptr} $w^l$,
\emph{SAW-tree ptr} $w^r$, \emph{SAW-tree ptr} $w$)}
\STATE \MLCOMMENT{Two SAWs are joined together, head to tail. Merge
  $w^l$ and $w^r$ into $w$,
  i.e. $w \Leftarrow \text{merge}(w^l, q, w^r)$. Pointers are not altered. See
  Eq.~\ref{eq:merge} and Fig.~\ref{fig:merge}. }
\STATE
\begin{tabular}{lcl}
$n$ &$\Leftarrow$& $n^l + n^r$\\
$\mathbf{B}$&$ \Leftarrow$&$ \mathbf{B}^l \cup
  \left(\mathbf{X}_{\mathrm{e}}^l + q \mathbf{B}^r \right)$\\
$\mathbf{X}_{\mathrm{e}} $&$\Leftarrow$&$ \mathbf{X}_{\mathrm{e}}^l + q \mathbf{X}_{\mathrm{e}}^r$\\
$\mathbf{X} $&$\Leftarrow$&$ \mathbf{X}^l + q \mathbf{X}^r + n^r \mathbf{X}_{\mathrm{e}}^l$\\
$X_2 $&$\Leftarrow$&$ X_2^l + X_2^r + 2 \mathbf{X}_{\mathrm{e}}^l
\cdot \left(q \mathbf{X}^r\right) + n^r \mathbf{X}_{\mathrm{e}}^l
\cdot \mathbf{X}_{\mathrm{e}}^l$\\
\end{tabular}
\ENDPROC{\textbf{Merge}}
\end{ncalgorithmic}

\tikzstyle{level 1}=[level distance=1.3cm, sibling distance=1.5cm]
\tikzstyle{level 2}=[level distance=1.3cm, sibling distance=1.5cm]
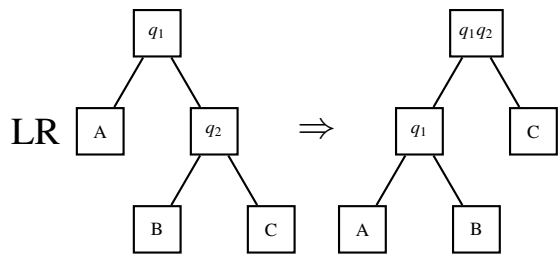
\begin{figure}[!ht]
\begin{center}
  \begin{tikzpicture}[
      thick,
      q/.style={rectangle,minimum height=1.75em,minimum width=1.75em,minimum width=1.75em,draw},
      leaf/.style={circle,minimum size=0mm,inner sep=2pt,draw}
    ]
    \begin{scope}[xshift=-1.6cm]
      \node at (0,-1.3) {{\Large LR}};
    \end{scope}
    \node[q] {\scriptsize $q_1$}
    child {
      node[q] {\scriptsize A}
    }
    child {
      node[q] {\scriptsize $q_2$}
      child { node[q] {\scriptsize B} }
      child { node[q] {\scriptsize C} }
    };
    \begin{scope}[xshift=2.1cm]
      \node at (0,-1.3) {{\Large $\Rightarrow$}};
    \end{scope}
    \begin{scope}[xshift=4.2cm]
      \node[q] {\scriptsize $q_1 q_2$}
      child {
        node[q] {\scriptsize $q_1$}
        child { node[q] {\scriptsize A} }
        child { node[q] {\scriptsize B} }
      }
      child {
        node[q] {\scriptsize C}
      };
    \end{scope}
  \end{tikzpicture}
\end{center}
\caption{Left tree-rotation applied to a SAW-tree, where A, B, and C
  are arbitrary SAW-trees. LHS and RHS are different
  representations of
  the same self-avoiding walk. 
}
\label{fig:LR}
\end{figure}

\begin{ncalgorithmic}
\STATE \textbf{Procedure:}
\PROC{\textbf{LR}(\emph{SAW-tree ptr} $w$)}
\STATE \MLCOMMENT{Left tree-rotation applied to $w$. Update of pointers to
  parents not shown, and note that only one merge operation is
  necessary. 
The pseudocode is faithful to the implementation used in the present
work and
\cite{Clisby2010}, but the definition of this operation is likely to
change in future implementations. 
In Fig.~\ref{fig:LR}, $w$
refers to the node with symmetry $q_1$ on the LHS, and $q_1 q_2$ on
the RHS. In future, we will adopt the convention 
that $w$ always refers to the same node with respect to left-right
ordering. By this convention, $w$ would refer to node with symmetry
$q_1$ on the RHS.}
\STATE
\begin{tabular}{lcl}
$w_t $ &$\Leftarrow $ &$w^r$\\
$w^r $ &$\Leftarrow $ &$w_t \ptr w^r $\\
$w_t \ptr w^r $ &$\Leftarrow $ &$w_t \ptr w^l$\\
$w_t \ptr w^l $ &$\Leftarrow $ &$w^l$\\
$w^l $ &$\Leftarrow $ &$w_t$\\
$q_t $ &$\Leftarrow $ &$q$\\
$q $ &$\Leftarrow $ &$q_t q^l$\\
$q^l$ &$\Leftarrow $ &$q_t$\\
\end{tabular}
\STATE \textbf{Merge}($w^{ll}$,$w^{lr}$,$w^l$)
\ENDPROC{\textbf{LR}}
\end{ncalgorithmic}

\tikzstyle{level 1}=[level distance=1.3cm, sibling distance=1.5cm]
\tikzstyle{level 2}=[level distance=1.3cm, sibling distance=1.5cm]
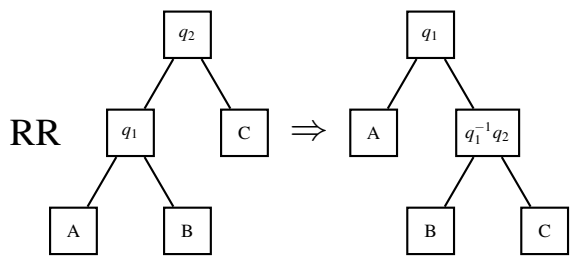
\begin{figure}[!ht]
\begin{center}
  \begin{tikzpicture}[
      thick,
      q/.style={rectangle,minimum height=1.75em,minimum width=1.75em,draw},
      leaf/.style={circle,minimum size=0mm,inner sep=2pt,draw}
    ]
    \begin{scope}[xshift=-2.0cm]
      \node at (0,-1.3) {{\Large RR}};
    \end{scope}
    \node[q] {\scriptsize $q_2$}
    child {
      node[q] {\scriptsize $q_1$}
      child { node[q] {\scriptsize A} }
      child { node[q] {\scriptsize B} }
    }
    child {
      node[q] {\scriptsize C}
    };
    \begin{scope}[xshift=1.6cm]
      \node at (0,-1.3) {{\Large $\Rightarrow$}};
    \end{scope}
    \begin{scope}[xshift=3.2cm]
      \node[q] {\scriptsize $q_1$}
      child {
        node[q] {\scriptsize A}
      }
      child {
        node[q] {\scriptsize $q_1^{-1}q_2$}
        child { node[q] {\scriptsize B} }
        child { node[q] {\scriptsize C} }
      };
    \end{scope}
  \end{tikzpicture}
\end{center}
\caption{Right tree-rotation applied to a SAW-tree, where A, B, and C
  are arbitrary SAW-trees. LHS and RHS are different
  representations of
  the same self-avoiding walk. 
}
\label{fig:RR}
\end{figure}

\begin{ncalgorithmic}
\STATE \textbf{Procedure:}
\PROC{\textbf{RR}(\emph{SAW-tree ptr} $w$)}
\STATE \MLCOMMENT{Right tree-rotation applied to $w$. Update of pointers to
  parents not shown. Note that only one merge operation is necessary. 
The pseudocode is faithful to the implementation used in the present work and 
\cite{Clisby2010}, but the definition of this operation is likely to
change in future implementations. 
In Fig.~\ref{fig:RR} $w$
refers to the node with symmetry $q_2$ on the LHS, and $q_1$ on
the RHS. In future, we will adopt the convention 
that $w$ always refers to the same node with respect to left-right
ordering. By this convention, $w$ would refer to node with symmetry
$q_1^{-1}q_2$ on the RHS.}
\STATE
\begin{tabular}{lcl}
$w_t $&$\Leftarrow$&$ w^l$\\
$w^l $&$\Leftarrow$&$ w_t \ptr w^l $\\
$w_t \ptr w^l $&$\Leftarrow$&$ w_t \ptr w^r$\\
$w_t \ptr w^r $&$\Leftarrow$&$ w^r$\\
$w^r $&$\Leftarrow$&$ w_t$\\
$q_t $&$\Leftarrow$&$ q$\\
$q $&$\Leftarrow$&$ q^r$\\
$q^r$&$\Leftarrow$&$ q^{-1} q_t $\\
\end{tabular}
\STATE \textbf{Merge}($w^{rl}$,$w^{rr}$,$w^r$)
\ENDPROC{\textbf{RR}}
\end{ncalgorithmic}

\medskip

\begin{ncalgorithmic}
\STATE \textbf{Function:}
\STATE \textbf{Find\-\_node}(\emph{integer} $n_t$, \emph{SAW-tree ptr} $w$)
\STATE \MLCOMMENT{Returns a pointer to the $n_t^{\mathrm{th}}$ node
  from the left in 
  the SAW-tree $w$. This may either be implemented as a numerical function, if
  the address of 
  the correct node can be easily determined (such as if the nodes of $w$
  are arranged in memory in pre-order fashion), or returned from a
  look-up array. This look-up array is static, and so will not need to
  be updated after it is created.
  Note: we require this function to take time $O(1)$,
  and thus it is inappropriate to use a binary search implementation
  which may take time $O(\log N)$. This function is required by
  \textbf{Attempt\-\_pivot\-\_fast}.}  
\end{ncalgorithmic}

\subsection{User level operations}
\label{sec:implementationuser}

Included in this subsection are user level operations which would
typically be called from the main program.

\begin{ncalgorithmic}
\STATE \textbf{Procedure:}
\STATE \textbf{Generate\-\_SAW-tree}(\emph{integer} $n$)
\STATE \MLCOMMENT{Allocates memory for the SAW-tree, and creates the
  initial arrangement of the tree in memory. We use a strictly
  balanced tree layout, which guarantees $O(\log N)$ behavior for
  basic operations. We tested pre-order and van Emde Boas layouts;
  given that performance  for large $N$ was memory bound, we were surprised to find
  that the pre-order layout was fastest, although we will 
  experiment more with this in the future (the van Emde Boas tree
  layout~\cite{vanEmdeBoas1975} is an example of a  cache-oblivious
  data structure~\cite{Frigo1999,Kumar2003}).}
\STATE
\STATE \textbf{Function:}
\STATE \textbf{Random\-\_integer\-\_uniform}(\emph{integer} $a$, \emph{integer} $b$)
\STATE \MLCOMMENT{Returns an integer in the interval $[a,b)$ selected
  uniformly at random.}
\STATE
\STATE \textbf{Function:}
\STATE \textbf{Random\-\_integer\-\_log}(\emph{integer} $a$, \emph{integer} $b$)
\STATE \MLCOMMENT{Returns an integer $i$ selected in the interval $[a,b)$
  with probability 
\begin{align}
P(i) &= \frac{\log(1+\frac{1}{i-a+1})}{\log(b-a+1)}.
\end{align}
This
  probability distribution has the property that if we set $a=1$ for
  convenience and choose a length
  scale $1 \leq x < 2x \leq b$, then 
\begin{align}
P(i\in[x,2x)) &= \frac{\log(2x) - \log x}{\log b} \nonumber \\
&= \frac{\log 2}{\log b},
\end{align}
i.e. the probability of $i$ lying in the semi-open interval $[x,2x)$
  is independent of $x$.}
\STATE
\STATE \textbf{Function:}
\STATE \textbf{Random\-\_symmetry}()
\STATE \MLCOMMENT{Symmetry selected uniformly at random, excluding
  identity. Other choices are possible.}
\STATE
\STATE \textbf{Procedure:}
\STATE \textbf{Accumulate\-\_statistics}(\emph{SAW-tree} $w$,
\emph{Boolean} success)
\STATE \MLCOMMENT{Accumulate statistics for Euclidean-invariant
  moments $R_x^{2k}$ with $x \in
  \{\mathrm{e,g,m}\}$, and $1 \leq k \leq 5$.
When the pivot attempt is unsuccessful there is no need to recalculate
observables; in fact, by using a counter, updates to storage variables
only need to be made when $\mathrm{success} = $TRUE,
i.e. when the last pivot attempt was successful.}
\end{ncalgorithmic}

\tikzstyle{level 1}=[level distance=1cm, sibling distance=1.75cm]
\tikzstyle{level 2}=[level distance=1cm, sibling distance=1cm]
\tikzstyle{level 3}=[level distance=1cm, sibling distance=1cm]
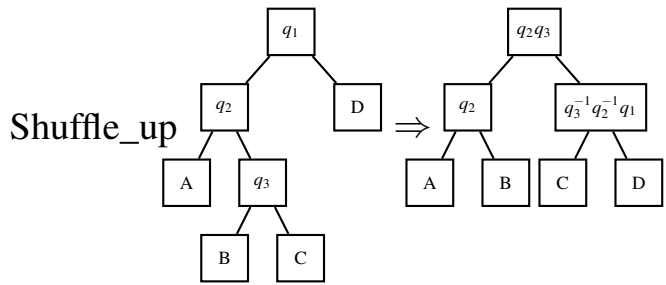
\begin{figure}[!ht]
\begin{center}
  \begin{tikzpicture}[
      thick,
      q/.style={rectangle,minimum height=1.75em,minimum width=1.75em,draw},
      leaf/.style={circle,minimum size=0mm,inner sep=2pt,draw}
    ]
    \begin{scope}[xshift=-2.6cm]
      \node at (0,-1.3) {{\Large Shuffle\-\_up}};
    \end{scope}
    \node[q] {\scriptsize $q_1$}
    child {
      node[q] {\scriptsize $q_2$}
      child { node[q] {\scriptsize A} }
      child { 
        node[q] {\scriptsize $q_3$} 
        child { node[q] {\scriptsize B} }
        child { node[q] {\scriptsize C} }
      }
    }
    child {
      node[q] {\scriptsize D}
    };
    \begin{scope}[xshift=1.6cm]
      \node at (0,-1.3) {{\Large $\Rightarrow$}};
    \end{scope}
    \begin{scope}[xshift=3.2cm]
      \node[q] {\scriptsize $q_2 q_3$}
      child {
        node[q] {\scriptsize $q_2$}
        child { node[q] {\scriptsize A} }
        child { node[q] {\scriptsize B} }
      }
      child {
        node[q] {\scriptsize $q_3^{-1} q_2^{-1} q_1$}
        child { node[q] {\scriptsize C} }
        child { node[q] {\scriptsize D} }
      };
    \end{scope}
  \end{tikzpicture}
\end{center}
\caption{Shuffle up operation applied to node with symmetry $q_3$ in a
  SAW-tree, via a sequence of left and right tree-rotations, with A,
  B, C, and D arbitrary SAW-trees. LHS and RHS
  are different representations of the same self-avoiding walk.}
\label{fig:shuffle_up}
\end{figure}

\begin{ncalgorithmic}
  \STATE \textbf{Procedure:}
  \PROC{\textbf{Shuffle\-\_up}(\emph{integer} $n_0$, \emph{SAW-tree ptr} w)}
  \STATE \MLCOMMENT{Brings node $n_0$ to the root of the SAW-tree, via a series
    of tree-rotation operations. See Fig.~\ref{fig:shuffle_up}.}
  \IF{$n_0 = n^l$}
  \STATE \MLCOMMENT{Do nothing, node already at root.}
  \ELSIF{$n_0 < n^l$}
  \STATE \textbf{Shuffle\-\_up}($n_0$,$w^l$)
  \STATE \textbf{RR}($w$)
  \ELSIF{$n_0 > n^l$}
  \STATE \textbf{Shuffle\-\_up}($n_0-n^l$,$w^r$)
  \STATE \textbf{LR}($w$)
  \ENDIF
  \STATE Return
  \ENDPROC{\textbf{Shuffle\-\_up}}
\end{ncalgorithmic}

\medskip

\begin{ncalgorithmic}
  \STATE \textbf{Procedure:}
  \PROC{\textbf{Shuffle\-\_down}(\emph{SAW-tree ptr} w)}
  \STATE \MLCOMMENT{Restores node at root to its correct place in the balanced
    SAW-tree, via a series of tree-rotation operations. Note that this
    operation is only necessary if one wishes that the SAW-tree remains
    balanced, and must be paired with \textbf{Shuffle\-\_up}.}
  \STATE $n_t \Leftarrow \lfloor (n+1)/2 \rfloor$
  \IF{$n_t = n^l$}
  \STATE \MLCOMMENT{Do nothing, node is in correct place.}
  \ELSIF{$n_t < n^l$}
  \STATE \textbf{RR}($w$)
  \STATE \textbf{Shuffle\-\_down}($w^r$)
  \ELSIF{$n_t > n^l$}
  \STATE \textbf{LR}($w$)
  \STATE \textbf{Shuffle\-\_down}($w^l$)
  \ENDIF
  \STATE Return
  \ENDPROC{\textbf{Shuffle\-\_down}}
\end{ncalgorithmic}

\medskip

The function \textbf{Intersect} is at the heart of 
our implementation of the pivot algorithm; we recommend the reader consult
the start of Sec.~\ref{sec:complexitysuccess} for an 
explanation of how \textbf{Intersect} works.

\begin{ncalgorithmic}
\STATE \textbf{Function:}
\PROC{\textbf{Intersect}
(\emph{vector} $x^l_{\mathrm{abs}}$, \emph{matrix} $q^l_{\mathrm{abs}}$, \emph{SAW-tree ptr} $w^l$, \\
\hspace{0.01mm} \phantom{Intersect(}
\emph{vector} $x^r_{\mathrm{abs}}$, \emph{matrix} $q^r_{\mathrm{abs}}$, \emph{SAW-tree ptr} $w^r$)}
\STATE \MLCOMMENT{Returns TRUE if $w^l$ and $w^r$ intersect, FALSE
  otherwise. $x^l_{\mathrm{abs}}$ and $x^r_{\mathrm{abs}}$ are the
  \emph{absolute} positions of the anchor points of the walks $w^l$ and $w^r$,
  and $q^l_{\mathrm{abs}}$ and $q^r_{\mathrm{abs}}$ are overall symmetry
  group elements.} 
\vspace{2mm}
\STATE \MLCOMMENT{First, calculate the absolute positions of the bounding
  boxes. If they do not intersect, then $w^l$ and $w^r$ cannot intersect.}
\STATE $B^l_t \Leftarrow x^l + q^l B^l$
\STATE $B^r_t \Leftarrow x^r + q^r B^r$
\IF{$B^l_t \cap B^r_t = \emptyset$}
\STATE Return FALSE
\ENDIF
\IF{$n^l \leq 2$ \textbf{and} $n^r \leq 2$}
\STATE \MLCOMMENT{The bounding boxes of $w^l$ and $w^r$ intersect, and
  thus $w^l$ and $w^r$ must intersect as each walk has two or fewer sites.
  The cut-off is dependent on the shape of the
  bounding box, and is the maximum number of sites a SAW can have
  which guarantee that the bounding box contains only the sites of the
  SAW and no others. For $\Z^d$ the value of the cut-off for the
  rectangular prism 
  is two as given here; see
  Sec.~\ref{sec:implementationdefinitions} for further 
  discussion.} 
\STATE Return TRUE
\ENDIF
\IF{$n^l \geq n^r$}
\STATE \MLCOMMENT{Split the left SAW-tree; compare the SAW-trees which
  are closest together on the chain first, as they are the most likely
  to intersect.} 
\IF{\textbf{Intersect}($x^l_{\mathrm{abs}} +
q^l_{\mathrm{abs}} \mathbf{X}_{\mathrm{e}}^{ll}$, $q^l_{\mathrm{abs}} q^l$, $w^{lr}$,
$x^r_{\mathrm{abs}}$, $q^r_{\mathrm{abs}}$, $w^r$)}
\STATE Return TRUE
\ENDIF
\STATE Return \textbf{Intersect}($x^l_{\mathrm{abs}}$, $q^l_{\mathrm
  abs}$, $w^{ll}$, $x^r_{\mathrm{abs}}$, $q^r_{\mathrm{abs}}$, $w^r$)
\ELSE
\STATE \MLCOMMENT{Split the right SAW-tree; compare the SAW-trees which
  are closest together on the chain first, as they are the most likely
  to intersect.} 
\IF{\textbf{Intersect}($x^l_{\mathrm{abs}}$, $q^l_{\mathrm
  abs}$, $w^{l}$, $x^r_{\mathrm{abs}}$, $q^r_{\mathrm{abs}}$, $w^{rl}$)}
\STATE Return TRUE
\ENDIF
\STATE Return \textbf{Intersect}($x^l_{\mathrm{abs}}$, $q^l_{\mathrm
  abs}$, $w^{l}$, $x^r_{\mathrm{abs}} + q^r_{\mathrm{abs}} \mathbf{X}_{\mathrm{e}}^{rl}$,
$q^r_{\mathrm{abs}} q^r$, $w^{rr}$)
\ENDIF
\ENDPROC{\textbf{Intersect}}
\end{ncalgorithmic}

\medskip

\textbf{Shuffle\-\_intersect} is akin to the procedure developed by
Madras and Sokal~\cite{Madras1988} for intersection testing: by
building new walks incrementally moving outwards from the pivot
site they achieved a speed-up from $O(N)$ to $O(N^{1-p})$ for
unsuccessful pivot attempts.
In Sec.~\ref{sec:complexity} we present a heuristic argument that the
speed-up achieved for the SAW-tree implementation is from $O(\log
N)$ to $O(1)$.

\begin{ncalgorithmic}
\STATE \textbf{Function:}
\PROC{\textbf{Shuffle\-\_intersect}(\emph{SAW-tree ptr} $w$,
  \emph{symmetry} $q_0$, \emph{integer} $\mathrm{was\_left\_child}$,\\
\hspace{0.01mm} \phantom{Shuffle\-\_intersect(}
  \emph{integer} $\mathrm{is\_left\_child}$)} 
\STATE \MLCOMMENT{Hybrid algorithm which combines elements of
  \textbf{Shuffle\-\_up} and
  \textbf{Intersect}. $w$ is the pivot node, and $q_0$ is the pivot
  operation which acts on the part of the walk to the right of $w$
  (both ancestors and descendants). was\_left\_child and
  is\_left\_child are integer flags specifying the local tree
  structure. Function returns true if 
  the left- and right-hand walks intersect, FALSE otherwise. 
  The pseudo-code here reproduces the method, but
  some small-scale optimizations have been omitted for the sake of clarity. 
  In particular, some computations are performed more often than
  strictly necessary, e.g. rotations of bounding boxes. For $\Z^3$,
  this results in the implementation here running approximately $40\%$
  slower than the optimized implementation.
  By using temporary variables to store the modified tree nodes, we
  guarantee that the original walk is left unmodified. Thus it is safe
  to execute concurrent versions of this function on the same SAW-tree
  in parallel.} 
\STATE \MLCOMMENT{Check if the left- and right- children of $w$
  intersect.}
\IF{$\mathrm{was\_left\_child} = 1$}
\STATE \MLCOMMENT{We have already verified that $w^{l}$ and $w^{rl}$ do
  not intersect; check if $w^l$ and $w^{rr}$ intersect.}
\IF{\textbf{Intersect}($0$, $I$, $w^l$, $\mathbf{X}_{\mathrm{e}}^{l} + q q_0 \mathbf{X}_{\mathrm{e}}^{rl}$,
  $q q_0 q^r$, $w^{rr}$)}
\STATE Return TRUE
\ENDIF
\ELSIF{$\mathrm{was\_left\_child} = 0$}
\STATE \MLCOMMENT{We have already verified that $w^{lr}$ and $w^r$ do
  not intersect; check if $w^{ll}$ and $w^{r}$ intersect.}
\IF{\textbf{Intersect}($0$, $I$, $w^{ll}$, $\mathbf{X}_{\mathrm{e}}^{l}$,
  $q q_0$, $w^r$)}
\STATE Return TRUE
\ENDIF
\ELSIF{$\mathrm{was\_left\_child} = -1$}
\STATE \MLCOMMENT{First call of this function, and so we have no prior
  information; check if $w^l$ and $w^r$ intersect.}
\IF{\textbf{Intersect}($0$, $I$, $w^l$, $\mathbf{X}_{\mathrm{e}}^{l}$,
  $q q_0$, $w^r$)}
\STATE Return TRUE
\ENDIF
\ENDIF
\STATE \MLCOMMENT{Check to see if we have reached the top of the tree.}
\IF{$w^p = \mathrm{NULL}$}
\STATE Return FALSE
\ENDIF
\STATE \MLCOMMENT{Not at top of tree, and so we need to perform a left or
  right rotation. First we need to determine if $w^p$ is a left or
  right child.}
\IF{$w^{pp} = \mathrm{NULL}$}
\STATE \MLCOMMENT{$w^{p}$ is the root of the tree so
  $\mathrm{is\_left\_child\_new}$ will not be needed. }
\ELSIF{$w^{ppl} = w^p$}
\STATE $\mathrm{is\_left\_child\_new} \Leftarrow$ TRUE
\ELSE
\STATE $\mathrm{is\_left\_child\_new} \Leftarrow$ FALSE
\ENDIF
\STATE \MLCOMMENT{Create a temporary SAW-tree node $w_t$ which is a
  copy of $w^p$ so we do not alter the original walk.}
\STATE $w_t \Leftarrow w^p$
\IF{$\mathrm{is\_left\_child}$}
\STATE \MLCOMMENT{right tree-rotation needed.}
\STATE \textbf{RR}($w_t$)
\ELSE
\STATE \MLCOMMENT{left tree-rotation needed.}
\STATE \textbf{LR}($w_t$)
\ENDIF

\STATE Return
\textbf{Shuffle\-\_intersect}($w_t$,$q_0$,$\mathrm{is\_left\_child}$,
$\mathrm{is\_left\_child\_new}$)
\ENDPROC{\textbf{Shuffle\-\_intersect}}
\end{ncalgorithmic}

\subsection{High level functions}
\label{sec:implementationhigh}

Here follows a pseudo-code description of three routines
which provide high level functionality for typical usage of the pivot
algorithm.

We define two versions of the function \textbf{Attempt\-\_pivot}, one
which is conceptually simple, while in Sec.~\ref{sec:complexity} we argue that the
other version is asymptotically faster
for $\Z^2$ and $\Z^3$.

\begin{ncalgorithmic}
\STATE \textbf{Function:}
\PROC{\textbf{Attempt\-\_pivot\-\_simple}(\emph{SAW-tree ptr} $w$, \emph{integer}
  $n_t$, \emph{symmetry} $q_t$)}
\STATE \MLCOMMENT{Attempt to apply a pivot to $w$ at site $n_t$ with
  symmetry $q_t$; if successful update the walk, otherwise leave walk
  unchanged. Return a boolean value to indicate whether the pivot
  operation successfully changed the walk.}
\STATE \textbf{Shuffle\-\_up}($n_t$,$w$)
\STATE $q \Leftarrow q q_t$
\STATE intersection $\Leftarrow$ \textbf{Intersect}($0$, $I$, $w^l$, $\mathbf{X}_{\mathrm{e}}^{l}$,
  $q$, $w^r$)
\IF{intersection}
\STATE \MLCOMMENT{Reject pivot, restore original symmetry.}
\STATE $q \Leftarrow q q_t^{-1}$
\ENDIF
\STATE \textbf{Shuffle\-\_down}($w$)
\STATE \MLCOMMENT{Pivot was successful if sub-walks \emph{did not} intersect.}
\STATE Return !intersection
\ENDPROC{\textbf{Attempt\-\_pivot\-\_simple}}
\end{ncalgorithmic}

\medskip

\begin{ncalgorithmic}
\STATE \textbf{Function:}
\PROC{\textbf{Attempt\-\_pivot\-\_fast}(\emph{SAW-tree ptr} $w$, \emph{integer}
  $n_t$, \emph{symmetry} $q_t$)}
\STATE \MLCOMMENT{Attempt to apply a pivot to $w$ at site $n_t$ with
  symmetry $q_t$; if successful update the walk, otherwise leave walk
  unchanged. Return a boolean value to indicate whether the pivot
  operation successfully changed the walk.}
\STATE $w_t \Leftarrow$ \textbf{Find\-\_node}($n_t$)
\IF{$w_t^{pl} =$ NULL}
\STATE \MLCOMMENT{Dummy argument, $\mathrm{is\_left\_child}$ will not
  be required.}
\STATE $\mathrm{is\_left\_child} \Leftarrow$ TRUE
\ELSIF{$w_t^{pl} = w_t$}
\STATE $\mathrm{is\_left\_child} \Leftarrow$ TRUE
\ELSE
\STATE $\mathrm{is\_left\_child} \Leftarrow$ FALSE
\ENDIF
\STATE intersection $\Leftarrow$ \textbf{Shuffle\-\_intersect}($w_t$,$q_t$,-1,$\mathrm{is\_left\_child}$)
\IF{!intersection}
\STATE \MLCOMMENT{Accept pivot, perform symmetry operation.}
\STATE \textbf{Shuffle\-\_up}($n_t$,$w$)
\STATE $q \Leftarrow q q_t$
\STATE \textbf{Shuffle\-\_down}($w$)
\ENDIF
\STATE \MLCOMMENT{Pivot was successful if sub-walks \emph{did not} intersect.}
\STATE Return !intersection
\ENDPROC{\textbf{Attempt\-\_pivot\-\_fast}}
\end{ncalgorithmic}

\medskip

\begin{ncalgorithmic}
\STATE \textbf{Procedure:}
\PROC{\textbf{Pseudo\-\_dimerize}(\emph{SAW-tree} $w$)}
\STATE \MLCOMMENT{Uses merge and pivot operations to generate an
  initial $N$-step SAW in time $\Theta(N)$ which is difficult to
  distinguish from a SAW sampled 
  uniformly at random. See 
  Sec.~\ref{sec:initialization} for discussion about why it is
  highly desirable to have such a procedure, and also for analysis of
  the algorithmic complexity.} 
\STATE \MLCOMMENT{Generate initial left- and right-hand sub-walks.}
\STATE\textbf{Pseudo\-\_dimerize}($w^l$)
\STATE\textbf{Pseudo\-\_dimerize}($w^r$)
\STATE \MLCOMMENT{Perform pivot operations on each of the sub-walks
  while attempting to merge them. When the two sub-walks are
  mutually avoiding, exit loop.}
\DOWHILE
\STATE $n_t \Leftarrow n^l - $ \textbf{Random\-\_integer\-\_log}($1$,$n^l$)
\STATE $q_t \Leftarrow$ \textbf{Random\-\_symmetry}()
\STATE \textbf{Attempt\-\_pivot}($w^l$, $n_t$, $q_t$)
\STATE $n_t \Leftarrow $ \textbf{Random\-\_integer\-\_log}($1$,$n^r$)
\STATE $q_t \Leftarrow$ \textbf{Random\-\_symmetry}()
\STATE \textbf{Attempt\-\_pivot}($w^r$, $n_t$, $q_t$)
\STATE $q \Leftarrow$ \textbf{Random\-\_symmetry}()
\ENDDOWHILE{\textbf{Intersect}($0$, $I$, $w^l$, $\mathbf{X}_{\mathrm{e}}^{l}$, $q$, $w^r$)}
\STATE \textbf{Merge}($w^l$,$w^r$,$w$)
\STATE \MLCOMMENT{Perform additional pivots on the SAW, preferentially
sampling close to the joint, in an attempt to reduce any sampling bias. We can
attempt $o(n/\log n)$ pivots without changing the asymptotic behavior of the
algorithm; we choose to attempt $n^{1/2}$ pivots. We do not believe
that these additional pivots are strictly necessary.}
\FOR{$i=1$ to $n^{1/2}$}
\STATE $n_t \Leftarrow n^l - $ \textbf{Random\-\_integer\-\_log}($1$,$n^l$)
\STATE $q_t \Leftarrow$ \textbf{Random\-\_symmetry}()
\STATE \textbf{Attempt\-\_pivot}($w$, $n_t$, $q_t$)
\STATE \MLCOMMENT{Now the walks are combined, we have to shift
  node label by $n^l$.}
\STATE $n_t \Leftarrow  n^l + $ \textbf{Random\-\_integer\-\_log}($0$,$n^r$)
\STATE $q_t \Leftarrow$ \textbf{Random\-\_symmetry}()
\STATE \textbf{Attempt\-\_pivot}($w$, $n_t$, $q_t$)
\ENDFOR
\STATE Return
\ENDPROC{\textbf{Pseudo\-\_dimerize}}
\end{ncalgorithmic}

\subsection{Main program}
\label{sec:implementationmain}

Here follows a pseudo-code description of 
the main program to generate
self-avoiding walks via the pivot algorithm.

\begin{ncalgorithmic}
\STATE \textbf{Procedure:}
\PROC{\textbf{Main}()}
\STATE \MLCOMMENT{Create SAW-tree structure.}
\STATE $w \Leftarrow$ \textbf{Generate\-\_SAW-tree}($n$)
\STATE \MLCOMMENT{Initialize SAW-tree.}
\STATE \textbf{Pseudo\-\_dimerize}($w$)
\STATE \MLCOMMENT{Warm up the Markov chain by discarding
  $n_{\text{discard}}$ time steps.}
\FOR{$i=1$ to $n_{\text{discard}}$}
\STATE $n_t \Leftarrow$ \textbf{Random\-\_integer\-\_uniform}($1$,$n$)
\STATE $q_t \Leftarrow$ \textbf{Random\-\_symmetry}()
\STATE \textbf{Attempt\-\_pivot}($w$, $n_t$, $q_t$)
\ENDFOR
\STATE \MLCOMMENT{Perform the computer experiment, collecting data on
  global observables at every time step.}
\FOR{$i=1$ to $n_{\text{sample}}$}
\STATE $n_t \Leftarrow$ \textbf{Random\-\_integer\-\_uniform}($1$,$n$)
\STATE $q_t \Leftarrow$ \textbf{Random\-\_symmetry}()
\STATE $\mathrm{success} =$ \textbf{Attempt\-\_pivot}($w$, $n_t$, $q_t$)
\STATE \textbf{Accumulate\-\_statistics}($w$,success)
\ENDFOR
\ENDPROC{\textbf{Main}}
\end{ncalgorithmic}

\section{Algorithmic complexity}
\label{sec:complexity}

Our goal in this section is to calculate the
mean time per attempted pivot, $T(N)$, which can in turn be
expressed in terms of the mean time for successful
pivots, $T_{\mathrm{S}}(N)$, and the mean time for unsuccessful pivots,
$T_{\mathrm{U}}(N)$. This is done as follows:
\begin{align}
T(N) &= \Pr(\text{successful}) T_{\mathrm{S}}(N) +
\Pr(\text{unsuccessful}) T_{\mathrm{U}}(N) \nonumber \\
&= O(N^{-p}) T_{\mathrm{S}}(N) +
O(1) T_{\mathrm{U}}(N) \nonumber \\
&= N^{-p} T_{\mathrm{S}}(N) + T_{\mathrm{U}}(N),
\end{align}
where the probability of a pivot being successful is $O(N^{-p})$ as
mentioned in Sec.~\ref{sec:introductionpivot}, and
constant factors have been dropped. Note that this expression is
valid for $\Z^2$ and $\Z^3$, while higher dimensions are discussed in
Sec.~\ref{sec:complexityhighdim}. 

The thread of argument is complicated by the fact that we have two
implementations we 
wish to characterize, \textbf{Attempt\-\_pivot\-\_simple} and
\textbf{Attempt\-\_pivot\-\_fast}. As the names indicate, we will find
that \textbf{Attempt\-\_pivot\-\_fast} is asymptotically faster, and so
$T(N)$ may be regarded as being the mean time per attempted pivot for
the procedure \textbf{Attempt\-\_pivot\-\_fast}.

We first introduce notation and do a minor calculation in
Sec.~\ref{sec:complexitybackground}. In
Sec.~\ref{sec:complexitysuccess} we give a detailed explanation
of the behavior of the procedure \textbf{Intersect}, and develop a 
heuristic argument that $T_{\mathrm{S}}(N)$ is $\Theta(\log N)$. In
Sec.~\ref{sec:complexitysimple} we show that for
\textbf{Attempt\-\_pivot\-\_simple} $T_{\mathrm{U}}(N) = \Theta(\log N)$, and
therefore $T(N) = \Theta(\log N)$, 
while in Sec.~\ref{sec:complexityfast} we provide a heuristic argument that for
\textbf{Attempt\-\_pivot\-\_fast} $T_{\mathrm{U}}(N) = O(1)$, which
leads to the prediction that $T(N) = O(1)$. 
In
Sec.~\ref{sec:complexityhighdim} we discuss the complexity of the
algorithm for dimensions greater than three, and finally in 
Sec.~\ref{sec:complexitynumerical} we examine the numerical
evidence. There is modest numerical evidence supporting our conclusion
that $T_{\mathrm{S}}(N)$ is $\Theta(\log N)$ for $\Z^2$ and $\Z^3$, while
the numerical evidence suggests $T_{\mathrm{U}}(N) = o(\log N)$ for
$\Z^2$ and $T_{\mathrm{U}}(N) = I(\log N)$ for $\Z^3$. This
is consistent with $T_{\mathrm{U}}(N) = O(1)$, but far from conclusive.

A far simpler version of the argument for our prediction that $T(N) =
O(1)$ for $\Z^2$ and $\Z^3$ is given in the companion
article~\cite{Clisby2010}. 

\subsection{Background}
\label{sec:complexitybackground}

We introduce the notation of the \emph{level}, $l$, of a node, which is the 
number of generations that separate the node from the leaves of the
tree. The leaves 
are at level zero, their parents are at level one, etc.. This is
non-standard notation, as usually the concept of depth is used to
describe trees, with the root at depth zero, its children at depth
one, etc.. For a perfectly
balanced tree with $2^k$ leaves (sites), the nodes at any
fixed level represent sub-walks of the same number of sites
independent of $k$, and for
this reason we find the level notation to be useful.

We will specialize to the case where we have a SAW-tree, $W$, with
$n = 2^k$ sites; this makes things somewhat simpler to deal with, but
makes no difference otherwise to the algorithmic complexity. $W$
therefore has $2^k$ leaves, $2^k - 1$ internal nodes, and the level of
the root is $k$.

The first step of the pivot algorithm is to select a pivot node
uniformly at random, from the $n-1$ internal nodes of $W$. The
average level of this node is
\begin{align}
\mathbb{E}(l) &= \frac{1}{2^k - 1}\sum_{j=1}^k j \; 2^{k-j} \nonumber \\
&= \frac{2^{k+1} - 2 - k}{2^k - 1} \nonumber \\
&= 2 - O\left(\frac{k}{2^k}\right).
\end{align}

We note that the discussion below is sufficiently general that it
applies to related models of polymers in the excluded volume limit,
such as the Domb-Joyce model, interacting self-avoiding walks, 
self-avoiding walks on other lattices, or self-avoiding walk models in
the continuum. 
SAW-tree implementations of these other models do not yet exist.
The discussion is valid for dimension $d \geq 2$, but there will be
some subtleties for $d \geq 4$ which will be explored in
Sec.~\ref{sec:complexityhighdim}. 

\subsection{Successful pivot}
\label{sec:complexitysuccess}

To estimate the expected running time of \textbf{Intersect}, we first
need to get a clear understanding of what the algorithm does.
\textbf{Intersect} recursively tests for the intersection of the
bounding boxes of the left and right SAW-trees, and searches until
either an intersection has been found, or until it has been
shown that no intersection is possible (i.e. no bounding boxes 
intersect). The search proceeds by 
recursively splitting the longer of the left and right SAW-trees, and 
although the action of \textbf{Shuffle\-\_up} on the SAW-trees
complicates things somewhat, 
the comparison typically occurs between left and right
SAW-trees which are of the same length, up to a factor of two. 
We note that the search method chosen
is \emph{depth-first search}, which minimizes
memory usage, and also that the
choice of splitting procedure is a key feature in the performance of
the algorithm. 
By choosing to split the larger walk, and then first
testing for intersection between the walks on the left and right hand
sides which are closest together on the chain, this means that
intersections are more likely to be found rapidly, allowing the search
to terminate after only a small portion of the walk has been examined.

For convenience we will only 
consider testing for intersection between the left- and right-hand sides
of $W$, i.e. we assume the pivot node is the root of the
SAW-tree. This will simplify the discussion of the algorithm, but  
the conclusions drawn will be valid for any balanced SAW-tree, with
any number of sites.

We will first examine the behavior of
\textbf{Intersect} in
the case that a pivot attempt is successful, when there is no
intersection. Later we will consider what happens when an intersection
is found.

When applied to $W$, \textbf{Intersect} compares sub-walks on the
left-hand side at level $j$ 
with sub-walks on the right-hand side at levels $j$ and $j+1$. When
bounding boxes overlap, the search proceeds to the next lowest level
in the tree, until there are no overlaps. The nodes that this search
procedure visits within $W$ induces a binary tree $W^\prime$ with internal
nodes that are sub-walks 
whose bounding boxes have been found to overlap sub-walks on the other
side, and whose leaves are sub-walks which do not overlap sub-walks on the
other side. As $W^\prime$ is a binary tree, the number of leaves is
exactly one more than the number of internal nodes.
The algorithmic complexity of \textbf{Intersect}, $T_{\mathrm{S}}(N)$, is the
number of 
intersection tests performed between sub-walks on the left- and
right-hand sides of 
$W^\prime$ (neglecting constant
factors). 

We seek to simplify the characterization of the counting problem, by
first noting that the number of intersection tests which involve
leaves of $W^\prime$ is at most four times the number of intersection
tests involving nodes of $W^\prime$, as we only test for intersection
between two sub-walks when both of their parents intersect. Thus $T_{\mathrm{S}}(N)$ is
asymptotically equal to the number of pairs of sub-walks with
overlapping bounding boxes, 
where sub-walks on the left-hand side are at level $j$, and sub-walks on the
right-hand side are at levels $j$ and $j+1$. We then note that the
number of pairs of sub-walks with overlapping bounding boxes with these
rules are at most 
twice the number of pairs of sub-walks  with overlapping bounding boxes on
the same level. $T_{\mathrm{S}}(N)$ is therefore given by the number
of pairs of sub-walks 
on the left- and right-hand sides of $W$ with overlapping bounding
boxes which are at the same level.

We introduce a new notation, describing the overlap between the
bounding boxes of two sub-walks at level $j$ in $W$ as a
``$j$-approach'', while any overlap between boxes of sub-walks which are at the same
level we denote as an ``approach''. Thus $T_{\mathrm{S}}(N)$ is given by the
expected number of 
\emph{approaches} between the left- and right-hand sides of $W$.

We give an example of a SAW with 16 sites in Fig.~\ref{fig:exampleb},
and show its corresponding SAW-tree in
Fig.~\ref{fig:sawtree_exampleb}, with arrows drawn between sub-walks
on the left- and right-hand sides 
which approach each other.

\begin{figure}[!ht]
\begin{center}
  \begin{tikzpicture}[
      thick
    ]
    \draw (0,2) circle (3pt);
    \draw[dashed] (0,2) -- (1,2);
    \walk
  \end{tikzpicture}
\end{center}
\caption{$\omega_b$, a self-avoiding walk of 16 sites.}
\label{fig:exampleb}
\end{figure}

\tikzstyle{level 1}=[level distance=2cm, sibling distance=4.4cm]
\tikzstyle{level 2}=[level distance=2cm, sibling distance=2.15cm]
\tikzstyle{level 3}=[level distance=2cm, sibling distance=1.05cm]
\begin{figure}[!ht]
\begin{center}
  \begin{tikzpicture}[
      thick,
      q/.style={rectangle,minimum height=1.75em,minimum width=1.75em,draw},
      leaf/.style={circle,minimum size=0mm,inner sep=2pt,draw}
    ]
    \node[q] (w) {
      \begin{tikzpicture}[thick,scale=0.22]
      \walk 
      \end{tikzpicture}
    }
    child {
      node[q] (l) {
        \begin{tikzpicture}[thick,scale=0.22]
          \path (1,-1) -- (5,3);
          \walkl 
        \end{tikzpicture}
      }
      child {
        node[q] (ll) {
          \begin{tikzpicture}[thick,scale=0.22]
            \path (1,-1) -- (5,3);
            \walkll 
          \end{tikzpicture}
        }
        child {
          node[q] (lll) {
            \begin{tikzpicture}[thick,scale=0.15]
              \path (1,-1) -- (5,3);
              \walklll 
            \end{tikzpicture}
          }
        }
        child {
          node[q] (llr) {
            \begin{tikzpicture}[thick,scale=0.15]
              \path (1,-1) -- (5,3);
              \walkllr 
            \end{tikzpicture}
          }
        }
      }
      child {
        node[q] (lr) {
          \begin{tikzpicture}[thick,scale=0.22]
            \path (1,-1) -- (5,3);
            \walklr 
          \end{tikzpicture}
        }
        child {
          node[q] (lrl) {
            \begin{tikzpicture}[thick,scale=0.15]
              \path (1,-1) -- (5,3);
              \walklrl 
            \end{tikzpicture}
          }
        }
        child {
          node[q] (lrr) {
            \begin{tikzpicture}[thick,scale=0.15]
              \path (1,-1) -- (5,3);
              \walklrr 
            \end{tikzpicture}
          }
        }
      }
    }
    child {
      node[q] (r) {
        \begin{tikzpicture}[thick,scale=0.22]
          \path (1,-1) -- (5,3);
          \walkr 
        \end{tikzpicture}
      }
      child {
        node[q] (rl) {
          \begin{tikzpicture}[thick,scale=0.22]
            \path (1,-1) -- (5,3);
            \walkrl 
          \end{tikzpicture}
        }
        child {
          node[q] (rll) {
            \begin{tikzpicture}[thick,scale=0.15]
              \path (1,-1) -- (5,3);
              \walkrll 
            \end{tikzpicture}
          }
        }
        child {
          node[q] (rlr) {
            \begin{tikzpicture}[thick,scale=0.15]
              \path (1,-1) -- (5,3);
              \walkrlr 
            \end{tikzpicture}
          }
        }
      }
      child {
        node[q] (rr) {
          \begin{tikzpicture}[thick,scale=0.22]
            \path (1,-1) -- (5,3);
            \walkrr 
          \end{tikzpicture}
        }
        child {
          node[q] (rrl) {
            \begin{tikzpicture}[thick,scale=0.15]
              \path (1,-1) -- (5,3);
              \walkrrl 
            \end{tikzpicture}
          }
        }
        child {
          node[q] (rrr) {
            \begin{tikzpicture}[thick,scale=0.15]
              \path (1,-1) -- (5,3);
              \walkrrr 
            \end{tikzpicture}
          }
        }
      }
    };
    \draw[ultra thick,<->] (l) -- (r);
    \draw[ultra thick,<->] (lr) -- (rl);
    \draw[ultra thick,<->] (ll) edge[out=-30,in=-150] (rr);
  \end{tikzpicture}
\end{center}
\caption{A SAW-tree representation of $\omega_b$, with arrows drawn
  between nodes on the left- and right-hand sides which approach each other.}
\label{fig:sawtree_exampleb}
\end{figure}

To this point our argument is exact, but we must now
resort to heuristic arguments to count the expected number of
approaches for $W$.

We first note that although there are typically $O(N)$ nearest
neighbor contacts for a SAW of length $N$, the
number of contacts between two halves of a SAW is typically $O(1)$, as shown
via renormalization group~\cite{Muller1998} and
Monte Carlo~\cite{Baiesi2001} methods. When we attempt to pivot part
of a SAW, it is guaranteed that each of the two sub-walks remain
self-avoiding, and hence we only need to determine if
the left- and right-hand sub-walks intersect. If the resulting walk is self-avoiding,
then we expect, on average, that there will be a constant number
of contacts between the
two sub-walks.

We now consider the renormalization group transformation along the
polymer chain described by
Kremer et al.~\cite{Kremer1981} in the context of a Monte Carlo
renormalization group calculation (see also
\cite{Gabay1978,Oono1979}), which maps a polymer chain with hard
sphere 
interactions to a new chain by rescaling the
number of monomers ($n$), the separation between monomers ($L$), and
the hard sphere diameter ($D$), so
as to keep the mean-square end-to-end distance fixed. 
The interaction
strength is defined as $\delta = D/L$. 
Kremer et
al. concluded that this renormalization group transformation converged
to a non-trivial (SAW) fixed point, with interaction strength
$\delta^\star > 0$. In the neighborhood of the fixed point the
renormalization transformations are
\begin{align}
n^\prime &= \frac{n}{s}, \\
L^\prime &= A s^\nu L, \text{\; and}\\
D^\prime &= A s^\nu D, 
\end{align}
for some positive constant $A$.

We can translate this to our SAW-tree representation of a
self-avoiding walk with $2^k$ sites if we regard the
sub-walks of each node at a particular level in the tree as the
monomers of a renormalized self-avoiding walk. 
We consider the
transformation that occurs as we pass from level $j$ to level $j+1$,
with $1 \ll 2^j \ll 2^k$. Sub-walks at level $j$ have $n_j = 2^{k-j}$
monomers, and as each sub-walk contains $2^{j}$ sites the mean
separation between monomers is therefore $L_j = 
A_1 (2^{j})^\nu$ for some constant $A_1$.
The
corresponding quantities for level $j+1$ are
\begin{align}
n_{j+1} &= 2^{k-j-1} = \frac{n_j}{2}, \\
L_{j+1} &= A_1 (2^{j+1})^\nu = 2^\nu L_j.
\end{align}
We note that the expectation of the perimeter of the bounding box of a
self-avoiding walk with
$n$ monomers is $A_2 n^\nu$ for some constant $A_2$ (see
Eq.~\ref{eq:perim} in
Sec.~\ref{sec:implementationdefinitions} for the definition of the perimeter).
If we take the (effective) hard sphere diameter at this particular
level $j$ to be 
some constant multiplied by the expected perimeter at level $j$,
$P_j$, $D_j = A_3 P_j = A_3 A_2 (2^{j})^\nu$, and we pass to level $j+1$,
then in order to keep the interaction strength fixed we must have
\begin{align}
D_{j+1} &= 2^\nu D_j \nonumber \\
&= 2^\nu A_3 P_j \nonumber \\
&= 2^\nu A_3 A_2 (2^{j})^\nu \nonumber \\
&= A_3 A_2 (2^{j+1})^\nu \nonumber \\
&= A_3 P_{j+1},
\end{align}
i.e. the new effective hard sphere diameter is given by the same
constant $A_3$ multiplied by $P_{j+1}$, and hence by induction $D_m =
A_3 P_m \; \forall m \geq j$.
Thus the nodes at level $j$  in a SAW-tree of total depth $k$, where
$1 \ll 2^j \ll 2^k$, may  be regarded as a self-avoiding walk with
$2^{k-j}$ monomers, separated by links of length $L_j = 
A_1 (2^{j})^\nu$, with hard sphere monomers of diameter $D_j = A_3
P_j$. This renormalization procedure becomes exact in the limit of
large $j$ and $k$ . 

We now note that \emph{approaches} for the SAW-tree correspond to
\emph{contacts} for the renormalized walks, where the minimum distance
at which two monomers of the walk are said to be in contact is a fixed
multiple of the hard sphere diameter. For fixed
values of the interaction strength parameter and contact distance 
the number of contacts between the left- and right-hand sides of a
walk is of $O(1)$, and we now know that as we proceed up
the tree from the leaves to the root we are converging to a fixed
value of the interaction strength. Thus the expected number of
contacts between the two halves of the walk also converges to a
constant.

Let the expected number of approaches at level $j$ for a walk with $n$
sites be $c_{n,j}$. The previous arguments imply that $\lim_{n
  \rightarrow \infty} c_{n,j} = c_j$ for some positive constant $c_j$, where
the sequence is not necessarily monotonic.
We then expect that the limit $\lim_{j \rightarrow
  \infty} c_j = c^\star$ exists, and that the $c_{n,j}$ are
bounded, i.e. there exists a
constant $c^\dagger$ such that $c_{n,j} < c^\dagger \; \forall
n,j$. 
Thus 
\begin{align}
T_{\mathrm{S}}(n) &= \sum_{j=0}^{\log_2 n} c_{n,j} \nonumber \\
 &< \sum_{j=0}^{\log_2 n} c^\dagger \nonumber \\
& = c^\dagger \log_2 n,
\end{align}
and so
$T_{\mathrm{S}}(N) = O(\log N)$. In fact, we expect that
 $c_{n,j} \approx c^\star$ for sufficiently large $n$ and $j$, and
so $T_{\mathrm{S}}(N) = \Theta(\log N)$.

\subsection{Algorithmic complexity of {\textbf{Attempt\-\_pivot\-\_simple}}}
\label{sec:complexitysimple}

On each iteration \textbf{Attempt\-\_pivot\-\_simple} executes the
procedures \textbf{Shuffle\-\_up} and \textbf{Shuffle\-\_down} as well as
the function \textbf{Intersect}.

The expected running time of both \textbf{Shuffle\-\_up} and
\textbf{Shuffle\-\_down} is independent of whether the pivot attempt is
successful or not. From Sec.~\ref{sec:complexitybackground}, the expected level of a
SAW-tree node selected uniformly at random is two. Thus
the expected number of tree-rotations that \textbf{Shuffle\-\_up}
must perform in order to bring a pivot node
to the top of a SAW-tree with $k$ levels, or \textbf{Shuffle\-\_down}
must perform to 
restore the node to its
correct place in the tree, is $k - 2 + O(k / 2^k)$. Given
that $k = \log_2 n$, the algorithmic complexity of both \textbf{Shuffle\-\_up}
and \textbf{Shuffle\-\_down} is therefore $O(\log N)$.

We already know \textbf{Intersect} takes time $\Theta(\log N)$ in the
case that the pivot attempt is successful, which combined with the
$\Theta(\log N)$ complexity of the shuffle operations gives
\begin{align}
  T_{\mathrm{S}}(N) &= \Theta(\log N).
\end{align}

When the pivot attempt is
unsuccessful and an intersection is found, \textbf{Intersect}
terminates early and thus takes time $O(\log N)$ (we will argue
in Sec.~\ref{sec:complexityfast} that it is in fact $O(1)$, but this
makes no difference to the argument here). Combined with the
$\Theta(\log N)$ complexity of the shuffle operations this results in
\begin{align}
  T_{\mathrm{U}}(N) &= \Theta(\log N).
\end{align}

For \textbf{Attempt\-\_pivot\-\_simple} we therefore have overall
algorithmic complexity of 
\begin{align}
T(N) &= N^{-p} T_{\mathrm{S}}(N) + T_{\mathrm{U}}(N) \nonumber \\
&= N^{-p} \Theta(\log N) + \Theta(\log N) \nonumber \\
&= \Theta(\log N).
\end{align}

\subsection{Algorithmic complexity of {\textbf{Attempt\-\_pivot\-\_fast}}}
\label{sec:complexityfast}

On each iteration \textbf{Attempt\-\_pivot\-\_fast} executes the
function \textbf{Shuffle\-\_intersect}, while it only executes the 
procedures \textbf{Shuffle\-\_up} and \textbf{Shuffle\-\_down} when a
pivot attempt is successful.

The key point to understand regarding the action of
\textbf{Shuffle\-\_intersect}, is that this function shuffles the pivot
node up the SAW-tree only as far as necessary to find an
intersection between the left- and right-hand sub-walks. In the case
that the pivot attempt is
successful, \textbf{Shuffle\-\_intersect} must therefore shuffle the
pivot node up 
to the top of the SAW-tree, but in the case where the pivot attempt is
unsuccessful the pivot node is only shuffled up just far enough to be
able to find this first intersection.

When the pivot attempt is successful, \textbf{Shuffle\-\_intersect} must
search the whole SAW-tree for intersections, and so is exactly
equivalent to performing \textbf{Shuffle\-\_up}, \textbf{Intersect}, and
\textbf{Shuffle\-\_down}. As shown in
Sections~\ref{sec:complexitysuccess} and \ref{sec:complexitysimple},
all of these procedures take time $\Theta(\log N)$, and thus
\begin{align}
  T_{\mathrm{S}}(N) &= \Theta(\log N).
\end{align}

The situation is considerably more complicated when the pivot attempt
is unsuccessful and an intersection is found.
We suppose that the pivot node is the $j^{\mathrm{th}}$
(internal) node from the left, which has $j$ sites (leaves) to the left, and
$n-j$ sites to the right.
\textbf{Shuffle\-\_intersect} will then search $W$ 
until it finds
the intersection with $j - i_l$ being the label of the left-hand site and
$j + i_r$ the label of the right-hand site such that $\sup(i_l,i_r)$
is minimum. Note: this description
is approximate, as the SAW-tree structure complicates the issue of
exactly which pair of intersecting sites is found, and the actual
$i_l$ and $i_r$ which are found may be different from the optimal values.
However, we are confident that for the average case this argument is
correct up to a 
constant factor in the  size of $i_l$ and $i_r$.

The sites $j - i_l$ and $j+i_r$ are exactly the pair of intersecting
sites which would be discovered in the 
usual implementation of the pivot algorithm~\cite{Madras1988},
where the new walk is incrementally built by moving outwards from the pivot
site. Let us assume, without loss of generality, that $i_r > i_l$.
We define $\Pr(i)$ as
the probability that $i$ is the intersection that is found, i.e. the
conditional probability that the sub-walk with sites $[j-i,j+i]$ has a
self-intersection, but there are no intersections between sites
in the open interval $(j-i,j+i)$. 
Madras and Sokal~\cite{Madras1988} convincingly argued that
the expected value of $i_r$ is
\begin{align}
\mathbb{E}(i_r) &= \sum_{i_r=1}^{2^k-j} i_r \Pr(i_r) \nonumber \\
&= O(N^{1-p}),
\label{eq:intersection}
\end{align}
with $p$ positive and close to zero for $\Z^2$ and $\Z^3$. 

The interval $[j-i_l,j+i_r]$ corresponds to a self-avoiding loop, and
it may be quite
difficult to characterize the probability space of these
configurations. We strongly suspect that these configurations are sufficiently
close to self-avoiding walks so that the mean number of approaches
between the intervals $[j-i_l,j)$ and $[j,j+i_r]$ is $O(\log i_r)$.
However, we do not have an argument to support this
statement, and so we now make the \emph{assumption} that 
the mean time for 
\textbf{Shuffle\-\_intersect} to confirm that the sub-walk $(j-i_r,j+i_r)$ 
is non-intersecting is $O(\log i_r)$. To then find the
intersection between $j - i_l$ and $j + i_r$ will take 
additional time of $O(\log i_r)$.

We now take $j$ to be a typical node, which places it near
the middle of the walk with $j \approx 2^{k-1}$, and at level two in
the SAW-tree.
To determine the expected time to find the intersection between $i_l$
and $i_r$, we need to determine the probability distribution for
$i_r$, given $j = 2^{k-1} = N/2$ (fixing $j$ makes the argument simpler, but
does not change the result). 
Eq.~\ref{eq:intersection} suggests that for sufficiently large $i_r$,
$\Pr(i_r) = O(i_r^{-1-p})$. 
We therefore neglect sub-dominant terms and take the leading order
approximation
$\Pr(i_r) \approx A / i_r^{1+p}$;
this may be a very bad
approximation for small $i_r$.

Using this approximation, we can now proceed to determine $T_{\mathrm{U}}(N)$:
\begin{align}
T_{\mathrm{U}}(N) &= \sum_{i_{\mathrm{r}}=1}^{N/2} \text{(Time to
  find intersection)} \Pr(i_r) \nonumber \\
&= \sum_{i_{\mathrm{r}}=1}^{N/2} \log i_r \frac{A}{i_r^{1+p}}.
\end{align}
We observe that for large $N$ $T_{\mathrm{U}}(N)$ approaches a constant from below,
i.e. $T_{\mathrm{U}}(\infty) = O(1)$. To determine the rate of approach, we examine
$T_{\mathrm{U}}(\infty) - T_{\mathrm{U}}(N)$:
\begin{align}
 \nonumber \\
T_{\mathrm{U}}(\infty) - T_{\mathrm{U}}(N) &\approx \int_{N/2}^{\infty} \log i_r \frac{A}{i_r^{1+p}} di_r \nonumber \\
&= \frac{A}{p^2} \left(\frac{N}{2}\right)^{-p} \left( p \log\frac{N}{2} + 1 \right)
\end{align}
Without detailed understanding 
of the exact form of $\Pr(i_r)$, it is impossible to estimate
$T_{\mathrm{U}}(\infty) - T_{\mathrm{U}}(N)$ for any particular value
of $N$.
However, we can gain a rough
idea of the rate of approach to the constant $T_{\mathrm{U}}(\infty)$
by determining a natural length scale, $L$, from the solution of
\begin{align}
\frac{T_{\mathrm{U}}(\infty) - T_{\mathrm{U}}(L)}{T_{\mathrm{U}}(\infty) -
T_{\mathrm{U}}(1)} &= \frac{1}{e}.
\label{eq:naturallength}
\end{align}
i.e. $L$ is the length at
which our approximation for the deviation from the limiting value has decayed to
$1/e$ of its initial value.
For $\Z^2$, $p \approx 0.19$ which gives $L \approx 1.6\e{5}$, while for
$\Z^3$, $p \approx 0.11$ and so $L \approx 5.8\e{8}$.
We can clearly see that convergence to constant behavior may indeed be
very slow, particularly for $\Z^3$. 

For \textbf{Attempt\-\_pivot\-\_fast} we therefore have overall
algorithmic complexity of 
\begin{align}
T(N) &= N^{-p} T_{\mathrm{S}}(N) + T_{\mathrm{U}}(N) \nonumber \\
&= O\left(N^{-p} \log N + 1\right) \nonumber \\
&= O\left(1\right).
\end{align}
In practice, sub-leading terms may result in
behavior which is quite different from $O(1)$ for lengths of the order
of millions or even billions of steps.

\subsection{$d > 3$}
\label{sec:complexityhighdim}

For $d > 3$ the time spent on successful pivots is non-negligible, and
this changes the overall mean time per attempted pivot.

As discussed by Madras and Sokal~\cite{Madras1988}, we expect that
the exponent associated with the acceptance fraction, $p$, is of the
same order of magnitude as the critical exponent $\gamma$, with a
heuristic argument that $p 
\lesssim \gamma - 1$.
It appears to be the case for
$\Z^2$ and $\Z^3$ that $p < \gamma - 1$; if $p \leq \gamma - 1$ holds for $d >
3$, then this implies that $p$ is zero for $d \geq 4$, with
perhaps a logarithmic correction for $d=4$.

For $d=4$, the number of self-avoiding walks of length $N$, $c_N$, is
given asymptotically by $c_N \sim A \left(\log N\right)^{1/4}
\mu^N$. Therefore, the probability of being able to successfully
merge two SAWs (see Sec.~\ref{sec:implementationsaw} for definition of
the merge operation)
of $N$ steps ($N+1$ sites) to form a SAW of length $2 N + 1$ ($2N+2$ sites)
is 
\begin{align}
\frac{c_{2N+1}}{c_N^2} &\approx \frac{A \left(\log (2N+1)\right)^{1/4}
  \mu^{2N+1}}
     {A^2  \left(\log N\right)^{1/2} \mu^{2N}} \nonumber \\
&= O\left(\left(\log N\right)^{-1/4}\right).
\label{eq:concatd4}
\end{align}
We have already observed that $\gamma - 1 \neq p$ for $\Z^3$, and
so we expect that
the exponent of $\log N$ in Eq.~\ref{eq:concatd4} is unlikely to be
the same as for  
the probability of a successful pivot. A plausible guess for the
probability is
$\Pr(\text{successful})= O((\log N)^{-\kappa})$, with $0 < \kappa <
1$. We have performed a 
preliminary analysis of the data for the acceptance fraction, $f$, for
$\Z^4$, which suggests that $\kappa \lesssim 0.42$; however, the plot of
$\log(\log(-f))$ versus $\log \log N$ has not settled down to
linearity by $N = 2^{28} - 1 \approx 2.7\e{8}$, and so the error on this 
estimate may well be quite large. 
Assuming this functional form is correct, we obtain
\begin{align}
\Pr(i_r) &= -\frac{d}{di_r} \left(\log i_r\right)^{-\kappa} \nonumber \\
&= O\left(i_r^{-1} \left(\log i_r\right)^{-\kappa - 1}\right)
\end{align}
\begin{align}
T_{\mathrm{U}}(N) &\approx \int_{1}^{N} \log i_r \; i_r^{-1} \left(\log
i_r\right)^{-\kappa - 1} di_r \nonumber \\
&= \int_{1}^{N} i_r^{-1} \left(\log i_r\right)^{-\kappa} di_r \nonumber \\
&= O\left(\left(\log N \right)^{1-\kappa}\right).
\end{align}
Assuming that $0 < \kappa < 1$, we have
\begin{align}
T(N) &= \Pr(\text{successful}) T_{\mathrm{S}}(N) +
\Pr(\text{unsuccessful}) T_{\mathrm{U}}(N) \nonumber \\
&= \left(\log N \right)^{-\kappa} T_{\mathrm{S}}(N) + T_{\mathrm{U}}(N) \nonumber \\
&= O\left(\left(\log N \right)^{-\kappa} \log N + \left(\log N \right)^{1-\kappa}\right) \nonumber \\
&= O\left(\left(\log N \right)^{1-\kappa}\right).
\end{align}
Note that the time spent on successful pivots is of precisely the same
order as the time spent on unsuccessful pivots. 
Once again, we mention that the above expression is based
on a plausible but untested assumption. We are confident that
$T(N) = \upomega(1)$, 
and also $T(N)=o(\log N)$ (this is the expression reported in
Tables~\ref{tab:performance} and \ref{tab:performanceb}). 

For $\Z^2$ and $\Z^3$, we could see that as $p \rightarrow 0$, the
natural length scale $L$ obtained from Eq.~\ref{eq:naturallength}
diverges. For this reason, we expect that deviations 
from the leading order behavior of $T(N)$ for $\Z^4$, where $p=0$ with
logarithmic corrections, may be large even for extremely long walks
with $10^9$ steps or more.

For $d > 4$, we can use the mean-field approximation and ignore long
range correlations within the chain, by assuming that a sub-walk of
length $N$ consists of sites selected uniformly at random from a
sphere of radius $r = O(N^{\nu}) = O(N^{1/2})$. Assuming that we already
have a self-avoiding sub-walk with sites in the interval $(j - i_r,
j+i_r)$, then 
the probability that the site $j+i_r$ intersects with one of the
previous sites is
$\Pr(i_r) = O(i_r / r^d) = O(i_r^{1-d/2})$.
Therefore
\begin{align}
T_{\mathrm{U}}(N) &\approx \int_{1}^{N} \log i_r i_r^{1-d/2} di_r \nonumber \\
&= O(1).
\end{align}
Finally, we have
\begin{align}
T(N) &= \Pr(\text{successful}) T_{\mathrm{S}}(N) +
\Pr(\text{unsuccessful}) T_{\mathrm{U}}(N) \nonumber \\
&= O(1) T_{\mathrm{S}}(N) + O(1) T_{\mathrm{U}}(N) \nonumber \\
&= O(\log N + 1) \nonumber \\
&= O(\log N).
\end{align}
As $T_{\mathrm{S}}(N) = \Theta(\log N)$, this leads to the stronger
statement that $T(N) = \Theta(\log N)$.
Note that the complexity of the algorithm is now dominated by the time
spent on successful pivots.

\subsection{Numerical evidence}
\label{sec:complexitynumerical}

\begin{figure}[!ht]
\begin{center}
\resizebox{4.375in}{!}{\includegraphics{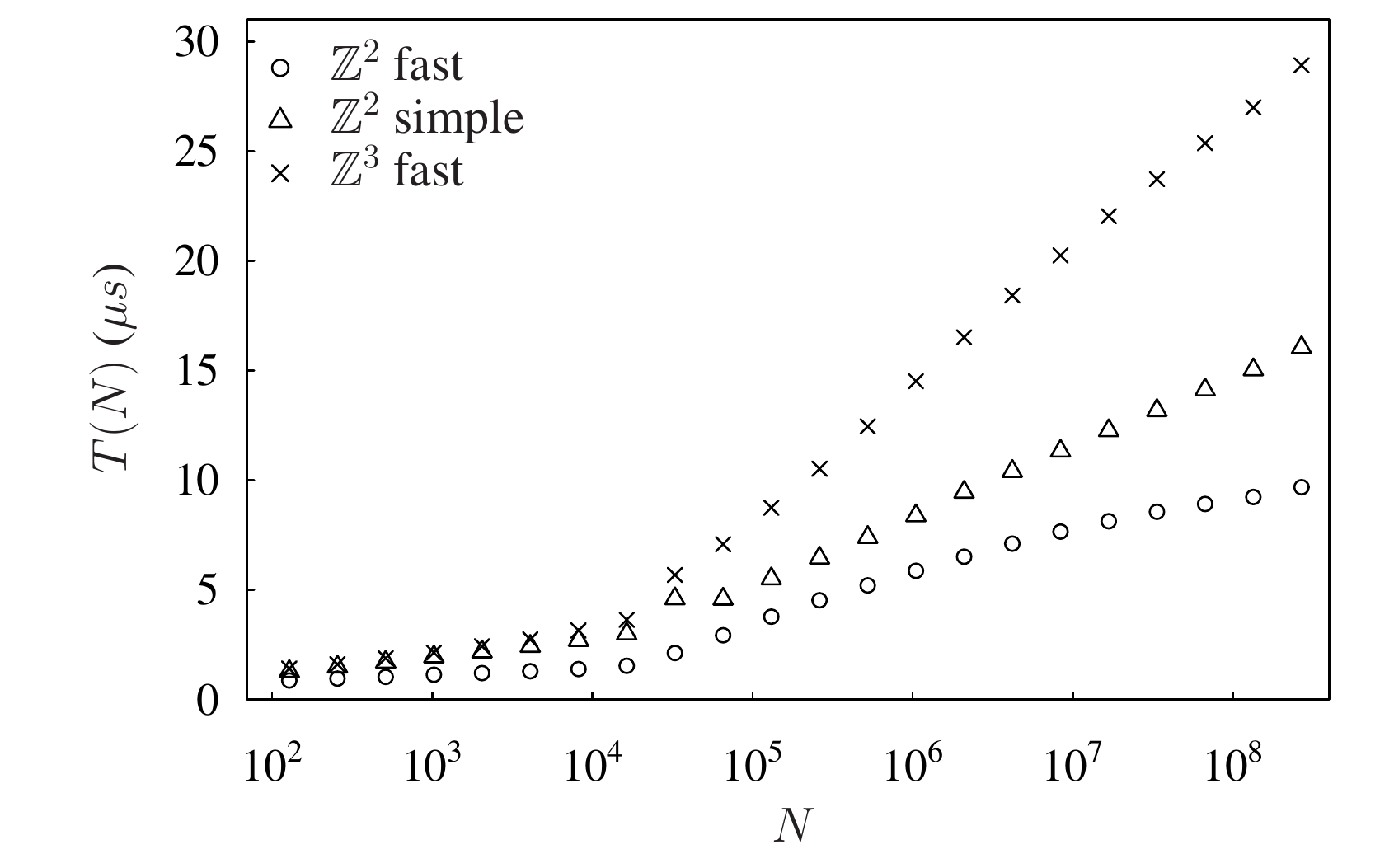}}
\end{center}
\caption{$T(N)$ for $\Z^2$,
  and
  $\Z^3$, where for $\Z^2$ data were
  collected for both
  the simple and fast versions of \textbf{Attempt\-\_pivot}.} 
\label{fig:cputimes}
\end{figure}

\begin{figure}[!ht]
\begin{center}
\resizebox{4.375in}{!}{\includegraphics{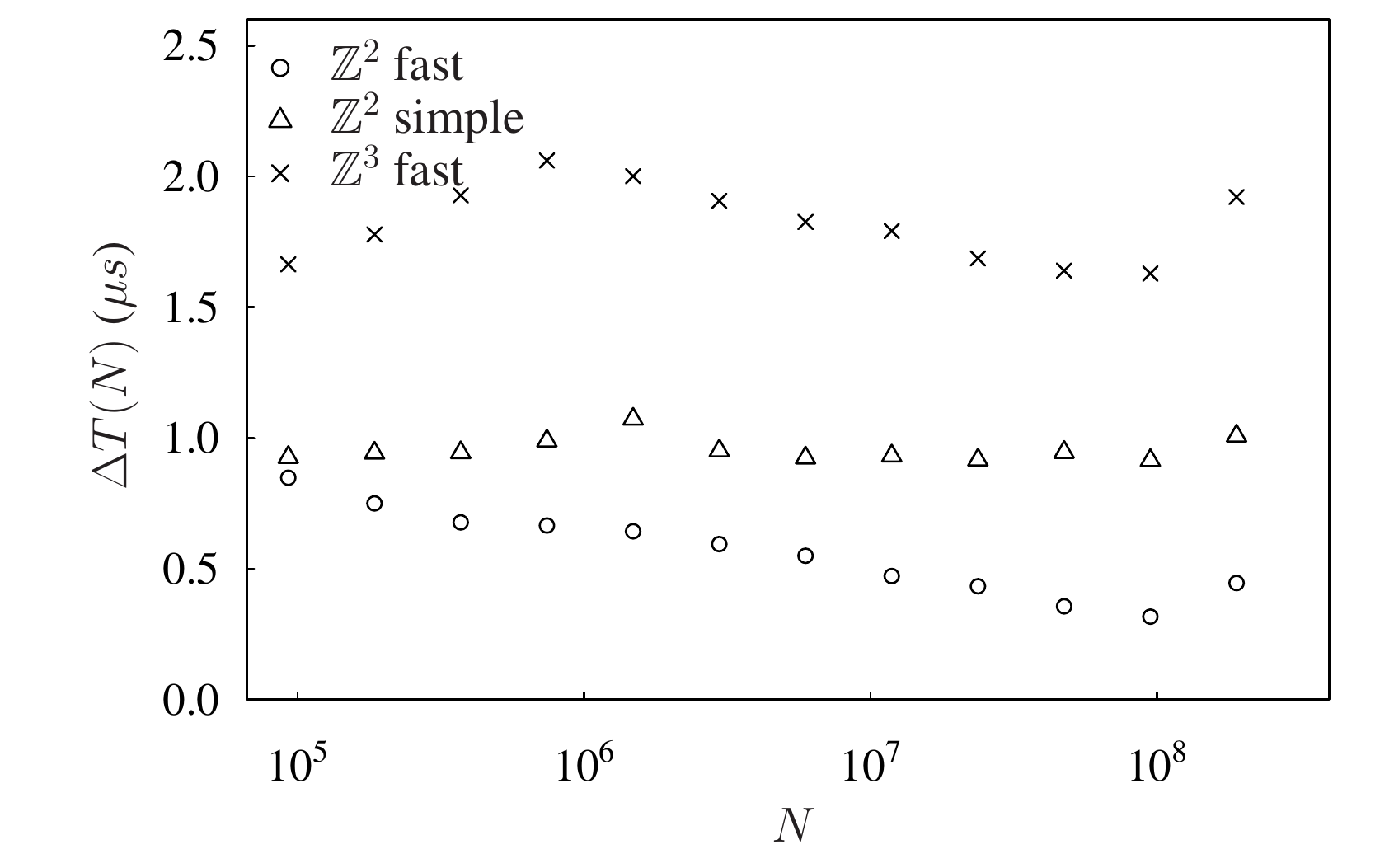}}
\end{center}
\caption{$\Delta T(N)$ for $\Z^2$ and $\Z^3$, where
  for $\Z^2$ data were 
  collected for both
  the simple and fast versions of \textbf{Attempt\-\_pivot}.
}
\label{fig:deltacputimes}
\end{figure}

In Fig.~\ref{fig:cputimes} we present $T(N)$ for
\textbf{Attempt\-\_pivot\-\_fast} ($\Z^2$ and $\Z^3$), and
\textbf{Attempt\-\_pivot\-\_simple} ($\Z^2$), for lengths from
  $N = 2^7 - 1$ to $N = 2^{28} - 1 \approx 2.7\e{8}$. In
Fig.~\ref{fig:deltacputimes} we present 
\begin{align}
\Delta T(N) \equiv T(N\sqrt{2}) - T(N/\sqrt{2});
\end{align}
the domain of $N$ for this plot is reduced due to the crossover in
algorithm performance which occurs at $N \approx 10^4$.
These estimates for $T(N)$
were obtained in a separate data run from the main computer experiment in the
  companion article~\cite{Clisby2010}. The computers used were
SunFire X4600M2 machines with 8 quad-core AMD
Barcelona CPUs with clock frequency 2.3GHz, and 64 GB memory. 

In Fig.~\ref{fig:cputimes}, the plot for
\textbf{Attempt\-\_pivot\-\_simple} appears to be linear, and this is
verified in Fig.~\ref{fig:deltacputimes} where $\Delta T$ appears to
be constant, providing support for the prediction that $T(N) = \Theta(\log N)$.
The plot for
\textbf{Attempt\-\_pivot\-\_fast} for $\Z^2$ appears to be consistent with
eventually approaching a constant; in Fig.~\ref{fig:cputimes} we see
that $T(N)$ is visibly curved for $N > 10^5$, and is plausibly
 $o(\log N)$, while in
Fig.~\ref{fig:deltacputimes} we observe that $\Delta T(N)$ appears to
be steadily 
declining. The final point increases somewhat, but there is an
anomalous increase for all three data sets which strongly suggests
this is an artifact. 
For the final data point a large fraction
of machine memory was used, and so it is very likely that
some limit
with the hardware was reached, causing degraded performance.
For $\Z^3$, there is no curvature visually
apparent in Fig.~\ref{fig:cputimes} which suggests $T(N) = O(\log N)$,
but if we examine 
Fig.~\ref{fig:deltacputimes} we see weak 
numerical evidence supporting $T(N) = o(\log N)$, due to
the decline in $\Delta T(N)$ for $N > 10^6$.

Rather than directly measuring the run-time of a computer experiment,
it is also possible to directly measure the mean number of
intersection tests required, $I(N)$, per attempted pivot for SAWs of
$N$ steps. For \textbf{Attempt\-\_pivot\-\_fast}, we can relate $T(N)$
to $I(N)$ as follows:
\begin{align}
T_{\mathrm{S}}(N) &= \log N +  I_{\mathrm{S}}(N), \\
T_{\mathrm{U}}(N) &= I_{\mathrm{U}}(N),
\end{align}
where $I_{\mathrm{S}}(N)$ and $I_{\mathrm{U}}(N)$ are the mean number of
intersection tests when a pivot attempt is successful and unsuccessful
respectively. 
Determining $I(N)$ is far cleaner than directly measuring $T(N)$,
which is affected by hardware, and it is also straightforward to
separately determine $I_{\mathrm{S}}(N)$ and $I_{\mathrm{U}}(N)$.

\begin{figure}[!ht]
\begin{center}
\resizebox{4.375in}{!}{\includegraphics{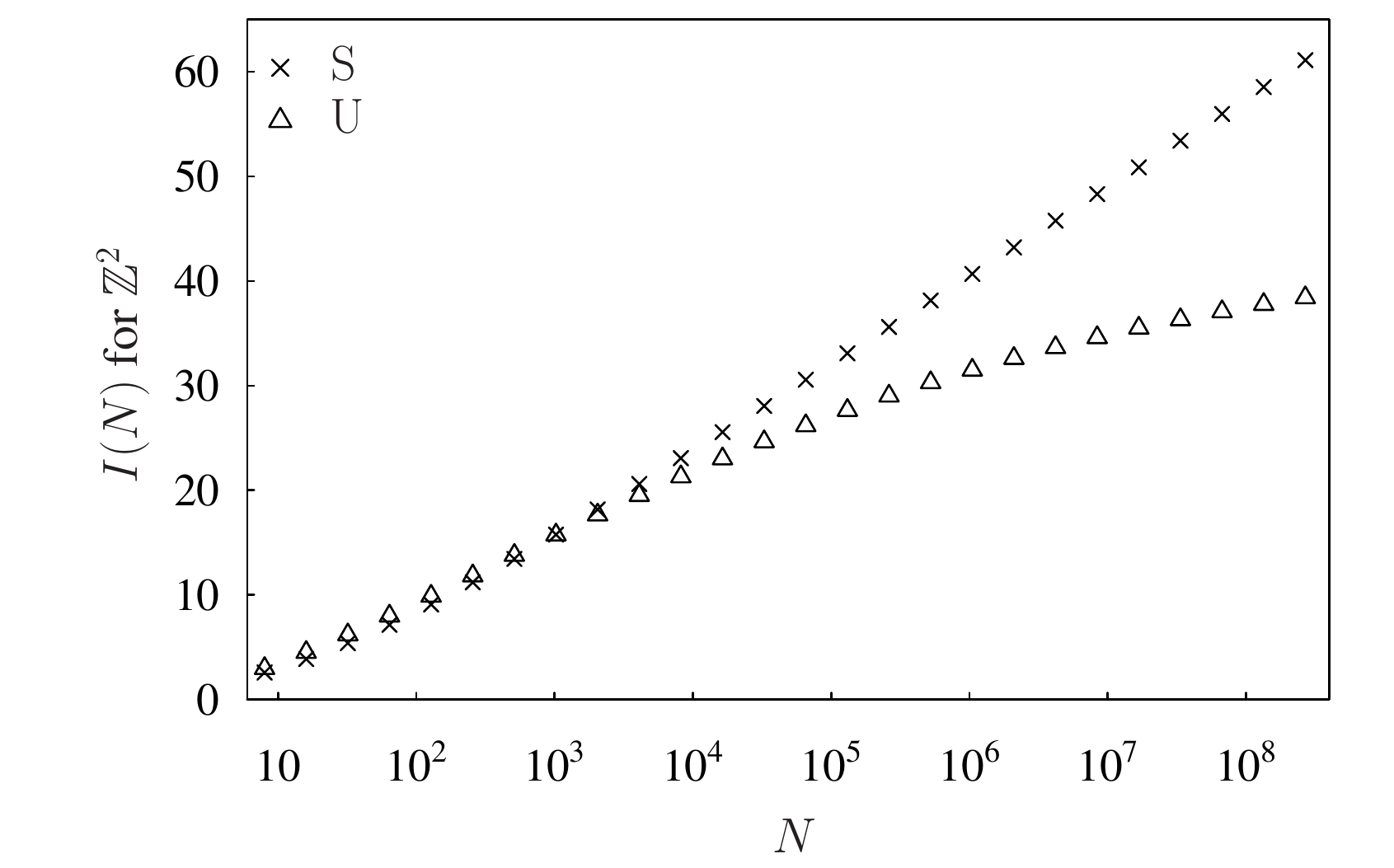}}
\end{center}
\caption{$I_{\mathrm{S}}(N)$ and $I_{\mathrm{U}}(N)$ for
  \textbf{Attempt\-\_pivot\-\_fast} on $\Z^2$.}
\label{fig:d2intersect}
\end{figure}

\begin{figure}[!ht]
\begin{center}
\resizebox{4.375in}{!}{\includegraphics{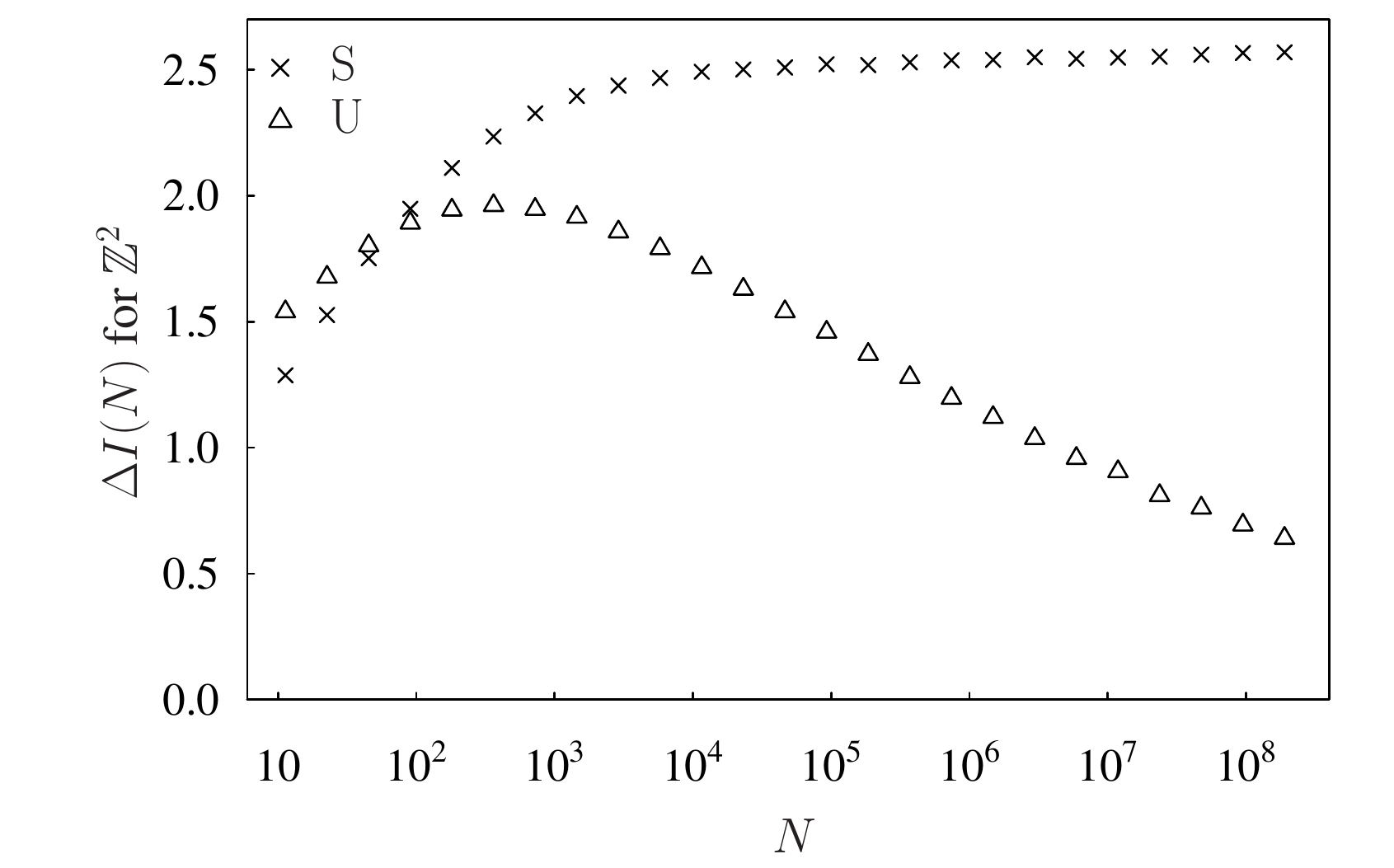}}
\end{center}
\caption{$\Delta I_{\mathrm{S}}(N)$ and $\Delta I_{\mathrm{U}}(N)$ for
  \textbf{Attempt\-\_pivot\-\_fast} on $\Z^2$.}
\label{fig:Dd2intersect}
\end{figure}

\begin{figure}[!ht]
\begin{center}
\resizebox{4.375in}{!}{\includegraphics{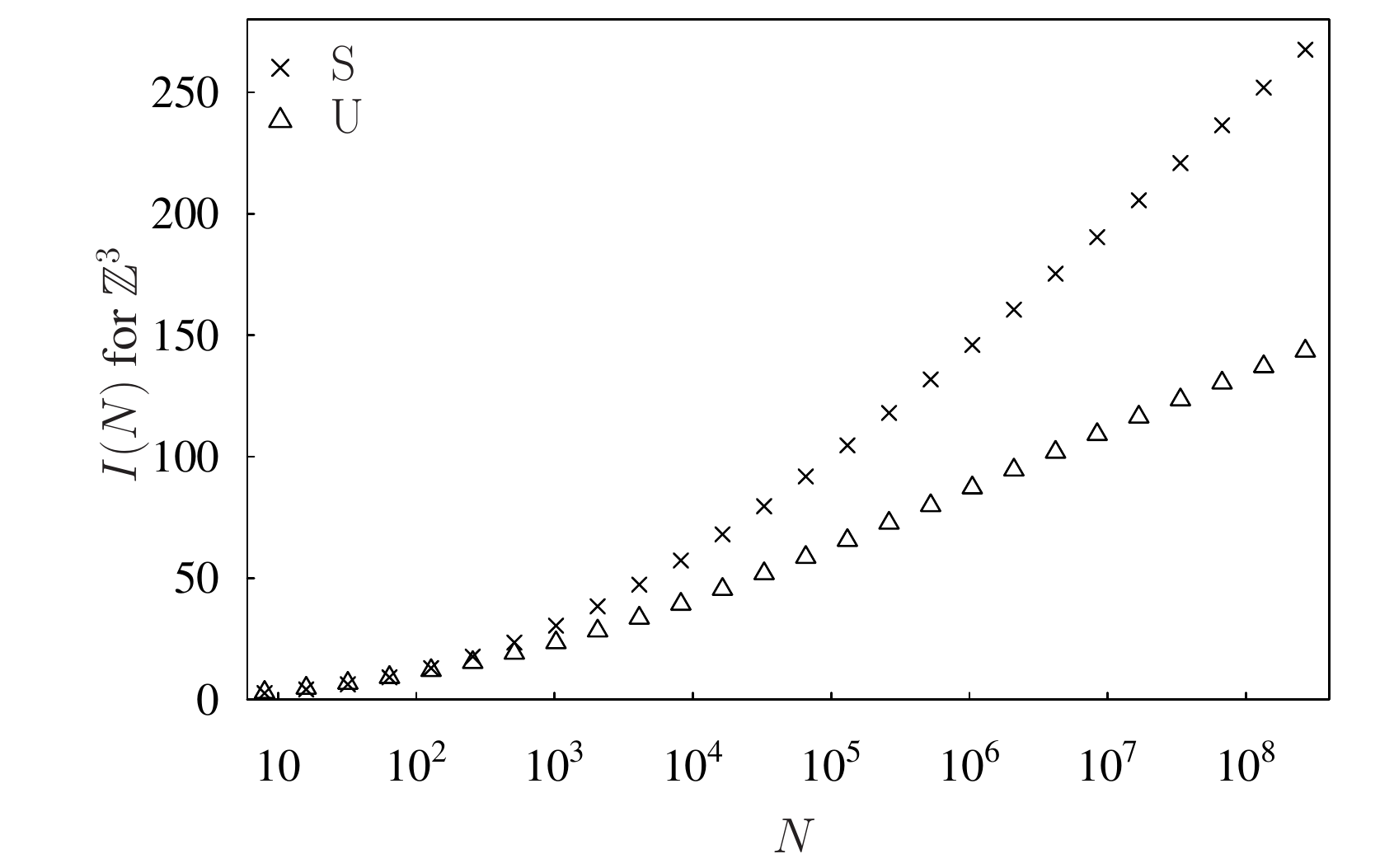}}
\end{center}
\caption{$I_{\mathrm{S}}(N)$ and $I_{\mathrm{U}}(N)$ for
  \textbf{Attempt\-\_pivot\-\_fast} on $\Z^3$.}
\label{fig:d3intersect}
\end{figure}

\begin{figure}[!ht]
\begin{center}
\resizebox{4.375in}{!}{\includegraphics{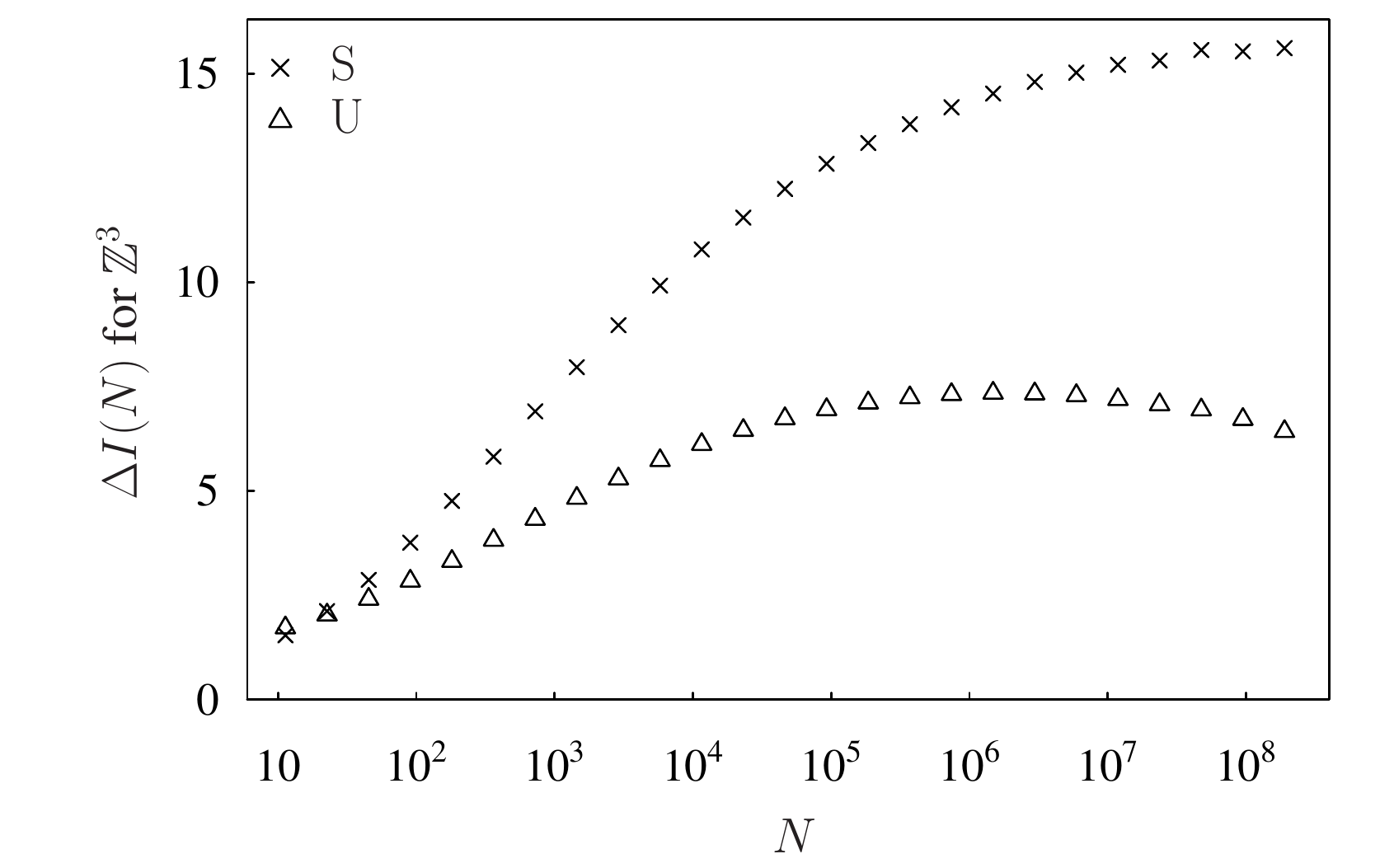}}
\end{center}
\caption{$\Delta I_{\mathrm{S}}(N)$ and $\Delta I_{\mathrm{U}}(N)$ for
  \textbf{Attempt\-\_pivot\-\_fast} on $\Z^3$.}
\label{fig:Dd3intersect}
\end{figure}

\begin{figure}[!ht]
\begin{center}
\resizebox{4.375in}{!}{\includegraphics{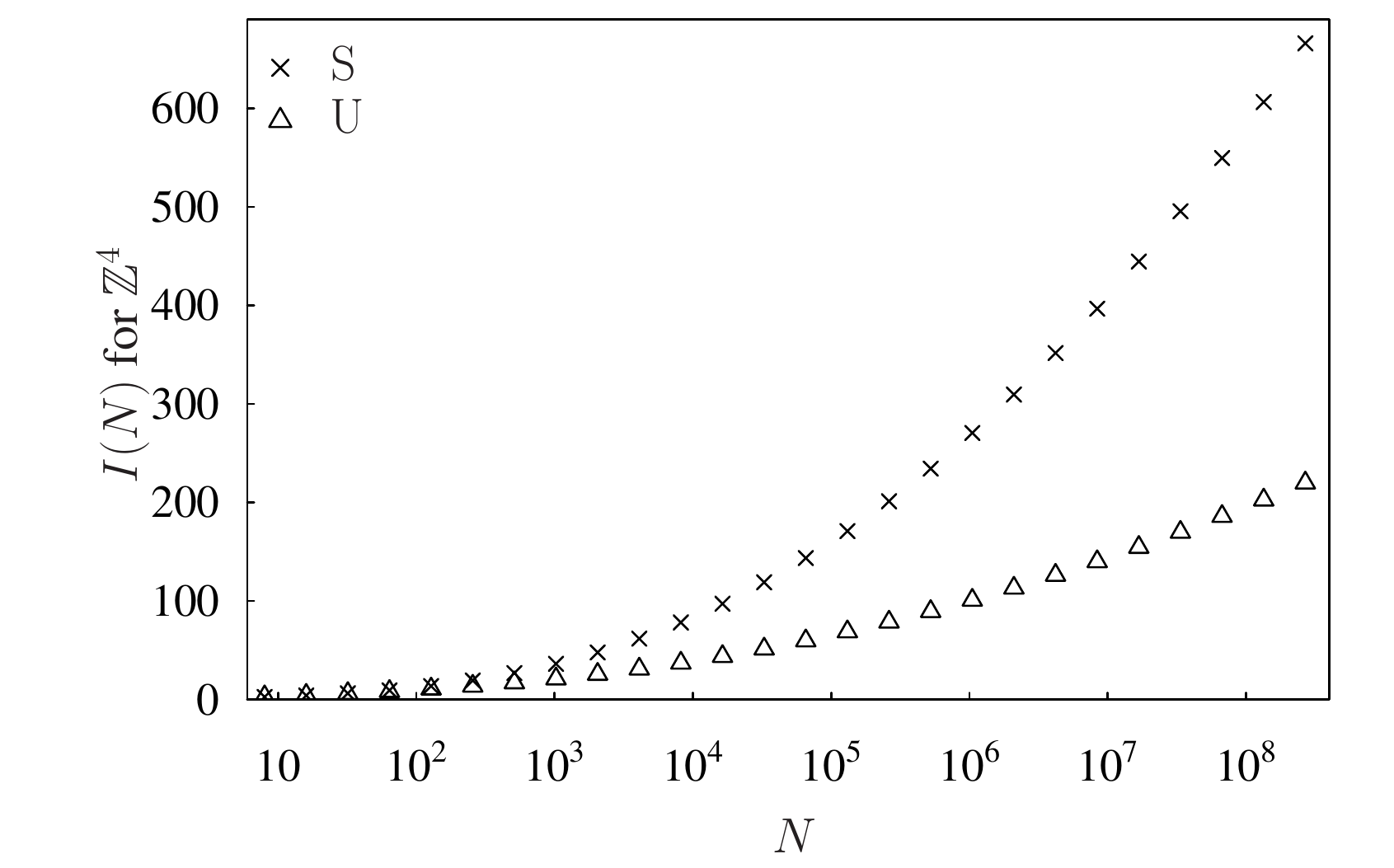}}
\end{center}
\caption{$I_{\mathrm{S}}(N)$ and $I_{\mathrm{U}}(N)$ for
  \textbf{Attempt\-\_pivot\-\_fast} on $\Z^4$.}
\label{fig:d4intersect}
\end{figure}

\begin{figure}[!ht]
\begin{center}
\resizebox{4.375in}{!}{\includegraphics{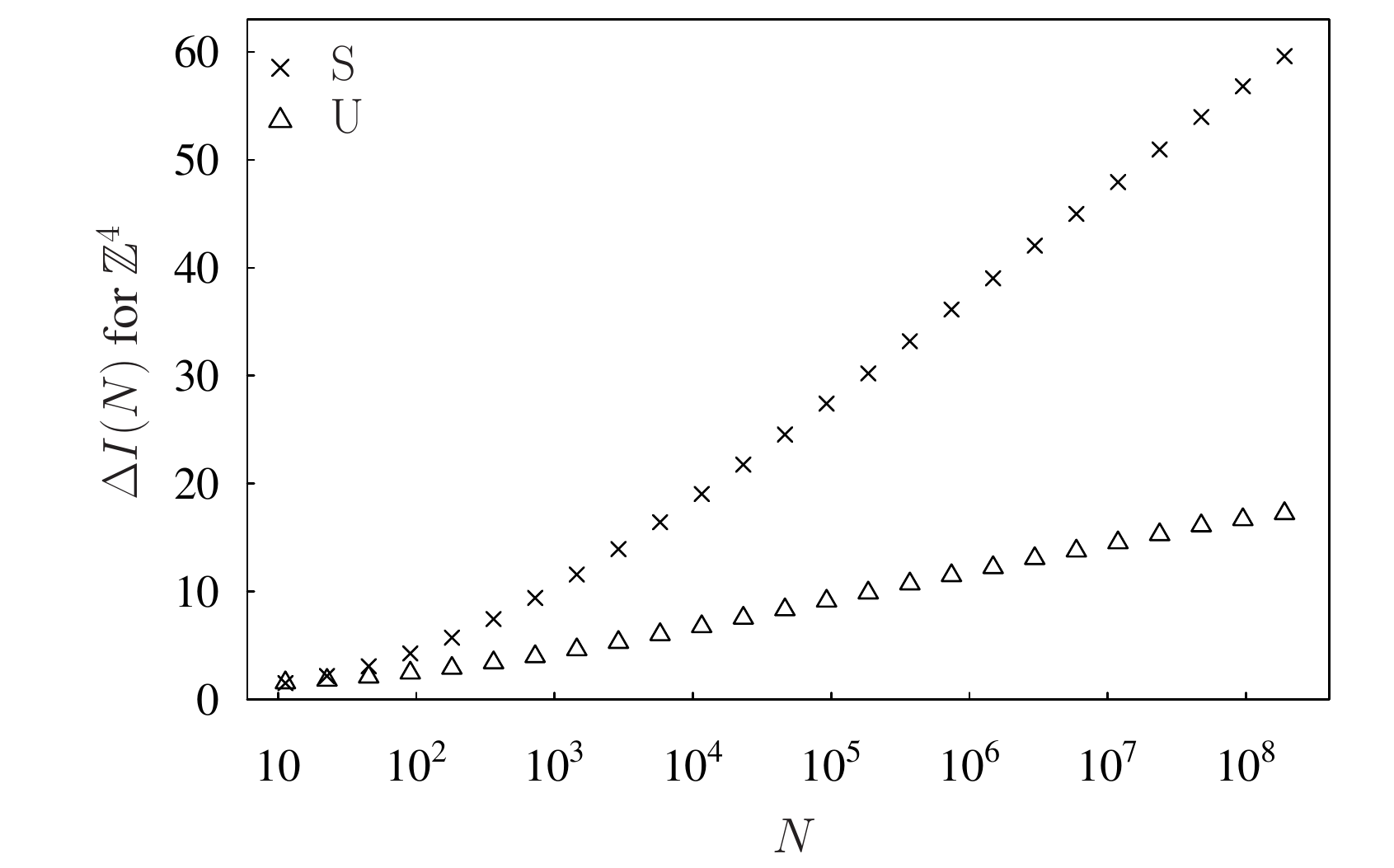}}
\end{center}
\caption{$\Delta I_{\mathrm{S}}(N)$ and $\Delta I_{\mathrm{U}}(N)$ for
  \textbf{Attempt\-\_pivot\-\_fast} on $\Z^4$.}
\label{fig:Dd4intersect}
\end{figure}

We have performed another computer experiment where we measured
$I_{\mathrm{S}}(N)$, $I_{\mathrm{U}}(N)$,  and $I(N)$, with lengths from
  $N = 7$ to $N = 2^{28} - 1 \approx 2.7\e{8}$. In each case the
initial $10^9$ configurations were discarded (independent of $N$),
and then 100 batches of 
$10^8$ configurations were generated.
The data from this computer experiment is presented visually in
Figs.~\ref{fig:d2intersect}-\ref{fig:Dd4intersect}, where 
\begin{align}
\Delta I(N) \equiv I(N\sqrt{2}) - I(N/\sqrt{2}).
\end{align}
This procedure does not properly initialize the Markov chain for large
$N$, but we have
checked the resulting time series for any
evidence of systematic errors, and we are confident that the
systematic error is small compared to the statistical error. We do not
show the statistical errors in
Figs.~\ref{fig:d2intersect}-\ref{fig:Dd4intersect}, but the largest
errors are smaller than the data point symbols; the small amount of
scatter visible in plots of $\Delta I(N)$ is consistent with these
errors. 

We note in passing that the mean number of intersection tests required
per attempted pivot is remarkably low: for $N = 2^{28}-1$, $I(N)$ is 39 for $\Z^2$, 158
for $\Z^3$, and 449 for $\Z^4$.

Examining the evidence for $\Z^2$ from Figs.~\ref{fig:d2intersect} and
~\ref{fig:Dd2intersect}, we can see clearly that $I_{\mathrm{S}}(N)$
is convex up to $N = 2^{28}-1$, and as $N \rightarrow \infty$
the lin-log plot of $I_{\mathrm{S}}(N)$ smoothly approaches a straight
line strongly supporting the statement $I_{\mathrm{S}}(N) =
\Theta(\log N)$. This leads to
$T_{\mathrm{S}}(N) = \Theta(\log N)$, in accordance with our conclusion in
Sec.~\ref{sec:complexitysuccess}. The evidence for $I_{\mathrm{U}}(N)$
is less conclusive: $I_{\mathrm{U}}(N)$ is convex for small $N$, but
becomes concave once a threshold is reached. 
We do not feel that it is likely that there is an additional length
scale beyond $N = 2^{28}-1$ where the behavior of $I_{\mathrm{U}}(N)$
would change, and therefore believe that the most likely scenario is 
that $\Delta I_{\mathrm{U}}(N)$ will decay smoothly to zero, which
would imply $I_{\mathrm{U}}(N) = o(\log N)$. This argument is by no means
conclusive, but it is consistent with the prediction from
Sec.~\ref{sec:complexityfast} that $I_{\mathrm{U}}(N) = O(1)$.

The evidence for $\Z^3$ from Figs.~\ref{fig:d3intersect} and
~\ref{fig:Dd3intersect} is less clear. In Sec.~\ref{sec:complexityfast}
we argued that for $\Z^3$ the asymptotic regime would be reached at
far greater lengths than for $\Z^2$, and this certainly appears to be
the case. The graphs for $I_{\mathrm{S}}(N)$ suggest that
$I_{\mathrm{S}}(N) = \Theta(\log N)$, but we do not believe the evidence is
particularly strong, as there is no extended flat region for
$\Delta I_{\mathrm{S}}(N)$ (as can be seen for $\Z^2$). The plots for
$I_{\mathrm{U}}(N)$  show that $I_{\mathrm{U}}(N)$ is convex for $N$
up to approximately $N = 10^6$, before becoming concave.
If the downward trend
in $\Delta I_{\mathrm{U}}(N)$ were to continue in the same manner as
for $\Z^2$, then this would imply $I_{\mathrm{U}}(N) = o(\log N)$.
However, the numerical evidence supporting $I_{\mathrm{U}}(N) = o(\log
N)$ is weak at best, while a stronger case can be made on the basis of
the numerics that $I_{\mathrm{U}}(N) = O(\log N)$.

For $\Z^4$, the evidence from Figs.~\ref{fig:d4intersect} and
~\ref{fig:Dd4intersect} appears to support the statements $I_{\mathrm{S}}(N) =
\upomega(\log N)$ and $I_{\mathrm{U}}(N) = \upomega(\log N)$, 
as the plots for both $I_{\mathrm{S}}(N)$ and $I_{\mathrm{U}}(N)$ are convex at $N =
2^{28}-1$. This directly contradicts
the predictions that $I_{\mathrm{S}}(N) =
\Theta(\log N)$ and $I_{\mathrm{U}}(N) = o(\log
N)$.
However, this is not too surprising, as for $\Z^4$ we have $p = 0$
with logarithmic corrections, and so the asymptotic regime may not be
reached until $N$ is truly large, i.e. $N \gg 2^{28}$.

In summary, the numerical evidence supports our heuristic argument that
$T_{\mathrm{S}}(N) = \Theta(\log N)$ for $\Z^2$ and to lesser extent for
$\Z^3$. We argue
that the data for $I_{\mathrm{U}}(N)$ and $T_{\mathrm{U}}(N)$ 
imply $T_{\mathrm{U}}(N) = o(\log N)$ for $\Z^2$, and $T_{\mathrm{U}}(N) = O(\log N)$ 
for $\Z^3$; the numerics are
consistent with predictions that $T_{\mathrm{U}}(N) = O(1)$. 
However, the asymptotic regime has not yet been reached at $N = 2^{28}-1$, and
so no strong conclusion regarding $T_{\mathrm{U}}(N)$ can be made
based on the numerics. For $\Z^4$ 
the numerical evidence does not support the conclusions from our
heuristic arguments; we believe that this is due to the asymptotic
regime being well beyond $N = 2^{28}-1$.

We wish to make one final point: although the performance of an $O(1)$
implementation must
eventually be far superior to the performance of an $\Theta(\log N)$
implementation, we note that for $N$ as high as $2.7\e{8}$ the
speed-up gained for $\Z^2$ is only a factor of 1.65 for
\textbf{Attempt\-\_pivot\-\_fast} versus
\textbf{Attempt\-\_pivot\-\_simple}. This factor can be expected  
to grow, but even for walks on $\Z^2$ with $10^{12}$ steps,
\textbf{Attempt\-\_pivot\-\_fast} will be only a factor of two
faster than \textbf{Attempt\-\_pivot\-\_simple}.

\subsection{Summary}
\label{sec:complexitysummary}

Although $T(N)$ is asymptotically smaller for $\Z^2$ and $\Z^3$ than
for higher dimensions,
this does \emph{not} mean that our implementation is more
efficient for $\Z^2$ and $\Z^3$ than for higher dimensions, as it is
the integrated autocorrelation time in CPU units for global
observables which is important. The integrated autocorrelation time in
physical units (pivot attempts),
$\tau_{\mathrm{int}}(N)$, 
for observables such as $R_{\mathrm{e}}^2$ is of approximately the
same order as the 
time needed to achieve a successful pivot, perhaps with an additional
logarithmic factor~\cite{Madras1988}.
The best available evidence suggests that there is no logarithmic
factor for $\Z^2$ and $\Z^3$~\cite{Madras1988,Li1995}, but there is as
yet no evidence on this question for $\Z^d$ with $d \geq 4$.
We define 
${\widetilde{\tau}}_{\mathrm{int}}(N)$ as the integrated autocorrelation
time in CPU units, which can be expressed as
${\widetilde{\tau}}_{\mathrm{int}}(N) = \tau_{\mathrm{int}}(N) T(N)$.
Neglecting constant
factors, the relevant expressions
for $T(N)$ and ${\widetilde{\tau}}_{\mathrm{int}}(N)$ are therefore
\begin{align}
T(N) &= \Pr(\text{successful}) T_{\mathrm{S}}(N) + \Pr(\text{unsuccessful})
  T_{\mathrm{U}}(N); \\
{\widetilde{\tau}}_{\mathrm{int}}(N) &= \tau_{\mathrm{int}}(N) T(N)
\\ \nonumber
&= \frac{1}{\Pr(\text{successful})} T(N) \\ \nonumber
&= T_{\mathrm{S}}(N) + \frac{\Pr(\text{unsuccessful})}{\Pr(\text{successful})}
  T_{\mathrm{U}}(N).
\end{align}
This leads to the estimates for the complexity of $T(N)$ and
${\widetilde{\tau}}_{\mathrm{int}}(N)$ for $\Z^d$ given in Table~\ref{tab:performanceb}.
\begin{table}[!ht]
\caption{$T(N)$ and ${\widetilde{\tau}}(N)$ for $N$-step SAWs on
  $\Z^d$. The expressions for ${\widetilde{\tau}}_{\mathrm{int}}(N)$
  may have an additional logarithmic factor~\cite{Madras1988}. }
\label{tab:performanceb}
\begin{center}
\begin{tabular}{ccccc}
\hline\hline 
  & $\Z^2$ & $\Z^3$ & $\Z^4$ & $\Z^d$, $d > 4$ \tstrut \bstrut \\ \hline 
$T(N)$& $O(1)$ & $O(1)$ & $o(\log N)$ & $\Theta(\log N)$ \tstrut\\
${\widetilde{\tau}}_{\mathrm{int}}(N)$& $O(N^{0.19})$ & $O(N^{0.11})$ &
  $O(\log N)$ & $\Theta(\log N)$ \bstrut\\
\hline\hline
\end{tabular}
\end{center}
\end{table} 

We observe that our implementation of the pivot algorithm has a
crossover at dimension $d=4$: for $d < 4$, most CPU time is spent on
unsuccessful pivots, while for $d > 4$, most CPU time is spent on
successful pivots.

\section{Initialization}
\label{sec:initialization}

As discussed in \cite{Madras1988}, the pivot algorithm has short
integrated autocorrelation time of $O(N^p)$, but long exponential 
autocorrelation time of $O(N^{1+p})$ ($p$ positive, but close to
zero). As it is infeasible to initialize the Markov chain for large
$N$ by directly sampling from the equilibrium distribution via
dimerization~\cite{Alexandrowicz1969} (for $d \geq 4$
  dimerization is sufficiently efficient that it can be used even for
  very long walks), there will necessarily be
a systematic bias introduced from the initialization.
To ensure that the systematic error is much less than the statistical
error, it is necessary to discard a number of time steps which is
significantly larger than the exponential autocorrelation time. Madras
and Sokal~\cite{Madras1988} argued that the exponential
autocorrelation time is  
$O(N/f)$, where $f$ is the acceptance fraction of the pivot algorithm, and
discarded the first $20 N/f$ time steps. We adopt the same procedure
here. 
We note that for sufficiently large $N$ the time for initialization
can dominate the running time of the algorithm; for the longest walks
studied, initialization took approximately 2 weeks of computer time.

There is an additional complication for our implementation:
although we expect the mean time per pivot attempt for the SAW-tree
implementation to be $O(1)$, this is not necessarily true for
atypical SAWs. In particular, if the Markov chain is initialized with
a walk with long 
straight segments, such as a straight rod, then there can be 
$O(N)$ nearest neighbor contacts between two halves of a walk, and
so in the worst case a successful pivot may take time $O(N)$,
which is the 
same as the average-case performance of the implementation of Madras
and Sokal, and far worse than $O(\log N)$.
We have not precisely characterized the average-case behavior of the
SAW-tree 
implementation when initialized with a straight rod, but 
it is clear that pivot attempts are far slower when the walk has many
straight segments.
It may be of interest to study the typical performance of the
algorithm when initialized with a straight rod, e.g. to characterize
how rapidly the CPU time per pivot attempt decays to the equilibrium
value, and 
this will be done in a future computer experiment.

\begin{figure}[!ht]
\begin{center}
\resizebox{4.375in}{!}{\includegraphics{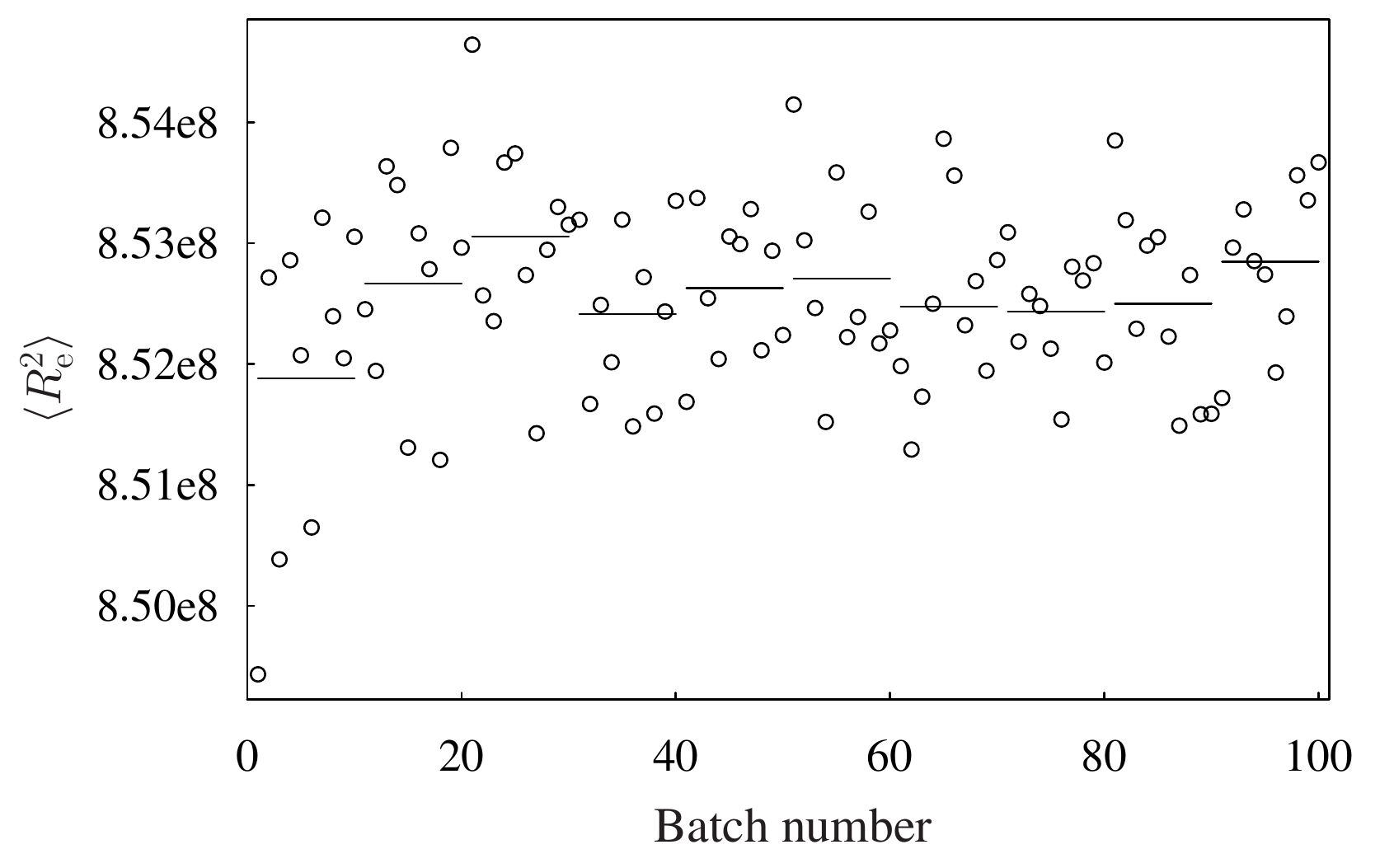}}
\end{center}
\caption{Initialization for 33554431-step SAWs on $\Z^3$, in batches of
$10^8$; means for batches of $10^9$
  configurations are shown as lines. The initial configuration was
  generated using \textbf{Pseudo\-\_dimerize}.}
\label{fig:initialization}
\end{figure}

To overcome the difficulty involving straight rods, we developed the
pseudo-dimerization procedure defined in
Sec.~\ref{sec:implementationuser}, 
\textbf{Pseudo\-\_dimerize}. This 
procedure utilizes repeated pivot operations to generate SAWs in time
$\Theta(N)$  
 which are quite difficult to distinguish from SAWs sampled from the uniform
distribution. 
For $N = 33554431$, $20 N/f$ corresponds to approximately 50 batches of
$10^8$. In
Fig.~\ref{fig:initialization}, 
initialization bias is visually apparent for (at most)
the first 10 batches, which suggests that discarding $20 N/f$
configurations is quite conservative when initializing the Markov
chain by using the
pseudo-dimerization procedure. 
If we could argue that the pseudo-dimerization samples from a
distribution which is in some sense close to the uniform distribution,
and if we could quantify this, then it might be possible to spend less
time on initialization. In the absence of any such argument, we
strongly recommend a cautious approach, i.e. that 
$20 N/f$ configurations should be discarded at the beginning of each data run.

Although this has not been carefully tested, it seems that there is no
significant increase in the time per attempted pivot during the
initialization period when the initial SAW is generated via the
pseudo-dimerization procedure.
It may appear that \textbf{Pseudo\-\_dimerize} is not strictly
necessary, 
but in practice it significantly extends the length of walks which can be
studied compared to initialization with a straight rod.

\subsection{Algorithmic complexity of \textbf{Pseudo\-\_dimerize}}

We will now provide a theoretical argument as to why pseudo-dimerization
takes time $\Theta(N)$. 

The mean CPU time needed for \textbf{Pseudo\-\_dimerize} to generate a SAW of
$n$ sites may be expressed recursively from the time required to
generate two SAWs of $n/2$ sites. We use the number of steps $N = n - 1$ in
the following expressions for consistency with other sections, and
denote the mean CPU 
time as $T_{\mathrm{PD}}(N)$. Thus we have:
\begin{align}
T_{\mathrm{PD}}(N) &= 2 T(N/2) + \text{(CPU time to merge
  walks)} \nonumber \\
&= 2 T_{\mathrm{PD}}(N/2) + (\text{CPU time per pivot attempt}) \nonumber \\
& \quad \quad \quad \quad \times (\text{Expected
  number of pivot attempts})
\end{align} 

In Sec.~\ref{sec:complexity} we predicted that the CPU time per pivot
attempt for $\Z^2$ and $\Z^3$ is
$O(1)$, and $O(\log N)$ for $\Z^d$ with $d \geq 2$. We will use the
weaker bound $O(\log N)$ for our argument here.
Calculating the expected number of pivot attempts is, however,  
quite subtle. Naively, we can assume that each of the
$n/2$ site walks has been sampled uniformly from SAWs of length $n/2$;
we expect that this approximation is ``good enough'', and will only
result in a small error that will not change our conclusions. 
The number of SAWs of $N$ steps for $\Z^2$ and $\Z^3$ is given
asymptotically by $c_n \sim A \mu^N N^{\gamma -1}$, and so
the probability of successfully merging two independent SAWs with $n/2$
sites is given by
\begin{align}
\Pr(\text{merge}) &=
\frac{c_{n-1}}{c_{n/2-1}^2} \nonumber \\
&\approx \frac{A \mu^{n-1} (n-1)^{\gamma - 1} }{\left(A \mu^{n/2-1}
  (n/2-1)^{\gamma - 1} \right)^2} \nonumber \\ 
&= O(N^{1-\gamma})
\label{eq:mergeprobability}
\end{align}
However, successive SAWs in the Markov chain are highly correlated,
and we account for this correlation by introducing
an observable $C$, defined as
\begin{align}
C(\omega^l,\omega^r) &= \begin{cases}
1 &\text{$\omega^l$ and $\omega^r$ can be merged}\\
0 & \text{$\omega^l$ and $\omega^r$ cannot be merged}.
\end{cases}
\end{align}

The expected number of pivot attempts before the left and right
sub-walks are successfully merged is $\tau_{\mathrm{int}}(C)/{\langle
  C \rangle}$, and we know that 
\begin{displaymath}
\langle C \rangle \equiv \Pr(\text{merge}) =
O(N^{1-\gamma}).
\end{displaymath}
In our implementation of \textbf{Pseudo\-\_dimerize} we perform an
additional $O(N^{1/2})$ pivot attempts after the sub-walks have been
merged, and hence
\begin{align}
T_{\mathrm{PD}}(N) &= 2 T_{\mathrm{PD}}(N/2) + O(\log N) (N^{\gamma -
  1} \tau_{\mathrm{int}}(C) + N^{1/2}) \nonumber \\
&= 2 T_{\mathrm{PD}}(N/2) + O(N^{\gamma - 1} \log N)
\tau_{\mathrm{int}}(C) + O(N^{1/2} \log N).
\end{align}

Eq.~\ref{eq:mergeprobability} leads to the conclusion that the mean distance of
the first intersection between the left- and right-hand sub-walks from
the joint where the walks meet is
$O(N^{2-\gamma})$. 
However, there
may well be other length scales which are also important for the
calculation of $\tau_{\mathrm{int}}(C)$; in particular, the shape of
the walks near the joint at distance scales from $O(1)$
up to $O(N^{2-\gamma})$ will strongly affect the probability of
successful merging.
This will be
discussed in much greater detail in a subsequent paper, in the context
of our calculation of the critical exponent $\gamma$ for self-avoiding
walks.

We expect that $\tau_{\mathrm{int}}(C)$ is of at most the same order as the time
needed to achieve a successful pivot on all possible length scales,
i.e. the time to achieve a successful pivot in each of the ranges, in
terms of distance from the joint, of
$[1,2),[2,4),[4,8),\cdots,[n/2,n-1)$.
By choosing pivot nodes using \textbf{Random\-\_integer\-\_log}, pivot
sites are chosen uniformly at random on a logarithmic scale in terms
of their distance from the joint, and so in
time $O(\log N)$ pivot \emph{attempts} are made on all length scales.
When selecting the pivot site uniformly at random (on a linear scale),
the probability of 
a successful pivot is $O(N^{-p})$. 
We expect that a pivot attempt is
more likely to be successful if the pivot site is
near one end rather than close to the middle, and thus
for pivots selected on a logarithmic scale
the probability of a successful pivot will remain $O(N^{-p})$.
Altogether, when using \textbf{Random\-\_integer\-\_log} to select pivot
sites, this implies that 
$\tau_{\mathrm{int}}(C) = O(N^p \log N)$.
Thus, our final recurrence relation is
\begin{align}
T_{\mathrm{PD}}(N) &= 2 T_{\mathrm{PD}}(N/2) + O(N^{\gamma - 1 + p}
\log^2 N) + O(N^{1/2} \log N). 
\end{align}

For $\lambda < 1$, the solution of the recurrence relation 
$T_{\mathrm{PD}}(N) = 2 T_{\mathrm{PD}}(N/2) + O(N^\lambda \log^2 N)+ O(N^{1/2} \log N)$ is
$T_{\mathrm{PD}}(N) = \Theta(N)$. 
The sequence of approximations used leads to the estimate $\lambda
\approx \gamma - 1 + p \approx 0.53$ for $\Z^2$ and $\lambda
\approx 0.27$ for $\Z^3$; we are confident that our
approximations are sufficiently 
accurate that the correct $\lambda$ will be less than one, and hence
that the expected running time of \textbf{Pseudo\-\_dimerize} is indeed
$\Theta(N)$. This argument can be straightforwardly repeated for $\Z^d$, $d
\geq 4$, leading to the same conclusion that $T_{\mathrm{PD}}(N) =
\Theta(N)$. 

We have clear numerical evidence from informal computer
experiments that this result is correct.

\section{Error estimates and the autocorrelation function}
\label{sec:autocorrelation}

Following \cite{Li1995}, and given the variance of an observable,
$\mathrm{var}(A) = \langle A^2 \rangle - \langle A \rangle^2$,
we define the autocorrelation function for 
the time series measurement of an observable $A$ as
\begin{align}
\rho_{AA}(t) &= \frac{\langle A_s A_{s+t} \rangle - \langle A
  \rangle^2}{\mathrm{var}(A)}.
\end{align}

\begin{figure}[!ht]
\begin{center}
\resizebox{4.375in}{!}{\includegraphics{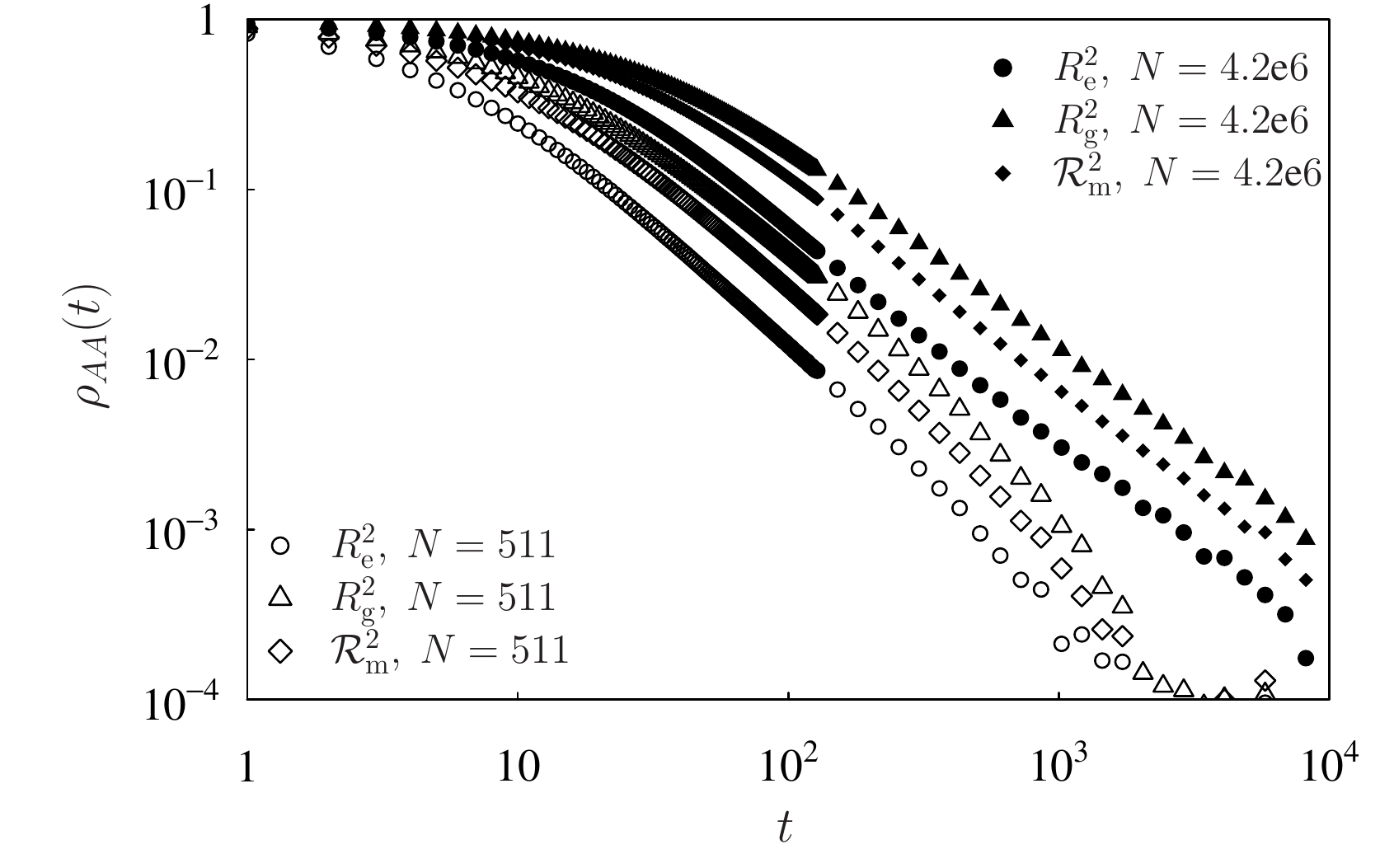}}
\end{center}
\caption{Autocorrelation function for observables $R_{\mathrm{e}}^2$,
  $R_{\mathrm{g}}^2$, and $\mathcal{R}_{\mathrm{m}}^2$. See Eq.~\ref{eq:calrm} for
  a definition of $\mathcal{R}_{\mathrm{m}}^2$; the key point is that
  $\langle \mathcal{R}_{\mathrm{m}}^2 \rangle = \langle
  R_{\mathrm{m}}^2 \rangle$ for all $N$.}
\label{fig:autocorrelation}
\end{figure}

We have calculated the autocorrelation function for the
Euclidean-invariant moments 
$R_x^{2k}$ with $x \in \{\mathrm{e,g,m}\}, 1 \leq k \leq 5$, for
$N=2^l - 1$, $9 \leq l \leq 22$, for times $t \leq 8192$. We invested
approximately 300 hours of CPU time in this endeavor, a relatively
small amount compared with the 16500 CPU hours spent on the computer
experiment to determine $\nu$ in \cite{Clisby2010}.
The autocorrelation functions for $R_x^{2}$, with $N = 511$ and $N =
2^{22}-1 = 4194303$, 
are shown in Fig.~\ref{fig:autocorrelation}. Error bars are
not shown on the graph, but the approximate size of the errors can be
inferred by
the degree of scatter from smooth behavior.

Despite the apparent linearity observed in 
Fig.~\ref{fig:autocorrelation} (particularly for $N=4.2\e{6}$), it is
surprisingly difficult to extract reliable estimates for the 
rate of decay of the tail of the autocorrelation function. Perhaps
this is because the tail is
not characterized by a single exponent. Madras and
Sokal~\cite{Madras1988} showed
that the pivot algorithm itself has a variety of exponents for the
acceptance fraction for different classes of lattice symmetries, and
this behavior may extend to the autocorrelation function itself.
This problem certainly deserves further study, but we do not have data
of sufficient quality to be able to accurately characterize the
autocorrelation function.

\begin{figure}[!ht]
\begin{center}
\resizebox{4.375in}{!}{\includegraphics{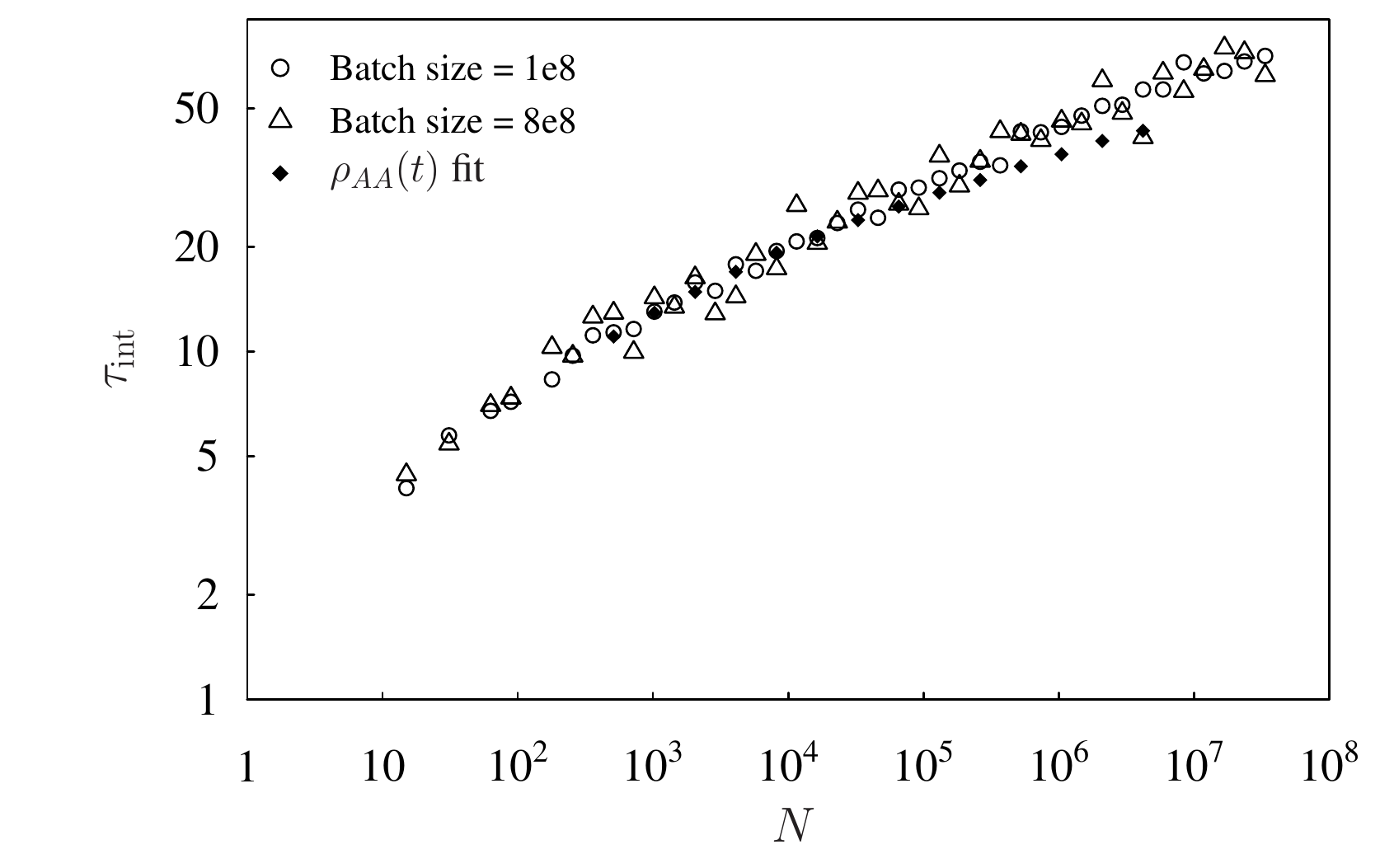}}
\end{center}
\caption{$\tau_{\mathrm{int}}$ for $R_{\mathrm{e}}^2$.}
\label{fig:tau_re}
\end{figure}

In Fig.~\ref{fig:tau_re}, we calculated the integrated autocorrelation time for
$R_{\mathrm{e}}^2$ using two different methods.

For the direct method, we calculated
\begin{align}
\tau_\mathrm{int}(A) &= \frac{1}{2} + \sum_{t=1}^{\infty} \rho_{AA}(t)
\end{align}
by using direct summation for short times ($t \leq 128$), and fitting
the intermediate regime
with a power law truncated at exactly $t=N/f$ (where $f$ is the fraction of
pivot attempts which are successful). The accuracy of this method
relies on the assumptions 
that the exponential autocorrelation time is of $O(N/f)$, and that the
intermediate regime of the autocorrelation function can be adequately fitted by a
single power law. We neglect the regime $t >  N /f$ where $\rho(t)$ is
presumed to decay exponentially, as the contribution of this tail to
$\tau_{\mathrm{int}}$ is negligible compared to the error introduced by
other approximations, e.g. the choice of $t = N/f$ rather than $t =
c N/f$ with $c \neq 1$.

The indirect method used the batch estimates for the observables, and
corresponding confidence intervals $\mathrm{stdev}(\bar{A})$, by solving
\begin{align}
\mathrm{stdev}(\bar{A}) &= \left(\frac{2\tau_{\mathrm{int}}(A)\mathrm{var}(A)}{n_{\text{sample}}}\right)^{\frac{1}{2}}
\end{align}
for $\tau_{\mathrm{int}}(A)$. The accuracy of this technique relies on the
assumption that the batch error estimate is accurate, which in turn
relies upon the degree of correlation between successive batches being
negligible. Provided the exponential autocorrelation time is finite
(guaranteed for a finite system), then this
condition will be satisfied for sufficiently large batch size.
If this condition were not satisfied then estimates of
$\mathrm{stdev}(\bar{A})$, and consequently $\tau_{\mathrm{int}}(A)$,
would be
systematically low.
To estimate $\tau_{\mathrm{int}}$ for $R_x^{2}$, with $x \in \{\mathrm{e,g,m}\}$,
we used data from the companion article~\cite{Clisby2010}, with
1000 batches of 
$10^8$ pivot attempts, and 125 batches of $8\e{8}$ pivot attempts.

The two indirect estimates shown in Fig.~\ref{fig:tau_re} 
are indistinguishable, although there is more scatter for the batches
of $8\e{8}$ because of the smaller number of batches used for the
estimate of $\mathrm{stdev}(\bar{A})$.
This
is strong evidence that a batch size of $10^8$ is sufficiently large 
so that the degree of correlation between successive batches is
negligible up to at least $N=2^{25}-1 \approx 3.36\e{7}$. Hence, we
expect the confidence intervals for estimates of observables in
\cite{Clisby2010} to be accurate, and recommend the batch method
for use with the pivot algorithm as
a simple and reliable method for estimating confidence intervals.

The direct
estimates for $\tau_{\mathrm{int}}$ are close to the indirect
estimates, but for sufficiently large $N$ the direct estimates
are systematically low. This suggests that either the tail fitting
procedure breaks down for large $N$, which we consider
unlikely as it is clear from Fig.~\ref{fig:autocorrelation} that the
tail has little curvature for large $N$. Or, for large $N$ the truncated
part of the tail still contributes non-negligibly to the integrated
autocorrelation time, i.e. the exponential autocorrelation time is
greater than $O(N/f)$. We consider this latter explanation to be more
probable, and will explore the asymptotic behavior of the exponential
autocorrelation time in future work.

We refer the interested reader to \cite{Madras1988,Li1995} for
more information on the 
autocorrelation function for the pivot algorithm.

\section{Performance: comparison with other implementations}
\label{sec:comparison}

In this section we present detailed comparison of the performance of
the SAW-tree implementation with the implementations of Madras and
Sokal~\cite{Madras1988} and Kennedy~\cite{Kennedy2002a}, for $N$-step
SAWs on $\Z^2$, $\Z^3$, and $\Z^4$, with $N$ ranging from 3 to 33554431.

For this section we will use the shorthand notations {\tt S-t} for
the SAW-tree implementation, {\tt M\&S} for the hash table implementation
of Madras and Sokal, and {\tt K} for Kennedy's implementation. 

\subsection{Experimental details}

For testing {\tt M\&S} we wrote our
own version using the programming language C. This implementation has
\emph{not} been extensively polished for maximum efficiency, and so it
is highly likely that there exist other implementations which are
faster by a (small) constant factor.

For testing {\tt K}, we used Kennedy's C++ program SAW\_pivot v1.0,
which has been released under the GNU General Public Licence.
For information about this implementation, we recommend you consult
the relevant article~\cite{Kennedy2002a}, as well as the source
code\footnote{Available at 
\if0\jsp 
http://math.arizona.edu/{\textasciitilde}tgk/.
\else \url{http://math.arizona.edu/~tgk/}. \fi}. 
We used the default settings for SAW\_pivot, which meant that
updates to the data structure were performed every
$N_{\mathrm{pivot}} = \lfloor (N/40)^{1/2} \rfloor$ successful pivots
(we set $N_{\mathrm{pivot}} = 1$ for $N \leq 40$). 
Kennedy~\cite{Kennedy2002a} indicates that
the performance 
of the algorithm is relatively insensitive to the precise choice of
$N_{\mathrm{pivot}}$, and so we expect that tuning
$N_{\mathrm{pivot}}$ would result in, at most, only modest improvement
in performance. 

We have observed that {\tt K} is faster when walks
are rod-like, presumably because the intersection testing algorithm is
more efficient when SAWs are spread out\footnote{Interestingly, this is in
stark contrast to the behavior of the SAW-tree implementation, where
long straight segments in rod-like walks result in significant
performance degradation, as explained in  
Sec.~\ref{sec:initialization}.}. For this reason the timing experiments
were initialized with SAWs sampled from the
equilibrium 
distribution, rather than straight rods; these SAWs were generated via
the pivot algorithm using the SAW-tree implementation.

The computer experiment was performed on a 64-bit Linux machine with Xeon
Barcelona 2.83GHz quad-core processors. The programs were compiled
with gcc version 4.1.2, with optimization flag ``-O3''; other
optimization flags made little if any difference.

\begin{figure}[bht]
\begin{center}
\resizebox{4.375in}{!}{\includegraphics{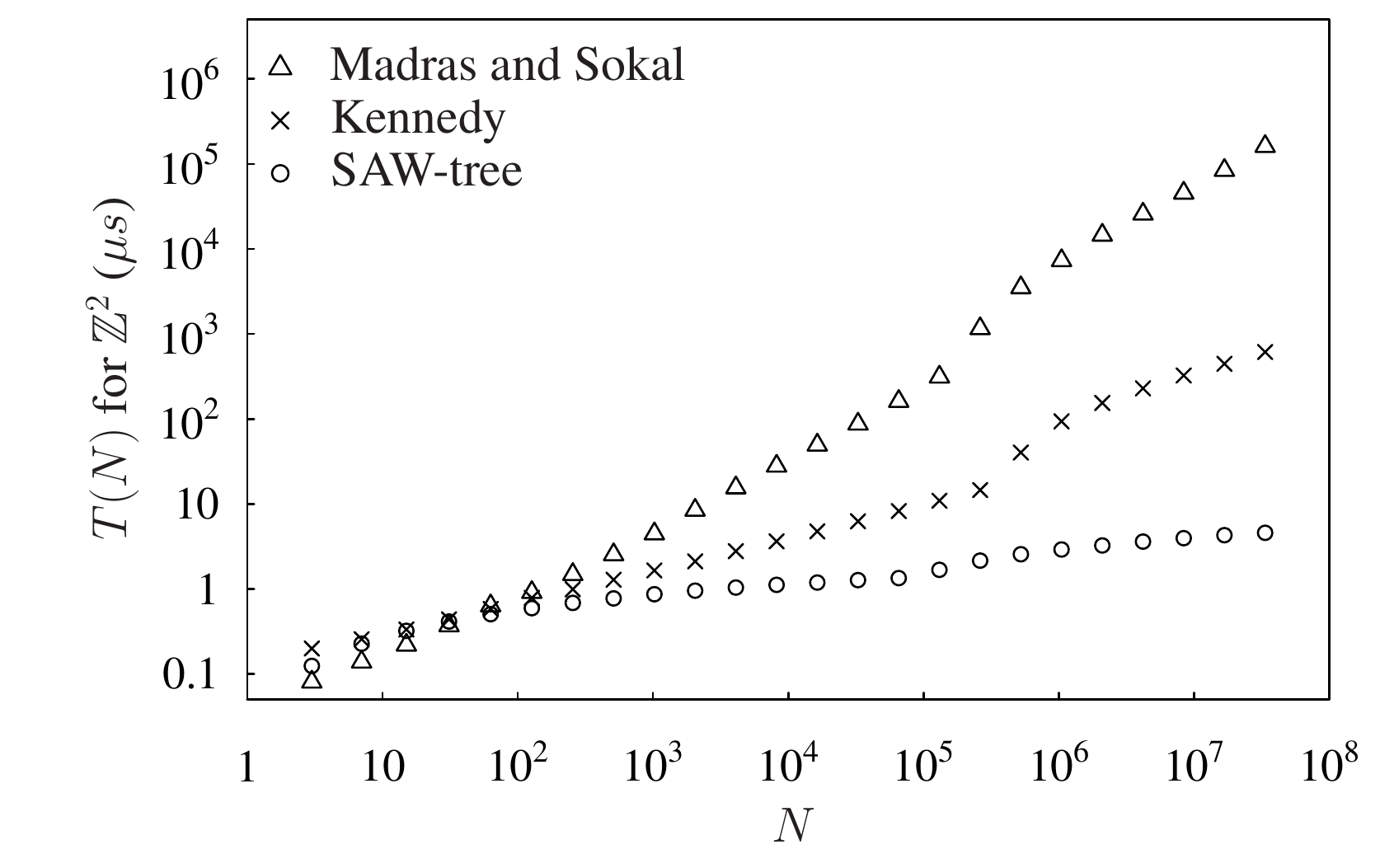}}
\end{center}
\caption{$T(N)$ for the Madras and Sokal, Kennedy,
  and SAW-tree implementations on $\Z^2$.}
\label{fig:cputimesd2}
\end{figure}

\begin{figure}[htb]
\begin{center}
\resizebox{4.375in}{!}{\includegraphics{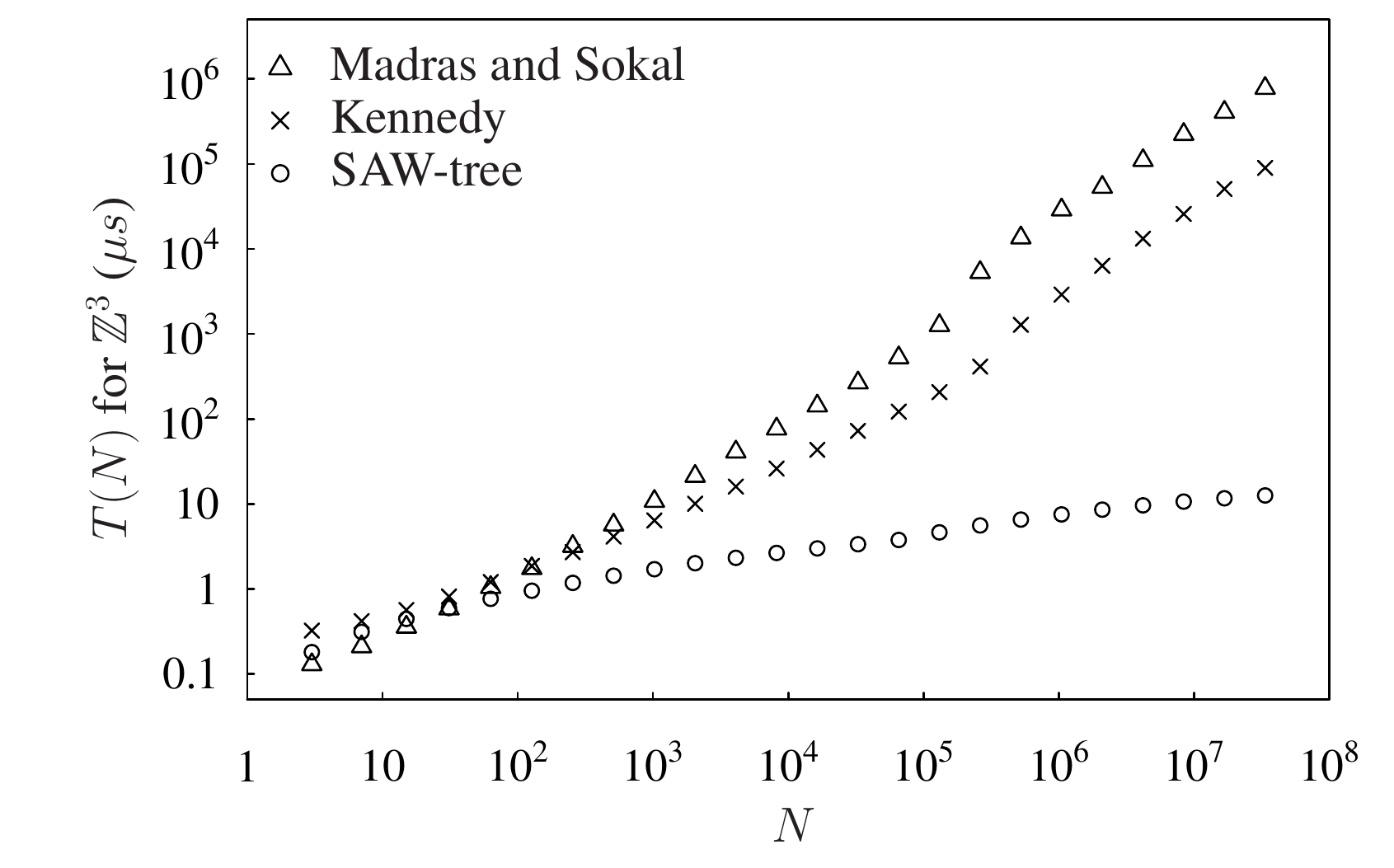}}
\end{center}
\caption{$T(N)$ for the Madras and Sokal, Kennedy,
  and SAW-tree implementations on $\Z^3$.}
\label{fig:cputimesd3}
\end{figure}

\begin{figure}[hbt]
\begin{center}
\resizebox{4.375in}{!}{\includegraphics{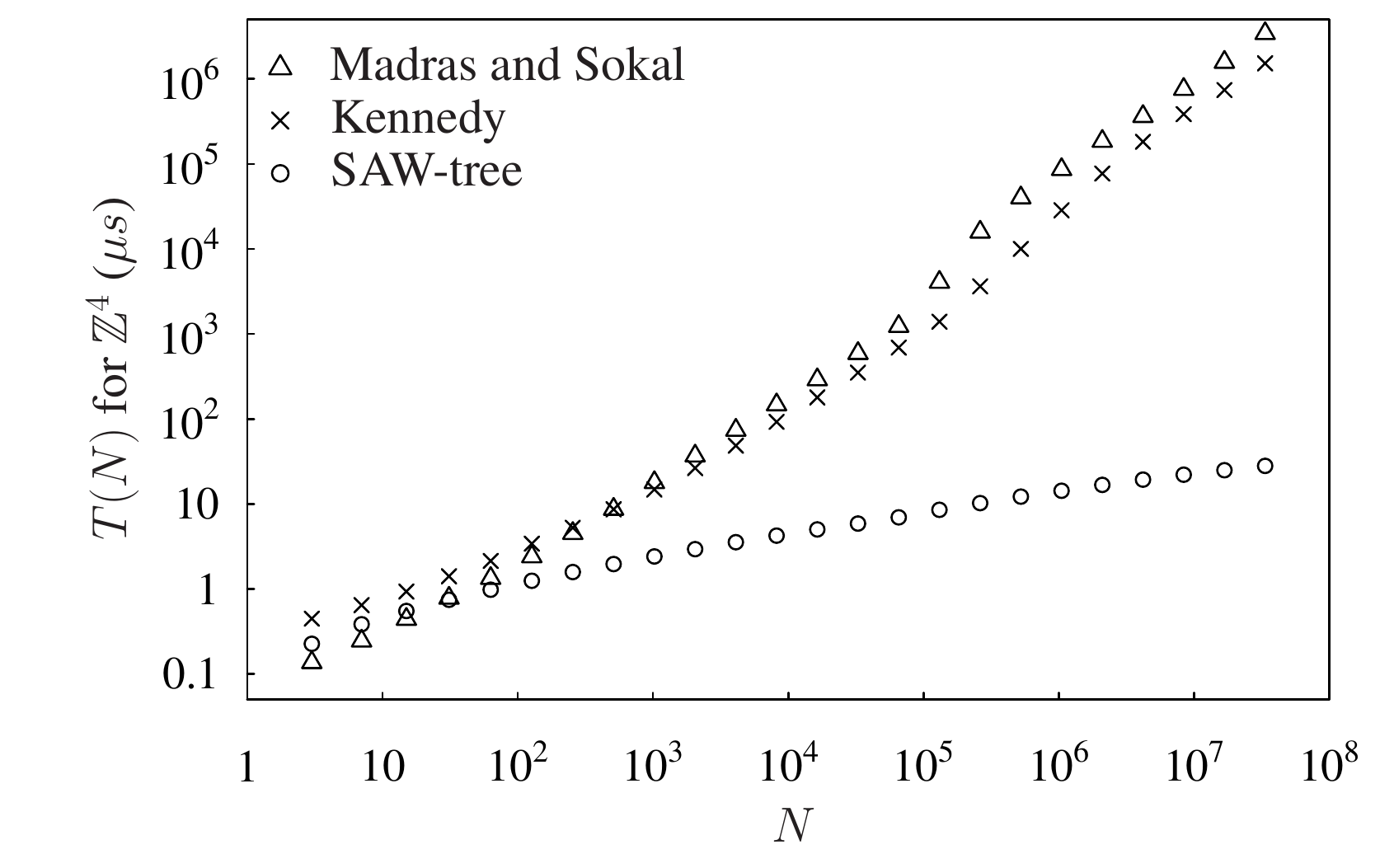}}
\end{center}
\caption{$T(N)$ for the Madras and Sokal, Kennedy,
  and SAW-tree implementations on $\Z^4$.}
\label{fig:cputimesd4}
\end{figure}

We do not include error bars for our measurements, but we
have repeated the experiment to ensure that these numbers are
reproducible, and have verified that the deviations between different
runs are quite small.
We present our results in Table~\ref{tab:comparison} of
Appendix~\ref{sec:comparisontable} and graphically in
Figs.~\ref{fig:cputimesd2}-\ref{fig:cputimesd4}. 


\subsection{Discussion}

In Fig.~\ref{fig:cputimesd2}, and to a lesser extent in
Figs.~\ref{fig:cputimesd3} and \ref{fig:cputimesd4}, a kink is visible
in each of the curves, indicating the length of walk where a hardware
limit is reached and the computer programs become memory bound. This
occurs at shorter lengths for {\tt S-t} and {M\&S}, as our
implementations of these 
algorithms use significantly more memory than {\tt K}.

In \cite{Clisby2010} we made the statement ``For SAWs of length $N =
10^6$ on 
the cubic lattice, the performance gain for our implementation is
approximately 200 when compared with Kennedy's, and over a 
thousand when compared with that of Madras and Sokal''.
This
statement was based on testing of an earlier version of our SAW-tree
implementation, on a different computer to the tests reported here,
and we regard the figures in
Table~\ref{tab:comparison} of Appendix~\ref{sec:comparisontable} as
more reliable. In this table, we
find that for SAWs on $\Z^3$ with $N = 1048576$, {\tt S-t} is 385 times
faster than {\tt K} and 3830  times faster than {\tt M\&S}.

For this computer experiment the only observable calculated was
$R_{\mathrm{e}}^2$; it is straightforward to extend this to other
observables such as $R_{\mathrm{g}}^2$ and $R_{\mathrm{m}}^2$ for the {\tt
  S-t} and {\tt M\&S} implementations for a 
constant factor penalty. It is likely also possible to do the same
thing for {\tt K}, but despite the clear performance advantage for
this algorithm over {\tt M\&S}, to the best of our knowledge this has
not been done.

We observe from the table and graphs that {\tt S-t} is lightweight,
as it is comparable with {\tt M\&S} for short walks with as few as 15
steps. For $\Z^2$, $\Z^3$, and $\Z^4$, {\tt S-t} is in fact faster
than the other implementations for 63 or more steps.

The difference between the implementations is particularly stark for
 $N \gtrsim 10^6$, where {\tt S-t} is significantly faster than the other
implementations in all dimensions; this improvement is quite dramatic
for $\Z^3$ and $\Z^4$.

On the basis of this computer experiment, where the precise timings
are  compiler
and machine dependent, we can nevertheless draw the following robust
conclusions: 
the SAW-tree implementation is efficient for short
walks, and much more powerful than other implementations for long
walks.
Compared with Kennedy's implementation, there is a large performance
boost for $\Z^2$, and a dramatic 
performance boost for $\Z^d$ with $d \geq 3$.

\appendix

\FloatBarrier

\section{Example SAW-trees}
\label{sec:examplesawtrees}
\tikzstyle{level 1}=[level distance=1.7cm, sibling distance=2.3cm]
\tikzstyle{level 2}=[level distance=1.7cm, sibling distance=2.3cm]
\tikzstyle{level 3}=[level distance=1.7cm, sibling distance=2.3cm]
\tikzstyle{level 4}=[level distance=1.7cm, sibling distance=2.3cm]
\tikzstyle{level 5}=[level distance=1.7cm, sibling distance=2.3cm]
\tikzstyle{level 6}=[level distance=1.7cm, sibling distance=2.3cm]
\begin{figure}[htb]
\begin{center}
  \begin{tikzpicture}[
      thick,scale=1.0,
      q/.style={rectangle,minimum height=4em,minimum
        width=4em,inner sep=2pt,draw,solid},
      leaf/.style={circle,minimum size=12mm,inner sep=1pt,draw,solid}
    ]
    \node[q] {\footnotesize 
      $q(1)$
    }
    child { node[leaf] {\footnotesize $\omega(0)$}}
    child {
      node[q] {\footnotesize $q(2)$}
      child { node[leaf] {\footnotesize $\omega(1)$}}
      child {
        node[q] {\footnotesize $q(3)$}
        child { node[leaf] {\footnotesize $\omega(2)$}}
        child[level distance=2.7cm, sibling distance=3.45cm] {
          node[q] {\footnotesize $q(n-2)$} edge from parent[dotted]
          child { node[leaf] {\footnotesize $\omega(n-3)$} edge from
            parent[solid]}
          child {
            node[q] {\footnotesize $q(n-1)$} edge from parent[solid]
            child { node[leaf] {\footnotesize $\omega(n-2)$}}
            child { node[leaf] {\footnotesize $\omega(n-1)$}}
          }
        }
      }
    };
  \end{tikzpicture}
\end{center}
\caption{SAW-tree which is precisely equivalent to the pivot sequence
  representation for a walk with $n$ sites. Note that $q(0)$ is an
  overall symmetry which is 
  applied to the whole walk, and cannot be directly included in the
  SAW-tree data structure.}
\label{fig:sawtree_sequence}
\end{figure}
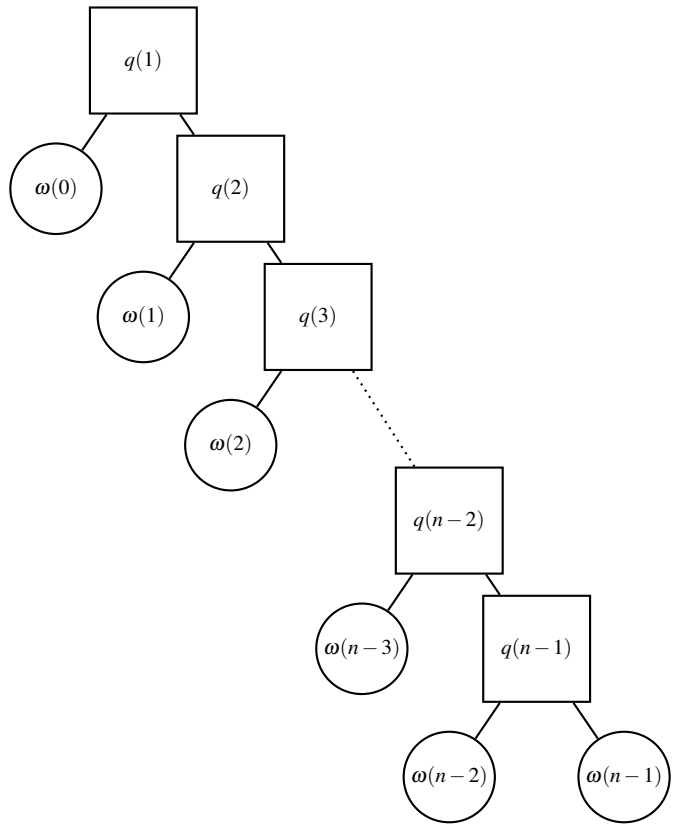

\tikzstyle{level 1}=[level distance=2.3cm, sibling distance=4.0cm]
\tikzstyle{level 2}=[level distance=2.3cm, sibling distance=2.7cm]
\tikzstyle{level 3}=[level distance=2.3cm, sibling distance=2.2cm]

\begin{figure}[htb]
\begin{center}
  \begin{tikzpicture}[
      thick,scale=1.0,
      q/.style={rectangle,minimum height=1.75em,minimum
        width=1.75em,inner sep=2pt,draw},
      leaf/.style={circle,minimum size=0mm,inner sep=1pt,draw}
    ]
    \node[q] {\footnotesize 
        \begin{tabular}{r@{=}l}
        $\mathbf{X}_{\mathrm{e}} $ & $ (3,0)$ \\ 
        $q $ & $\; q(3) = \left(\begin{array}{rr} 1 & 0 \\ 0 & 1 \\ \end{array}\right)$ \\
        $B $ & $ [1,3] \times [0,1]$ \\
       \end{tabular}
      }
    child {
      node[q] {\footnotesize 
        \begin{tabular}{r@{=}l}
          $\mathbf{X}^l_{\mathrm{e}}$ & $(2,1)$ \\ 
          $q^l $ & $\; q(2) = \left(\begin{array}{rr} 1 & 0 \\ 0 & 1 \\ \end{array}\right)$ \\
          $B^l $ & $ [1,2] \times [0,1]$ \\
        \end{tabular}
      }
      child {
        node[q] {\footnotesize
          \begin{tabular}{r@{=}l}
            $\mathbf{X}^{ll}_{\mathrm{e}} $ & $ (1,1)$ \\ 
            $q^{ll} $ & $ \; q(1) = \left(\begin{array}{rr} 0 & -1 \\ 1 & 0 \\ \end{array}\right)$ \\
            $B^{ll} $ & $ [1,1] \times [0,1]$ \\
          \end{tabular}
        }
        child { node[leaf] {\footnotesize $\omega_a(0)$} }
        child { node[leaf] {\footnotesize $\omega_a(1)$} }
      }
      child { node[leaf] {\footnotesize $\omega_a(2)$} }
    }
    child {
      node[q] {\footnotesize
        \begin{tabular}{r@{=}l}
          $\mathbf{X}^{r}_{\mathrm{e}} $ & $ (1,-1)$ \\ 
          $q^{r} $ & $ \; q(4) = \left(\begin{array}{rr} 0 & 1 \\ -1 & 0 \\ \end{array}\right)$ \\
          $B^{r} $ & $ [1,1] \times [-1,0]$ \\
        \end{tabular}
      }
      child { node[leaf] {\footnotesize $\omega_a(3)$} }
      child { node[leaf] {\footnotesize $\omega_a(4)$} }
    };
  \end{tikzpicture}
\end{center}
\caption{A SAW-tree representation of $\omega_a$ (from
  Fig.~\ref{fig:example}) involving proper 
  rotations only.}
\label{fig:sawtree_exampleaa}
\end{figure}
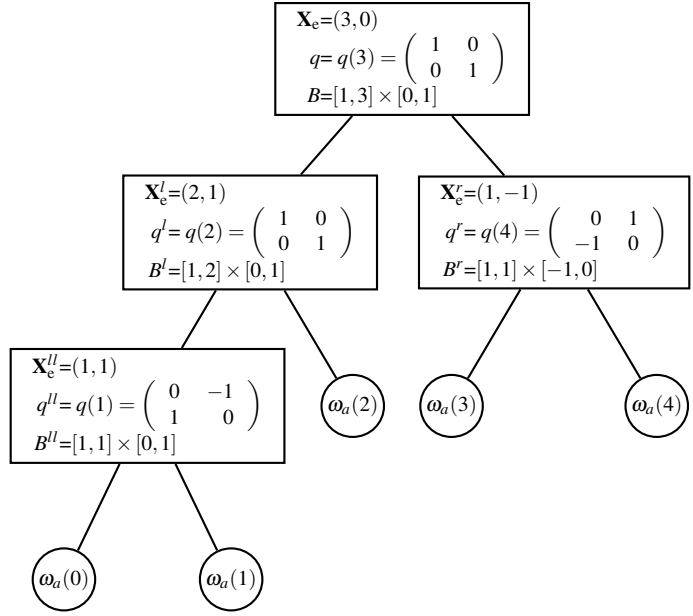

\begin{figure}[htb]
\begin{center}
  \begin{tikzpicture}[
      thick,scale=1.0,
      q/.style={rectangle,minimum height=1.75em,minimum
        width=1.75em,draw,inner sep=2pt},
      leaf/.style={circle,minimum size=0mm,inner sep=1pt,draw}
    ]
    \node[q] {\footnotesize 
        \begin{tabular}{r@{=}l}
        $\mathbf{X}_{\mathrm{e}} $ & $ (3,0)$ \\ 
        $q $ & $\; q(2) = \left(\begin{array}{rr} 1 & 0 \\ 0 & -1 \\ \end{array}\right)$ \\
        $B $ & $ [1,3] \times [0,1]$ \\
       \end{tabular}
      }
    child {
      node[q] {\footnotesize
        \begin{tabular}{r@{=}l}
          $\mathbf{X}^{l}_{\mathrm{e}} $ & $ (1,1)$ \\ 
          $q^{l} $ & $ \; q(1) = \left(\begin{array}{rr} 0 & 1 \\ 1 & 0 \\ \end{array}\right)$ \\
          $B^{l} $ & $ [1,1] \times [0,1]$ \\
        \end{tabular}
      }
      child { node[leaf] {\footnotesize $\omega_a(0)$} }
      child { node[leaf] {\footnotesize $\omega_a(1)$} }
    }
    child {
      node[q] {\footnotesize 
        \begin{tabular}{r@{=}l}
          $\mathbf{X}^r_{\mathrm{e}}$ & $(2,1)$ \\ 
          $q^r $ & $\; q(3) = \left(\begin{array}{rr} 1 & 0 \\ 0 & -1 \\ \end{array}\right)$ \\
          $B^r $ & $ [1,2] \times [0,1]$ \\
        \end{tabular}
      }
      child { node[leaf] {\footnotesize $\omega_a(2)$} }
      child {
        node[q] {\footnotesize
          \begin{tabular}{r@{=}l}
            $\mathbf{X}^{rr}_{\mathrm{e}} $ & $ (1,-1)$ \\ 
            $q^{rr} $ & $ \; q(4) = \left(\begin{array}{rr} 0 & 1 \\ -1 & 0 \\ \end{array}\right)$ \\
            $B^{rr} $ & $ [1,1] \times [-1,0]$ \\
          \end{tabular}
        }
        child { node[leaf] {\footnotesize $\omega_a(3)$} }
        child { node[leaf] {\footnotesize $\omega_a(4)$} }
      }
    };
  \end{tikzpicture}
\end{center}
\caption{A SAW-tree representation of $\omega_a$ (from
  Fig.~\ref{fig:example}) involving proper and improper 
  rotations.}
\label{fig:sawtree_exampleab}
\end{figure}
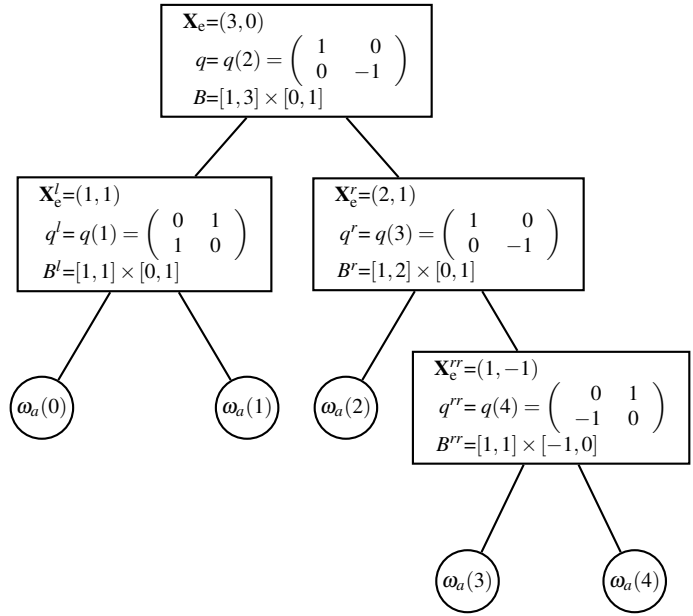

\FloatBarrier

\newpage

\section{Running times of SAW-tree, Madras and Sokal, and Kennedy
  implementations of the pivot algorithm}
\label{sec:comparisontable}

\begin{table}[!ht]
\centering
\hbox{
\rotcaption{\small $T(N)$ for the SAW-tree implementation ({\tt S-t})
  on $\Z^2$, $\Z^3$, and $\Z^4$,
  with relative 
  performance given for the implementations of Madras and
  Sokal ({\tt M\&S})~\cite{Madras1988}  and Kennedy ({\tt
    K})~\cite{Kennedy2002a}. These 
  numbers are given as a rough guide only, and are machine and
  compiler dependent.}
\label{tab:comparison}
\begin{sideways}
\begin{tabular}{r|rrr|rrr|rrr}
\hline\hline 
 & \multicolumn{3}{c|}{$\Z^2$} & \multicolumn{3}{c|}{$\Z^3$} &
\multicolumn{3}{c}{$\Z^4$ \tstrut} \\
$N$ & {\tt S-t} ($\mu s$) & {\tt M\&S}/{\tt S-t} & {\tt K}/{\tt S-t} &
{\tt S-t} ($\mu s$) &  {\tt M\&S}/{\tt S-t} & {\tt K}/{\tt S-t} & {\tt S-t}
($\mu s$) & {\tt M\&S}/{\tt S-t} & {\tt K}/{\tt S-t} \bstrut
\\ \hline 
3 & 0.12 & 0.647 & 1.6 & 0.18 & 0.715 & 1.79 & 0.23 & 0.602 & 1.98
\tstrut \\  
7 & 0.23 & 0.61 & 1.13 & 0.31 & 0.671 & 1.34 & 0.38 & 0.637 & 1.68 \\  
15 & 0.32 & 0.681 & 1.04 & 0.44 & 0.802 & 1.27 & 0.55 & 0.800 & 1.69 \\  
31 & 0.41 & 0.894 & 1.06 & 0.59 & 0.981 & 1.37 & 0.75 & 1.05 & 1.89 \\  
63 & 0.50 & 1.26 & 1.15 & 0.76 & 1.37 & 1.58 & 0.98 & 1.36 & 2.18 \\  
127 & 0.59 & 1.53 & 1.32 & 0.95 & 1.83 & 1.95 & 1.25 & 1.92 & 2.73 \\  
255 & 0.68 & 2.16 & 1.45 & 1.18 & 2.70 & 2.30 & 1.58 & 2.85 & 3.32 \\  
511 & 0.77 & 3.27 & 1.66 & 1.43 & 3.97 & 2.89 & 1.96 & 4.42 & 4.41 \\  
1023 & 0.87 & 5.15 & 1.90 & 1.71 & 6.31 & 3.75 & 2.41 & 7.41 & 6.13 \\  
2047 & 0.95 & 8.85 & 2.21 & 2.01 & 10.6 & 5.01 & 2.94 & 12.5 & 8.98 \\  
4095 & 1.04 & 14.9 & 2.68 & 2.32 & 17.7 & 6.90 & 3.55 & 20.8 & 13.7 \\  
8191 & 1.12 & 25.0 & 3.25 & 2.65 & 28.8 & 9.84 & 4.24 & 34.7 & 21.8 \\  
16383 & 1.19 & 41.5 & 4.00 & 3.00 & 47.5 & 14.4 & 5.01 & 57.8 & 35.8 \\  
32767 & 1.27 & 68.6 & 4.92 & 3.36 & 79.2 & 21.5 & 5.88 & 100 & 59.6 \\  
65535 & 1.34 & 120 & 6.13 & 3.77 & 139 & 32.3 & 6.95 & 177 & 99.3 \\  
131071 & 1.68 & 186 & 6.50 & 4.63 & 272 & 44.5 & 8.52 & 476 & 164 \\  
262143 & 2.15 & 536 & 6.76 & 5.58 & 949 & 74.0 & 10.23 & 1.54\e{3} & 355 \\  
524287 & 2.56 & 1.37\e{3} & 15.7 & 6.55 & 2.07\e{3} & 195 & 12.19 & 3.28\e{3} & 821 \\  
1048575 & 2.91 & 2.51\e{3} & 32.2 & 7.53 & 3.83\e{3} & 385 & 14.28 & 6.02\e{3} & 1.99\e{3} \\  
2097151 & 3.25 & 4.48\e{3} & 47.5 & 8.56 & 6.22\e{3} & 740 & 16.75 & 1.10\e{4} & 4.60\e{3} \\  
4194303 & 3.60 & 7.14\e{3} & 63.5 & 9.62 & 1.14\e{4} & 1.37\e{3} & 19.32 & 1.88\e{4} & 9.41\e{3} \\  
8388607 & 3.96 & 1.15\e{4} & 82.0 & 10.65 & 2.09\e{4} & 2.42\e{3} & 22.07 & 3.42\e{4} & 1.74\e{4} \\  
16777215 & 4.29 & 1.96\e{4} & 104 & 11.65 & 3.49\e{4} & 4.36\e{3} & 24.93 & 6.31\e{4} & 2.97\e{4} \\  
33554431 & 4.57 & 3.52\e{4} & 134 & 12.58 & 6.17\e{4} & 7.13\e{3} &
28.03 & 1.22\e{5} & 5.44\e{4} \bstrut \\  
\hline\hline
\end{tabular}
\end{sideways}
}
\end{table} 

\FloatBarrier

\if0\jsp \begin{acknowledgements}
\else \vspace{0.2cm} \noindent {\bf Acknowledgments} \fi
I thank Ian Enting, Tony Guttmann, Gordon Slade, Alan Sokal, and two
anonymous referees for useful comments on the manuscript.
I would also like to thank an
anonymous referee for comments on an earlier version of this article
which led to deeper consideration of the algorithmic complexity of 
\textbf{Shuffle\-\_intersect}. I am grateful to Tom Kennedy for releasing his
implementation of the pivot algorithm under the GNU GPLv2 licence.
Computations were performed using the
resources of the Victorian Partnership for Advanced Computing (VPAC).
Financial support from the Australian Research Council is gratefully
acknowledged. 
\if0\jsp 
\end{acknowledgements}
\else
{}
\fi


\begin{thebibliography}{10}
\providecommand{\url}[1]{{#1}}
\providecommand{\urlprefix}{URL }
\expandafter\ifx\csname urlstyle\endcsname\relax
  \providecommand{\doi}[1]{DOI~\discretionary{}{}{}#1}\else
  \providecommand{\doi}{DOI~\discretionary{}{}{}\begingroup
  \urlstyle{rm}\Url}\fi

\bibitem{Alexandrowicz1969}
Alexandrowicz, Z.: {M}onte {C}arlo of chains with excluded volume: a way to
  evade sample attrition.
\newblock J. Chem. Phys. \textbf{51}, 561--565 (1969)

\bibitem{Baiesi2001}
Baiesi, M., Orlandini, E., Stella, A.L.: Peculiar scaling of self-avoiding walk
  contacts.
\newblock Phys. Rev. Lett. \textbf{87}, 070602 (2001)

\bibitem{Caracciolo2005}
Caracciolo, S., Guttmann, A.J., Jensen, I., Pelissetto, A., Rogers, A.N.,
  Sokal, A.D.: Correction-to-scaling exponents for two-dimensional
  self-avoiding walks.
\newblock J. Stat. Phys. \textbf{120}, 1037--1100 (2005)

\bibitem{Clisby2010}
Clisby, N.: Accurate estimate of the critical exponent $\nu$ for self-avoiding
  walks via a fast implementation of the pivot algorithm.
\newblock Phys. Rev. Lett. \textbf{104}, 055702 (2010)

\bibitem{vanEmdeBoas1975}
van Emde~Boas, P.: Preserving order in a forest in less than logarithmic time.
\newblock In: Proceedings 16th Annual Symposium on Foundations of Computer
  Science, pp. 75--84 (1975)

\bibitem{Frigo1999}
Frigo, M., Leiserson, C.E., Prokop, H., Ramachandran, S.: Cache oblivious
  algorithms.
\newblock In: Proceedings 40th Annual Symposium on Foundations of Computer
  Science, pp. 285--297 (1999)

\bibitem{Gabay1978}
Gabay, M., Garel, T.: Renormalization along the chemical sequence of a single
  polymer chain.
\newblock J. Physique Lett. \textbf{39}, 123--125 (1978)

\bibitem{Guttman1984}
Guttman, A.: R-trees: {A} dynamic index structure for spatial searching.
\newblock In: B.~Yormark (ed.) SIGMOD '84, pp. 47--57. ACM Press, New York
  (1984)

\bibitem{Hara1992}
Hara, T., Slade, G.: Self-avoiding walk in five or more dimensions {I.} {T}he
  critical behaviour.
\newblock Commun. Math. Phys. \textbf{147}, 101--136 (1992)

\bibitem{Kennedy2002a}
Kennedy, T.: A faster implementation of the pivot algorithm for self-avoiding
  walks.
\newblock J. Stat. Phys. \textbf{106}, 407--429 (2002)

\bibitem{Klosowski1998}
Klosowski, J.T., Held, M., Mitchell, J.S.B., Sowizral, H., Zikan, K.: Efficient
  collision detection using bounding volume hierarchies of $k$-dops.
\newblock IEEE T. Vis. Comput. Gr. \textbf{4}, 21--36 (1998)

\bibitem{Kremer1981}
Kremer, K., Baumg\"artner, A., Binder, K.: {M}onte {C}arlo renormalization of
  hard sphere polymer chains in two to five dimensions.
\newblock Z. Phys. B: Condens. Matt. \textbf{40}, 331--341 (1981)

\bibitem{Kumar2003}
Kumar, P.: Cache Oblivious Algorithms, chap.~9, pp. 193--212.
\newblock Springer Berlin / Heidelberg (2003)

\bibitem{Lal1969}
Lal, M.: `{M}onte {C}arlo' computer simulation of chain molecules. {I}.
\newblock Mol. Phys. \textbf{17}, 57--64 (1969)

\bibitem{Lawler2004}
Lawler, G.F., Schramm, O., Werner, W.: On the scaling limit of planar
  self-avoiding walk.
\newblock In: Fractal Geometry and Applications: a Jubilee of {B}enoit
  {M}andelbrot, Part 2. Proc. Sympos. Pure Math., vol.~72, pp. 339--364. Am.
  Math. Soc., Providence (2004)

\bibitem{Li1995}
Li, B., Madras, N., Sokal, A.D.: Critical exponents, hyperscaling, and
  universal amplitude ratios for two- and three-dimensional self-avoiding
  walks.
\newblock J. Stat. Phys. \textbf{80}, 661--754 (1995)

\bibitem{Madras1993}
Madras, N., Slade, G.: The Self-Avoiding Walk.
\newblock Birkha\"user, Boston (1993)

\bibitem{Madras1988}
Madras, N., Sokal, A.D.: The pivot algorithm: {A} highly efficient {M}onte
  {C}arlo method for the self-avoiding walk.
\newblock J. Stat. Phys. \textbf{50}, 109--186 (1988)

\bibitem{Muller1998}
M{\"u}ller, S., Sch{\"a}fer, L.: On the number of intersections of
  self-repelling polymer chains.
\newblock Eur. Phys. J. B \textbf{2}, 351--369 (1998)

\bibitem{Nienhuis1982}
Nienhuis, B.: Exact critical point and critical exponents of {O($n$)} models in
  two dimensions.
\newblock Phys. Rev. Lett. \textbf{49}, 1062--1065 (1982)

\bibitem{Oono1979}
Oono, Y.: Renormalization along the polymer chain.
\newblock J. Phys. Soc. Jpn \textbf{47}, 683--684 (1979)

\bibitem{Sedgewick1998Books}
Sedgewick, R.: Algorithms in {C}, {P}arts 1--4, {T}hird edn.
\newblock Addison-Wesley (1998)

\bibitem{Sokal1994}
Sokal, A.D.: {M}onte {C}arlo methods for the self-avoiding walk.
\newblock arXiv:{\tt hep-lat/9405016} (1994)

\bibitem{Sokal1996}
Sokal, A.D.: {M}onte {C}arlo methods for the self-avoiding walk.
\newblock Nucl. Phys. B (Proc. Supp.) \textbf{47}, 172--179 (1996)

\end{thebibliography}

\if0\jsp

\else

\providecommand{\bysame}{\leavevmode\hbox to3em{\hrulefill}\thinspace}
\providecommand{\MR}{\relax\ifhmode\unskip\space\fi MR }
\providecommand{\MRhref}[2]{%
  \href{http://www.ams.org/mathscinet-getitem?mr=#1}{#2}
}
\providecommand{\href}[2]{#2}

\fi

\end{document}